\documentclass[a4paper,11pt]{article}
\usepackage{jcappub}
\usepackage{lineno}
\usepackage{comment}
\usepackage{bm}
\usepackage{braket}
\usepackage{xcolor}
\usepackage{booktabs}
\usepackage{adjustbox}
\arxivnumber{1234.56789} 

\title{\boldmath Modeling and measuring the anisotropic halo 3-point correlation function: a coordinated study}

\author[1,2,3]{A. Farina}
\author[2,3,1]{A. Veropalumbo}
\author[1,2,3]{E. Branchini}
\author[4,5,6]{M. Guidi}
\affiliation[1]{Dipartimento di Fisica, Università degli Studi di Genova, Via
Dodecaneso 33, I-16146, Genova, Italy}
\affiliation[2]{INAF - Osservatorio Astronomico di Brera, Via Brera 28, 20122 Milano, via E. Bianchi 46, 23807 Merate, Italy}
\affiliation[3]{INFN - Sezione di Genova, Via
Dodecaneso 33, I-16146, Genova, Italy}
\affiliation[4]{Dipartimento di Fisica e Astronomia “Augusto Righi”–Università di Bologna, via Piero Gobetti 93/2, I-40129 Bologna, Italy}
\affiliation[5]{INAF - Osservatorio di Astrofisica e Scienza dello Spazio di Bologna, via Piero Gobetti 93/3, I-40129 Bologna, Italy}
\affiliation[6]{Physics Department, Technion, 3200003 Haifa, Israel}

\emailAdd{antonio.farina@inaf.it}

\newcommand{\diff}{\mathrm{d}}
\newcommand{\hMpc}{\,h^{-1} \mathrm{Mpc}}
\newcommand{\hGpc}{\,h^{-3} \mathrm{Gpc}^{3}}
\newcommand{\meas}[1]{\tilde{#1}}

\abstract{Apparent anisotropies in the statistical properties of the spatial galaxy distribution encode precious cosmological information. Its extraction is commonly performed using 2-point clustering statistics. However, ongoing and future spectroscopic galaxy surveys will cover unprecedented volumes with a number of objects large enough to effectively probe clustering anisotropies through higher-order statistics.\\
In this work, we present a novel and efficient implementation of both a model for the multipole moments of the anisotropic 3-point correlation function (3PCF) and of their estimator. To evaluate the performance of our model, we compared its predictions against direct 3PCF measurements obtained with our estimator from a set of 298 dark matter halo catalogs drawn from the $z=1$ snapshots of $N$-body simulations. For the statistical analysis, we employed a covariance matrix estimated from an independent suite of 3000 mock halo catalogs at the same redshift.
We then repeated the analysis by combining the 2-point correlation function (2PCF) to the 3PCF, with and without including its anisotropic part.
In the 3PCF-only analysis, the addition of the anisotropic component of the 3PCF effectively
breaks the degeneracy between the growth rate $f$ and the linear bias $b_1$, significantly reducing their uncertainties. It also significantly improves the precision of the 
Alcock-Paczy\'{n}ski parameter $\varepsilon$ but does not reduce the $\sim 1$ \% offset we find in the estimate of the isotropic dilation parameter $\alpha$.
The joint 2PCF+3PCF analysis reduces, though does not fully remove, biases in the AP and isotropic dilation parameters and breaks the $f$–$b_1$–$\sigma_8$ degeneracy, leading to tighter constraints overall. The anisotropic 3PCF adds little to the joint analysis because the tree-level 3PCF model fails to capture the anisotropic information primarily encoded on small scales and in squeezed triangle configurations. A more advanced model, e.g. based on 1-loop perturbation theory, will be required to exploit this information fully.}

\begin{document}
\maketitle
\flushbottom

\section{Introduction}
\label{sec:intro}
Studying the clustering of galaxies has proven to be one of the most effective ways to trace the evolution of the Universe and that of cosmic structures and to investigate the nature of its components.
Consequently, several worldwide observational campaigns, such as {\it DESI} \citep{DESI:2016,DESI:2016b}, {\it Euclid} \citep{Laureijs:2011, Mellier:2024} and {\it Roman} Telescope \citep{Akeson:2019} have been, or are being, designed to 
fully exploit the potential of this cosmological probe. \\
The common goal of these projects is to map the three dimensional positions of galaxies and, thus, to probe the distribution of matter over
increasingly large regions of the Universe and across a significant fraction of its evolution history.\\
Precise 3D mapping of large galaxy samples requires measuring angular positions and spectroscopic redshifts.
These latter are used to infer radial distances, under the assumption that the measured line shifts are caused by the recession velocity of the source. However, extragalactic objects also possess peculiar velocities, whose line-of-sight (LOS) component contributes to the measured redshift through classical Doppler effect.
Ignoring peculiar velocities induces systematic distortions along the radial direction in the derived 3D galaxy distribution, commonly referred to as Redshift Space Distortions (RSD). These distortions break the statistical isotropy of the 3D galaxy distribution and modify its clustering statistics, as first illustrated in the seminal works of 
\cite{Jackson:1972, Kaiser:1987, Hamilton:1998}.\\
Despite considerably complicating clustering analyses, RSD are a very valuable cosmological probe. On large scales, their magnitude is directly proportional to the rate at which cosmic structures evolve, which in turn depends on the amount and type of Dark Energy present in the system (see e.g. \cite{Linder:2005}).
While RSD have been exploited to derive effective cosmological constraints for the first time by \cite{Peacock:2001}
using the 2dF galaxy survey \citep{Colless:2001}, 
their first application to measuring the linear growth rate of density fluctuations dates back to \cite{Guzzo:2008}, who analyzed the 
clustering of VVDS galaxies at a redshift close to unity \citep{Garilli:2008}.
Since then, RSD in 2-point galaxy clustering statistics, either in configuration (the 2-point correlation function, 2PCF) and Fourier space (the power spectrum), have become a standard cosmological probe.\\
If the matter density field were Gaussian, as it is indeed the case for the primordial Universe in many inflationary scenarios \citep{Starobinsky:1980, Guth:1980, Sato:1981, Linde:1982}
and observed in the CMB temperature map \citep{Akrami:2020}, then 
2-point statistics would be sufficient for a thorough statistical description of the distribution of matter.
However, the mass density field traced by galaxies is far from being Gaussian for several reasons: the possible presence of non-Gaussian features in the primordial field, the nonlinear evolution of density perturbations, and the galaxy-to-matter relation, commonly known as 'galaxy bias'. All these non-Gaussian effects contribute to transferring part of the cosmologically relevant information from 2-point to higher-order statistics, particularly to the 3-point one, which can be measured in configuration space as the galaxy 3-point correlation function (3PCF) or in Fourier space as the galaxy bispectrum.\\
Two and three-point statistics complement each other effectively: their combination 
removes degeneracy among parameters and consequently improves the precision in the estimate of the quantities of interests (see e.g. \cite{Sefusatti:2006, Moresco:2014, Yankelevich:2019}). As a result, 3-point statistics have been recognized as fundamental tools to extract cosmological information from the analysis of the Large Scale Structure (LSS) of the Universe. 
\\
So far, the majority of these analyses have focused on the bispectrum. The motivation for this preference is twofold. First of all, the matter and the galaxy bispectrum
can be effectively modeled using Eulerian perturbation theory 
(i.e. \cite{Bernardeau:2002, Crocce:2006, Bernardeau:2008, Bernardeau:2012}), Lagrangian perturbation theory \citep{Matsubara:2008a, Matsubara:2008b, Carlson:2013, Wang:2014}, and Effective Field Theory (EFT) \cite{Baumann:2012, Carrasco:2012, Porto:2014, Angulo:2015, Senatore:2014, Senatore:2015, Perko:2016, Lewandowski:2018}.
Second of all, efficient estimators have been developed to measure the bispectrum of large galaxy samples at a relatively small computational cost (see e.g. \cite{Bernardeau:2002} and references therein).\\
On the other hand, a common drawback of Fourier space-based analyses is the complex geometry of galaxy surveys, often characterized by intricate footprints with irregular boundaries, gaps, and holes. This complexity introduces couplings between Fourier modes, leading to distortions in the shape of the bispectrum.
To account for these couplings, one must either deconvolve the measured bispectrum or convolve the model with survey-specific window functions \cite{Philcox:2021, Pardede:2022}.
Both strategies are prone to uncertainties that can impact the estimation of cosmological parameters. \\
In contrast, footprint-related effects are automatically accounted for when the analysis is performed in configuration space. For the 2 and 3PCF the Szapudi-Szalay estimator \cite{Szapudi:1998} efficiently corrects for boundary and selection effects.\\
Despite these advantages, the limited computational efficiency and challenges in the modeling have long hindered 3PCF-based analyses \cite{Gaztanaga:2005, Pan:2005, McBride:2011, Moresco:2017}. This has changed over the past years, as highly efficient 3PCF estimators have reduced the computational cost from an  $N^3$ scaling to an $N^2$ one, making it possible to analyze catalogs with $N\sim O(10^6-10^7)$ galaxies \cite{Slepian:2015, Slepian:2018, Sugiyama:2019, Wang:2023}. 
This computational breakthrough has triggered a rapid development in the modeling of the 3PCF \cite{Slepian:2017, Kuruvilla:2020}, culminating to the next-to-leading order 3PCF model of \cite{Guidi:2023}, and has spurred a series of analyses \cite{Slepian:2015v, Slepian:2017L, Veropalumbo:2021, Moresco:2021} crowned by the first detection of the Baryon Acoustic Oscillations (BAO) feature in the 3PCF from BOSS DR12 data \cite{Slepian:2017B}.\\
However, the vast majority of the 3PCF analyses performed so far have exclusively dealt with  the isotropic component of this statistic, i.e. its average across all the possible LOS directions. The additional information contained in the higher-order multipoles, which is expected to be sensitive to all sources of anisotropies, including RSD, is yet to be fully exploited. Results from pioneering analyses have been obtained by \cite{Sugiyama:2019, Sugiyama:2021, Umeh:2021, Sugiyama:2023, Sugiyama:2023c} who proposed an optimal decomposition formalism, leveraging on the Tripolar Spherical harmonics (TripoSH) expansion \cite{Varshalovich:1988} and the FFTLog algorithm \cite{Hamilton:2000, Fang:2020}, for effectively computing and quantifying the anisotropic 3PCF.\\
One of the goal of this work is to further these recent studies by introducing and offering:

\begin{itemize}
    \item \texttt{Mod3L} \footnote{\href{https://gitlab.com/antoniofarina/mod3l}{https://gitlab.com/antoniofarina/mod3l}}, an original and fast implementation of the perturbation-theory model introduced by \cite{Sugiyama:2021} for the anisotropic 3PCF.
    
    \item \texttt{MeasCorr} \footnote{
    In this work, \texttt{MeasCorr} is applied to simulated datasets with periodic boundary conditions. However, the estimator is applicable to measuring the 3PCF of any sample of objects with arbitrary geometry and selection functions. The code is publicly available at 
    \href{https://gitlab.com/veropalumbo.alfonso/meascorr}{https://gitlab.com/veropalumbo.alfonso/meascorr}}, to the best of our knowledge, the first public implementation of the estimator proposed by \cite{Slepian:2018} in its pair-counting version.
\end{itemize}

\noindent To test the validity of these tools, and, as a second goal, to assess the performance of an anisotropic 3PCF analysis in configuration space, we rely on a large suite of simulated catalogs with an aggregate volume of about
$1000\hGpc$, far exceeding the volumes of any ongoing or planned cosmological spectroscopic survey.
This is the same simulated dataset used by \cite{Oddo:2020, Oddo:2021, Rizzo:2023}, who performed similar analyses in Fourier space. Using the same data ensures a meaningful comparison between the two approaches.\\
Following \cite{Veropalumbo:2022}, which this work extends methodologically, we first perform a 3PCF-only analysis and then a joint 2PCF+3PCF analysis, both with and without the inclusion of anisotropic multipoles. 
The article is organized as follows. In Sec.~\ref{sec:model} we describe the 2 and 3PCF models adopted in this work, we review the properties of the TripoSH expansion and lay out our computational strategy. In Sec.~\ref{sec:datasets} we provide a brief overview of the simulated datasets that we use to validate our models and estimators and evaluate their uncertainties. Estimators and measurements are presented in Sec.~\ref{sec:measurements}, alongside the covariance matrices utilized in the likelihood analysis of Sec.~\ref{sec:parinf}. The results of our 3PCF and 2+3PCF analyses are outlined in Sec.~\ref{sec:results}, while in Sec.~\ref{sec:conclusions} we draw our conclusions and discuss future developments.

\section{Theoretical 2PCF and 3PCF models}
\label{sec:model}
In this section we present the theoretical models for the halo 2 and 3PCF employed in our analysis and the strategies adopted to compute them.\\
To map the observed halo over-densities into their corresponding matter ones, we build upon the bias and redshift space expansions proposed respectively by \citep{Assassi:2014}
and \citep{Scoccimarro:1999}. Given this mapping, we first model 2PCF and 3PCF multipoles in Fourier space in the framework of the EFT of LSS \cite{Baumann:2012, Carrasco:2012, Porto:2014, Angulo:2015, Senatore:2014, Senatore:2015, Perko:2016, Lewandowski:2018} and, then, utilize FFTLog-based algorithms to efficiently compute their configuration space counterparts \cite{Hamilton:2000, Fang:2020, Umeh:2021, Guidi:2023}.
To do that, following \cite{Veropalumbo:2021}, we expand 2PCF and 3PCF at different perturbative orders. Specifically, while for the 2PCF we build upon a next-to-leading order expansion, in the 3PCF case we stop at the tree-level \citep{Scoccimarro:1999, Sugiyama:2021}.
Finally, to deal with the high dimensionality of three-point statistics in redshift-space we adopt the tripolar spherical harmonics (TripoSH) expansion, originally introduced by \citep{Varshalovich:1988} and extensively discussed by \citep{Sugiyama:2019}.\\

\subsection{Density field expansion}
\label{sub:2.1}
It has long been realized that using spectroscopic redshifts as a distance proxy
introduces anisotropic distortions in the clustering pattern that must be accurately taken into account \citep{Jackson:1972, Kaiser:1987, Hamilton:1998, Scoccimarro:1999}.
The observed position $\bm{s}$ of cosmic structures (halos in our case) is indeed displaced from the real one, $\bm{x}$, by the radial component of their peculiar velocity $\delta\bm{v}$:

\begin{equation}
\bm{s}=\bm{x}+\frac{\delta\bm{v}\cdot\bm{\hat n}}{a\,H(a)}\,\bm{\hat n}
\label{eq:RSpos}
\end{equation}

\noindent where $a$ is the expansion factor, $H$ the Hubble parameter and $\bm{\hat n}$ the LOS direction. By requiring that the halo number is conserved, this mapping allows us to relate the measured halo contrast $\delta_\mathrm{h}(\bm{s}, \bm{\hat{n}})$ to the real-space one, $\delta_\mathrm{h}(\bm{x})$:

\begin{equation}
\delta_\mathrm{h}(\bm{s}, \bm{\hat n})=\left[1+\delta_\mathrm{h}(\bm{x})\right]\,\left|\frac{d^3s}{dx^3}\right|^{-1}-1
\label{eq:delta_mapping}
\end{equation}

\noindent where $\left|\frac{d^3s}{dx^3}\right|$ is the Jacobian of the map defined by Eq.~(\ref{eq:RSpos}). 
Under the assumption that pair separations are smaller than their distances to the observer, we assume the plane parallel approximation and a common LOS to all pairs. Thus, Eq.~(\ref{eq:delta_mapping}) becomes:

\begin{equation}
\delta_\mathrm{h}(\bm{s}, \bm{\hat n})=\frac{\delta_\mathrm{h}(\bm{x})+f\partial_{\bm{\hat n}} \bm{u}}{1-f\partial_{\bm{\hat n}} \bm{u}}
\label{eq:delta_mapping2}
\end{equation}

\noindent where $\bm{u}=\frac{\delta\bm{v}}{f\,a\,H(a)}$ and $\partial_{\bm{\hat n}}$ indicates the partial derivative along the radial direction.\\
Finally, we must account for the relationship between halos and the underlying distribution of dark matter. To link the halo density contrast $\delta_h$ to the matter non-linear overdensity $\delta$ \citep{McDonald:2009, Baldauf:2012, Assassi:2014, Mirbabayi:2015, Desjacques:2018}, we adopt the renormalized bias expansion model proposed by \citep{Assassi:2014}:

\begin{equation}
    \delta_\mathrm{h} = b_1\delta + \frac{b_2}{2}\delta^2 + b_{\mathcal{G}_2}\mathcal{G}_2 + b_{\Gamma_3}\Gamma_3
\end{equation}

\noindent where we ignore all the higher-order terms that do not contribute to our 2PCF and 3PCF models. Here, the $\mathcal{G}_2$ and $\Gamma_3$ operators are defined as a combination of quadratic tidal tensors of the form $\partial_i\partial_j\phi$:

\begin{equation}
\begin{split}
    \mathcal{G}_2(\phi_g) &= \left( \partial_i\partial_j\phi_g\right)^2 - \left(\partial^2\phi_g\right) \\
    \Gamma_3(\phi_g, \phi_v) &= \mathcal{G}_2(\phi_g) - \mathcal{G}_2(\phi_v)  
\end{split}
    \label{eq:G2}
\end{equation}

\noindent where $\phi_g$ and $\phi_v$ represent the gravitational and velocity potential as defined in Eqs.~(2.3) and (2.4) of \citep{Assassi:2014}, respectively.

\subsection{2PCF model} 
To model the halo 2PCF, we start from the redshift-space halo power spectrum:

\begin{equation}
    \braket{\delta_\mathrm{h}(\bm{k_1}, \bm{\hat n})\delta_\mathrm{h}(\bm{k_2}, \bm{\hat n})} = (2\pi)^3\delta_D(\bm{k_1}+\bm{k_2})P(\bm{k_1}, \bm{\hat n})
\end{equation}

\noindent where $\delta_\mathrm{h}(\bm{k}, \bm{\hat n})=\int{d^3s\;\delta_\mathrm{h}(\bm{s},  \bm{\hat n})\,e^{-i\bm{k}\cdot\bm{s}}}$ is the Fourier transform of the redshift-space halo density contrast and $\delta_D$ indicates the Dirac delta distribution.
Using Eq.~(\ref{eq:delta_mapping2}) and following the prescriptions of the EFT of LSS, we can express the power spectrum at the next-to-leading order as:

\begin{equation}
\begin{split}
    P_{\mathrm{nlo}}(k,\mu) = Z_1^2(\bm{k}, \bm{\hat n})P_{\mathrm{lin}}(k)\; +\;
             &2\int{\frac{d^3q}{(2\pi)^3}\,Z_2^2(\bm{q}, \bm{k-q}, \bm{\hat n})P_{\mathrm{lin}}(|\bm{k-q}|)P_{\mathrm{lin}}(q)} \;+\\
             +\;&6\,Z_1(\bm{k},\bm{\hat n})P_{\mathrm{lin}}(k)\int{\frac{d^3q}{(2\pi)^3}Z_3(\bm{q}, \bm{-q}, \bm{k}, \bm{\hat n})P_{\mathrm{lin}}(q)}\; +\\
             -\;&2\left[c_0\;+\;c_2\mathcal{L}_2(\mu)\;\right]\,k^2P_{\mathrm{lin}}(k)\;+\\
             -\;&c_{\mathrm{nlo}}f^4\mu^4Z_1(\bm{k}, \bm{\hat n})k^4P_{\mathrm{lin}}(k)
\end{split}    
    \label{eq:Pk_NoIR}
\end{equation}

\noindent where $P_{\mathrm{lin}}(k)$ denotes the linear matter power spectrum, $\mathcal{L}_{\ell}(\mu)$ is the Legendre polynomial of order $\ell$ and $\mu=\frac{\bm{k}\cdot\hat n}{k}$ represents the cosine angle between the LOS and the wavevector. The $Z_n$ functions indicate the standard Eulerian perturbation theory kernels, which encode the information about non-linear evolution, RSD and the bias mapping of the halos. Their analytic expression
can be found in appendix A.2 of \cite{philcox:2022B}. 
Finally, $c_0$, $c_2$ and $c_{\mathrm{nlo}}$ represent the so-called EFT counter-terms, i.e. coupling coefficients introduced to describe departures from ideal behavior in the fluid equation, which correct for UV divergences in the loop integrals (see e.g. \cite{Lewandowski:2018} and references therein).\\
Despite being a large scale feature, the BAO are affected by nonlinear effects, which we account for by adopting the infrared resummation (IR)
approach \citep{Crocce:2008, Matsubara:2008a, Blas:2016, Ivanov:2018}. 
In particular, we rely on the strategy proposed by \citep{Ivanov:2018}, which separates the power spectrum in a smooth (no-wiggle) and a wiggly component and damps this latter. In practice, we implement this resummation scheme in Eq.~(\ref{eq:Pk_NoIR}) by replacing the linear matter power spectrum with 

\begin{equation}
    P_{\mathrm{lin}}(k) \longrightarrow P_{\mathrm{nw}}(k) + D^2(k,\mu)P_\mathrm{w}(k).
    \label{eq:Pk_IR}
\end{equation}

\noindent in the tree-level and counterterm components, and with

\begin{equation}
    P_{\mathrm{lin}}(k) \longrightarrow P_{\mathrm{nw}}(k) + \left[1+k^2\Sigma_{\mathrm{tot}}^2(\mu) \right]D^2(k,\mu)P_\mathrm{w}(k).
    \label{eq:Pk_IR2}
\end{equation}

\noindent within the loop integrals. The $D(k,\mu)$ function is a Gaussian damping factor that accounts for the non-linear degradation of the BAO features  \citep{Eisenstein:2007, Crocce:2008, Matsubara:2008a}:

\begin{equation}
    \begin{split}
        D(k,\mu) &= e^{-\frac12 k^2\Sigma_{\mathrm{tot}}^2(\mu)} \\
        \Sigma_{\mathrm{tot}}^2(\mu) &= \Sigma^2\left[1+f\mu^2(2+f )\right] + \delta\Sigma^2f^2\mu^2(\mu^2 - 1)
    \end{split}
    \label{eq:Dk}
\end{equation}

\noindent $\Sigma^2$ and $\delta\Sigma^2$ are smoothing parameters that we compute using the Zel'dovich approximation \cite{Zel'dovich:1970}:

\begin{equation}
    \begin{split}
        \Sigma^2 =&\; \frac{1}{6\pi^2}\int_0^{k_S}{dq\;P_{\mathrm{nw}}(q)\left[1- j_0(q\,r_{BAO}) + 2j_2(q\,r_{BAO}) \right]}\\
        \delta\Sigma^2=&\; \frac{1}{2\pi^2}\int_0^{k_S}{dq\;P_{\mathrm{nw}}(q)j_2(q\,r_{BAO})}
    \end{split}
    \label{eq:Sigmas}
\end{equation}

\noindent where $j_n(x)$ are spherical Bessel functions, $r_{BAO}=110\hMpc$ is the sound-horizon scale and the cutoff wavelength $k_S$ is set equal to $0.2\hMpc$
as in \citep{Blas:2016, Sanchez:2017, Guidi:2023}.\\
From a computational perspective, Eq.~(\ref{eq:Pk_NoIR}) shows that obtaining the linear and counterterm components of the power spectrum is relatively straightforward, whereas the evaluation of the loop integrals requires much greater care. Although recent years have seen the widespread adoption of efficient semi-analytical techniques based on FFTLog~\cite{Simonovic:2018}, we adopt a more conservative approach here. Indeed, we numerically integrate the next-to-leading-order terms up to the cutoff scale of $k_{\mathrm{cutoff}}=100\,h\,\mathrm{Mpc}^{-1}$ using \texttt{CUBA}~\cite{Hahn:2005}, as implemented in \texttt{MelCorr}\footnote{\href{https://gitlab.com/veropalumbo.alfonso/melcorr}{https://gitlab.com/veropalumbo.alfonso/melcorr}}, a \texttt{Python}-wrapped \texttt{C++} package for 2PCF calculations that we make publicly available.
Once the reference power spectrum model has been estimated, we expand it into multipoles using Legendre polynomials:

\begin{equation}
    P_{\ell}^{\mathrm{nlo}}(k) = \frac{2\ell+1}{2}\int_{-1}^{1}{d\mu\; P_{\mathrm{nlo}}(k, \mu)\mathcal{L}_{\ell}(\mu)}.
    \label{eq:Pl}
\end{equation}

\noindent

\noindent From these multipoles, we compute the corresponding 2PCF multipoles as

\begin{equation}
    \xi_{\ell}^{\mathrm{nlo}}(r) = i^{\ell} \int_0^\infty{\frac{dk}{2\pi^2}\,k^2P_{\ell}^{\mathrm{nlo}}(k)\,j_{\ell}(kr)}
    \label{eq:fftlog}
\end{equation}

\noindent using the FFTLog algorithm \citep{Hamilton:2000}, in the implementation of \cite{Karamanis:2021}.

\subsection{3PCF model}
Analogously to the 2PCF case, we model the halo 3PCF by starting from its Fourier-space counterpart. We define the halo bispectrum as

\begin{equation}
\langle \delta_\mathrm{h}(\bm{k_1, \bm{\hat n}})\delta_\mathrm{h}(\bm{k_2}, \bm{\hat n})\delta_\mathrm{h}(\bm{k_3}, \bm{\hat n})\rangle = (2\pi)^3\delta_D(\bm{k_1}+\bm{k_2}+\bm{k_3})\,B(\bm{k_1},\bm{k_2},\bm{\hat n})
\label{eq:bispectrum_def}
\end{equation}

\noindent where the wavevector $\bm{k_3}$ is implicitely defined by the triangle condition $\bm{k_1}+\bm{k_2}+\bm{k_3}=0$. 
Following \cite{Scoccimarro:1999}, we express the bispectrum at the tree-level perturbative order as:

\begin{equation}
B(\bm{k_1},\bm{k_2},\bm{\hat n})= 2\,Z_1(\bm{k_1},\bm{\hat n})Z_1(\bm{k_2},\bm{\hat n})Z_2(\bm{k_1},\bm{k_2},\bm{\hat n})P_{\mathrm{lin}}(k_1)P_{\mathrm{lin}}(k_2)+ \mathrm{cycl}.
\label{eq:bisp}
\end{equation}

\noindent with "$\mathrm{cycl.}$" denoting the sum over the cyclic permutations of $\{\bm{k_1},\bm{k_2},\bm{k_3}\}$.\\
Analogously to the power spectrum case, we model the degradation of the BAO signal through IR resummation. In particular, we adopt the IR-resummed redshift-space bispectrum proposed by \cite{Sugiyama:2021}:

\begin{equation}
    \begin{split}
        B(\bm{k_1},\bm{k_2},\bm{\hat n})=& \;2\,Z_1(\bm{k_1},\bm{\hat n})Z_1(\bm{k_2}, \bm{\hat n})Z_2(\bm{k_1},\bm{k_2},\bm{\hat n})\\
        &\left\{D(k_1,\mu_1)D(k_2, \mu_2)D(k_3, \mu_3)P_\mathrm{w}(k_1)P_\mathrm{w}(k_2)+\right.\\
        &\left.+D^2(k_1,\mu_1)P_\mathrm{w}(k_1)P_{\mathrm{nw}}(k_2)+D^2(k_2, \mu_2)P_\mathrm{w}(k_2)P_{\mathrm{nw}}(k_1)+\right. \\
        &\left. +P_{\mathrm{nw}}(k_1)P_{\mathrm{nw}}(k_2)\right\} + cycl \,.
    \end{split}
\label{eq:bispIR}
\end{equation}

\noindent This expression clearly shows that the redshift-space bispectrum (and therefore the corresponding 3PCF) depends on both shape and orientation of the triangles w.r.t. the LOS, identified by the three unit vectors $\bm{\hat k_1}$, $\bm{\hat k_2}$ and $\bm{\hat n}$. To deal with this triple angular dependence, we adopt the tripolar spherical harmonics (TripoSH) expansion \citep{Sugiyama:2019}. In this formalism, assuming statistical homogeneity, isotropy and parity symmetry, the bispectrum can be expressed as

\begin{equation}
B(\bm{k_1},\bm{k_2},\bm{\hat n})= \sum_{\ell_1+\ell_2+L\,\in\, 2 \mathbb{N}}{B_{\ell_1\,\ell_2\,L}(k_1,k_2)\,S_{\ell_1\,\ell_2\,L}(\bm{\hat k_1},\bm{\hat k_2}, \bm{\hat n})}
\label{eq:TripoSH}
\end{equation}

\noindent with

\begin{equation}
S_{\ell_1\,\ell_2\,L}(\bm{\hat k_1},\bm{\hat k_2}, \bm{\hat n})=\frac{1}{H_{\ell_1\,\ell_2\,L}}\sum_{m_1\,m_2\,M}{\begin{pmatrix} \ell_1 & \ell_2 & L \\ m_1 & m_2 & M \end{pmatrix}\;y_{\ell_1}^{m_1}(\bm{\hat k_1}) y_{\ell_2}^{m_2}(\bm{\hat k_2}) y_{L}^{M}(\bm{\hat n})}
\label{eq:TSHbasis}
\end{equation}

\noindent where $H_{\ell_1\,\ell_2\,L}=\left(\begin{smallmatrix} \ell_1 & \ell_2 & L \\ 0 & 0 & 0 \end{smallmatrix}\right)$ and the circle brackets with 6 multipole indices denote Wigner-$3j$ symbols. The $y_\ell^m$ functions are Schmidt semi-normalized spherical harmonics.\\
One can show that the TripoSH formalism represents a generalization of the Legendre polynomials expansion proposed by \cite{Szapudi:2004} 
in real-space. Its multipole moments
are equivalent to tripolar $L=0$ modes after averaging over all possible $\bm{\hat n}$ directions. 
This implies that the isotropic component of the 3-point statistics in redshift space is fully characterized by $L=0$ modes, whereas modes with  with $L>0$ capture the anisotropic information content. \\
As apparent from Eqs.~(\ref{eq:TripoSH}) and (\ref{eq:TSHbasis}), the TripoSH decomposition formalism does not depend on the choice of the coordinate system. This leaves us the freedom 
to choose different coordinate axes to describe the 3PCF. 
In this work, we adopt the coordinate system of \cite{Scoccimarro:1999}, for which $\bm{k_1} \parallel \bm{\hat z}$ and: 

\begin{equation}
\begin{cases} \bm{k_1}=(0,\;0,\;k_1) \\ \bm{k_2}=(k_2\sin{\theta},\;0,\;k_2\cos{\theta}) \\ \bm{\hat n}=(\sqrt{1-\mu_1^2}\,\sin{\varphi},\;\sqrt{1-\mu_1^2}\,\cos{\varphi},\;\mu_1) \end{cases}
\label{eq:Scoord1}
\end{equation}

\noindent where $\varphi$ is the azimuthal angle between $\bm{\hat x}$-axis and the projection of $\bm{\hat n}$ on the $\bm{\hat x}$-$\bm{\hat y}$ plane.\\
To measure the 3PCF multipole moments, instead, we adopt the coordinate system introduced by \cite{Slepian:2018}, in which  $\bm{\hat z}$ axis is aligned with the LOS. \\
An additional benefit of the TripoSH expansion is that it allows one to relate the multipole moments of the bispectrum to those of the 3PCF via a 2D Hankel transform:

\begin{equation}
\zeta_{\ell_1\,\ell_2\,L}(r_1,r_2)=i^{\ell_1+\ell_2}\int{\frac{dk_1}{2\pi^2}\frac{dk_2}{2\pi^2}\;k_1^2k_2^2\,j_{\ell_1}(k_1r_1)j_{\ell_2}(k_2r_2)\,B_{\ell_1\,\ell_2\,L}(k_1,k_2)}.
\label{eq:TriRelBtoZ}
\end{equation}

\noindent
Accurate numerical evaluation of this transform is computationally challenging due to the oscillatory nature of Bessel functions, which prevents efficient use of standard quadrature techniques.
Therefore, following \citep{Umeh:2021, Guidi:2023}, we adopt the 2D FFT-Log algorithm \citep{Fang:2020}, which enables the evaluation of Eq.~(\ref{eq:TriRelBtoZ}) with a computational cost scaling as  $\mathcal{O}(N^2\log{N})$, where $N^2$ is the total number of grid points used to sample the bispectrum multipoles.
To do this, we adopt a $256\times256$ grid, logarithmically sampling the interval $[k_{\mathrm{min}},k_{\mathrm{max}}]=[10^{-4},10]\hMpc$. 
To reduce the ringing effect, we zero-pad both small and large wavevectors sides with $N_{\mathrm{pad}}=200$. We then use the bin-averaged output of 2D FFT-Log \citep{Fang:2020} to compare model predictions with measurements. \\
Our theoretical predictions for the 3PCF are obtained with \texttt{Mod3L}, a new and publicly available\footnote{\href{https://gitlab.com/antoniofarina/mod3l}{https://gitlab.com/antoniofarina/mod3l}} \texttt{Python} package designed to provide a fast and flexible framework for computing the 3PCF. \texttt{Mod3L} takes as input a model bispectrum, which depends on five variables, decomposes it into Tripolar spherical harmonics, and then uses 2D FFTLog to compute the 3PCF multipoles.
To optimize performance and speed up the calculation, the TripoSH expansion is performed in two steps. In the first step, the bispectrum is expanded in standard spherical harmonics using \texttt{HEAlPix} \citep{Gorski:2005, Zonca:2019}, In the second step, the TripoSH multipole moments are evaluated through Eq.~(25) of \cite{Sugiyama:2019}.

\subsection{Alcock-Paczy\'{n}ski and isotropic dilation effects}
RSDs are not the only source of anisotropy in the spatial distribution of matter tracers. Another source of distortion is the Alcock-Paczyński (AP) and \textbf{isotropic dilation} effects \cite{Alcock:1979}, which arise when distances are inferred using an incorrect mapping from redshift space to physical space. These produce systematic deformations in the observed clustering pattern, which can be described by the following transformations of wavevectors and cosine angles \citep{Padmanabhan:2008}:

\begin{align}
    k' &= \left( \frac{1 + \varepsilon}{\alpha} \right) k
    \left[ 1 + \mu^2 \left( (1 + \varepsilon)^{-6} - 1 \right) \right]^{\frac{1}{2}}, \label{eq:k_prime} \\[3mm]
    \mu' &=\mu\; (1 + \varepsilon)^{-3}\left[ 1 + \mu^2 \left( (1 + \varepsilon)^{-6} - 1 \right) \right]^{-\frac{1}{2}}. \label{eq:mu_prime}
\end{align}
\vspace{0.1cm}

\noindent Geometric distortions are quantified by two parameters: the isotropic dilation factor $\alpha$, rescaling distances uniformly, and the anisotropic warping parameter $\varepsilon$, introducing direction-dependent distortions.
These parameters affect the clustering statistics by altering both the amplitude and shape of the power spectrum and bispectrum multipoles. \\
Given the coordinate transformations in Eqs.~(\ref{eq:k_prime}) and (\ref{eq:mu_prime}), the power spectrum and bispectrum multipole moments incorporating the AP and isotropic dilation effects are given by

\begin{equation} P_\ell^{\mathrm{obs}}(k) =\; \frac{2\ell+1}{\alpha^3}\int{d\mu\; P(k', \mu')\mathcal{L}_\ell(\mu)}, \end{equation}

\begin{equation} B_{\ell_1,\ell_2,L}^{\mathrm{obs}}(k_1, k_2) =\; \frac{4 \pi H_{\ell_1 \ell_2 L }^2}{\alpha^6} \int \frac{d^2 \hat{k}_1}{4 \pi} \int \frac{d^2 \hat{k}_2}{4 \pi} \int \frac{d^2 \hat{n}}{4 \pi} \; S_{\ell_1 \ell_2 L} (\hat{k}_1, \hat{k}_2, \hat{n}) B(\mathbf{k_1'}, \mathbf{k_2'}, \mathbf{\hat n}). \end{equation}

\noindent As seen in these expressions, $\alpha$ affects the overall normalization of the power spectrum and bispectrum, wheras $\varepsilon$ introduces anisotropic distortions that modify their angular dependence. These effects must be accounted for to accurately interpret clustering measurements in LSS analyses.

\subsection{2PCF and 3PCF models implementation}
\label{sec:mod_implementation}
To model the halo 2PCF and 3PCF, we adopt a template-fitting approach, fixing all cosmological parameters that characterize the linear matter power spectrum, $P_{\mathrm{lin}}({\bm{k}})$,
to the cosmology of the Minerva simulations \citep{Grieb:2016}, and allowing to vary: the amplitude of matter fluctuations ($\sigma_8$), the growth rate of structures
($f$), galaxy bias coefficents ($b_1, \, b_2, \,  b_{\mathcal{G}_2} $
and EFT counter-terms ($c_0, \, c_2$ and $c_{\mathrm{nlo}}$).
This template-fitting strategy, which involves pre-computing and tabulating all terms obtained by factoring out 
$\left\{\sigma_8,\, f,\, b_1,\,b_2,\,b_{\mathcal{G}_2},\,b_{\Gamma_3},\,c_0,\,c_2,\,c_{\mathrm{nlo}}\right\}$ in Eqs.~(\ref{eq:Pk_NoIR}),
accelerates the statistical inference of the model parameters described in Sec.~\ref{sec:parinf}. The analytic expressions for these terms can be found in Sec.~4.3 of \cite{Sugiyama:2021}.\\
However, as Eqs.~(\ref{eq:k_prime}) and (\ref{eq:mu_prime}) suggest, incorporating the AP and isotropic dilation effects into the analysis makes this procedure computationally prohibitive.
Geometrical distortions, indeed, introduce an angular dependence in $k'$, requiring the re-computation of the multipoles of the template fitting tables for each combination of $\alpha$ and $\varepsilon$ sampled during the fitting process. This significantly slows down parameter inference, particularly for the 3PCF.\\
To address this issue, we build a two-dimensional interpolator for each of the template fitting tables introduced above. Specifically, for every table we precompute the 2 and 3PCF multipoles on a $101\times51$ grid of equally spaced points spanning $\varepsilon\in[-0.025,\,0.025]$ and $\alpha\in[0.95,\,1.05]$. The results are stored in dedicated libraries, from which we construct separate bicubic spline interpolators. Thus, each template fitting table has its own interpolator defined over the $(\alpha,\,\varepsilon)$ grid. During sampling, whenever a new pair of these two parameters is drawn, the relevant interpolator is queried to provide smooth estimate for its corresponding table without the need for recomputation.
We assessed the precision of the interpolator by comparing its predictions with the exact model evaluated over a grid of $(\alpha, \varepsilon)$ values, explicitly excluding the nodes used to construct the interpolator. We found that the median residuals are  below the 0.001 \% level.\\
Finally, to assess the reliability of these models, we compared the output of our implementation with that of other publicly available numerical packages. For the anisotropic 2PCF, we compared with \texttt{CLASS-PT} \cite{Chudaykin:2020} using the same parameters and found that the two predictions agree to within machine precision.\\
For the anisotropic 3PCF, we compared our implementation with the output of the \texttt{HITOMI} package \cite{Sugiyama:2023}. We found good agreement between the two models across the entire range of scales and triangle configurations explored. This agreement is particularly notable given that the two 3PCF implementations are independent and differ in several respects.
For a more detailed discussion of the validation tests performed on our 3PCF model, we refer the reader to Appendix~\ref{app:A}.

\section{Datasets}
\label{sec:datasets}
In our analysis, we use numerical simulations to validate our models and estimators.
For this purpose we have considered two independent sets of simulated catalogs of dark matter halos extracted from two different numerical experiments.
\\
The first set consists of 298 mock catalogs extracted from the same number of Minerva N-body simulations \citep{Grieb:2016}.
As these are fully non-linear simulations, we use them for the estimate of the 2 and 3PCF.
The second set consists of 3000 halo catalogs extracted from the Pinocchio simulations, which are based on the Lagrangian perturbation theory approach (see e.g. \cite{Monaco:2013}).
As this second set of simulations provides an approximate treatment of the system’s dynamics, we use their outputs to estimate the covariance matrix of the 2PCF and 3PCF measurements.

\subsection{Minerva halo catalogs}
The first set of halo catalogs were extracted from the
$z=1$ output of 298 N-body Minerva simulations \citep{Grieb:2016} performed with the Gadget-II code \citep{Springel:2005}. Each simulation box has a size $L=1500\,\hMpc$ and contains $1000^3$ particles with 
mass $m_\mathrm{p} \simeq 2.67\times 10^{11}h^{-1}\mathrm{M_\odot}$.
The aggregate volume of all the boxes is about $1000 \hGpc$.
\\
Dark matter halos were identified using a friends-of-friends algorithm with a linking length $l=0.2$ in unit of mean inter-particle distance.
In this work, we only consider halos with masses larger than $1.12\times10^{13}\,h^{-1}\mathrm{M_\odot} $.
This choice guarantees that the selected halos are bound structures \citep{Oddo:2020,Veropalumbo:2022} and that their number density 
($\bar n=2.13\times10^{-4}\,h^3\mathrm{Mpc}^{-3}$) roughly matches the one expected for the H${\alpha}$
galaxies that are being targeted by the ongoing spectroscopic galaxy surveys at the same redshift \citep{Laureijs:2011}.

\subsection{Pinocchio halo catalogs}
The second dataset considered in this work is a suite of 3000 dark matter halo catalogs extracted from the $z=1$ outputs of numerical simulations performed with the Pinocchio method \citep{Monaco:2013, Monaco:2013a, Munari:2017}. The simulation box size and cosmological model used match exactly those of the Minerva simulations.
\\
Unlike standard N-body simulations, the Pinocchio algorithm relies on excursion set theory, ellipsoidal collapse, and Lagrangian Perturbation Theory (LPT) to identify halos in Lagrangian space. Halo positions are determined by applying a single third-order LPT displacement to their center of mass.\\
In the catalogs used in this work, we consider halos with masses larger than
 $1.06\times10^{13}\,h^{-1}\mathrm{M_\odot}$, a cut also adopted by \cite{Oddo:2020, Oddo:2021, Veropalumbo:2022, Rizzo:2023} to minimize differences with the halo power spectrum and bispectrum (and their covariance) measured in the Minerva simulations, both in configuration and redshift space.

\section{Statistical estimators}
\label{sec:measurements}
To measure the 2 and 3PCF of the halos in the mock catalogs, we use our own implementation of the 
unbiased, minimum variance Szapudi \& Szalay estimator \citep{Szapudi:1998}, which, written in compact form, reads:

\begin{equation}
    \label{eq:ssest}
    \meas{\xi}^{\rm N}\left(\bm{x_1}, \dots, \bm{x_{\rm N}}\right) = \frac{(D-R)^{\rm N}}{R^{\rm N}} \, ,
\end{equation}

\noindent where $\meas{\xi}^{\rm{N}}$ represents a generic $N$-point correlation function, while D and R indicate, respectively, the data and random sample, the latter being a set of unclustered objects distributed in the same volume and sharing the same selection effects of the data.
This estimator automatically subtracts the disconnected part of the 3PCF and corrects for selection and edge effects. Neither effect, however, is present in our catalogs since these  are contained in periodic boxes.\\
Hereafter, we use  $\meas{\xi}$ to indicate the measured 2PCF (corresponding to $N=2$ in Eq. \ref{eq:ssest}) and $\meas{\zeta}$ to indicate the connected part of the measured 3PCF (${\rm N} = 3$).\\ 
The Szapudi \& Szalay estimators rely on counting pairs and triplets of data and random objects in various combinations. The explicit expressions can be found in \cite{Szapudi:1998}.
To minimize the contribution of shot noise to the error budget, the random catalogs contain $50$ times more objects than the halo catalogs.\\ 
Central to this work is the ability to capture all clustering anisotropies. For this reason, in our estimators, we consider all orientations of pairs and triplets with respect to the line of sight, which we take to be the same to all objects and parallel to the $\bm{\hat{z}}-$axis.\\
Both estimators have been implemented, validated and are now publicly available in the \texttt{Python} package \texttt{MeasCORR} \footnote{\url{https://gitlab.com/veropalumbo.alfonso/meascorr}}. Details on both estimators are provided in the next two sections.

\subsection{2PCF estimator}
\label{sec:twopoint_est}
To estimate the anisotropic 2PCF, we 
measure the modulus  $r$ of the separation vector and 
the cosine angle relative to the line of sight direction $\mu = \hat{\bm{r}}\cdot \hat{\bm{z}}$.
Thanks to the periodic boundary conditions, the indexed pair counts
$DR(r,\mu)$ and $RR(r, \mu)$ can be estimated analytically.
This significantly decreases the computational cost of the estimator, reducing it to that of the "natural" one \citep{Peebles:1974}

\begin{equation}
    \label{eq:xi_natest}
    \tilde{\xi}(r, \mu) = \frac{DD(r, \mu)}{RR(r, \mu)} - 1.
\end{equation}
The computational budget is primarily dominated by the $DD$ counts that, unlike the $RR$ ones, must be estimated numerically. 
To reduce computational time we use linked list for efficient
data spatial partitioning and consider only pairs separated by less than $150\hMpc$, thus completely including the BAO feature.
Next, we compress information by expanding $\meas{\xi}(r, \mu)$ in Legendre polynomials and obtain its multipoles:

\begin{equation}
    \label{eq:xi_meas_leg}
    \meas{\xi}_{\ell}(r) = \frac{2\ell+1}{2} \int_{-1}^{1} \diff \mu\; \mathcal{L}_{\ell}(\mu) \meas{\xi}(r, \mu)
\end{equation}

\noindent of which we only consider the $\ell=0,\,2$ ones as these are routinely measured in current and future surveys.
\noindent Since we do not expect the 2PCF model to fully capture non-linearities on small scales, in the 2PCF-only analysis we only consider pairs with separation above
$20 \hMpc$. The criteria adopted to select the minimum scale $r_{\mathrm{min}}^{\mathrm{2pt}} $ in the joint 2+3PCF analysis presented in Sec.~\ref{sec:3PCF+2PCF} are extensively discussed in Appendix \ref{app:A1}.
Finally, we set the bin size to $\Delta r^{\mathrm{2pt}} = 5 \hMpc$ along the radial direction and $\Delta \mu = 0.01$ for the cosine angle. For every bin, we report the measurement at the bin center, and all results and figures presented in this work should be interpreted accordingly.

\subsection{3PCF estimator}
\label{sec:threepoint_est}
Our estimator follows the approach introduced by \cite{Slepian:2015, Slepian:2018} to efficiently measure 3PCF multipoles in large samples. 
The method exploits the properties of spherical harmonics to reduce the computational cost of triplet counting, achieving an $\mathcal{O}(N^2)$ scaling instead of the $\mathcal{O}(N^3)$ scaling of brute-force estimators.\\
This speedup enables the evaluation of the 3PCF for large samples of objects and, in particular, allows the numerical estimation of covariance matrices from suites of independent mock catalogs rather than relying on analytic approximations \citep{Slepian:2018}. Our 3PCF estimator measures the multipole moments of the following spherical-harmonic expansion

\begin{equation}
    \meas{\zeta}(\bm{r_1}, \bm{r_2}, \bm{x}) = \sum_{\ell\; m} \sum_{\ell^{\prime}\; m^{\prime}} \meas{\zeta}_{\ell\; \ell^{\prime}}^{m\; m^{\prime}} (r_1, r_2, \bm{x}) Y_{\ell}^{m}({\bm{\hat{r}_1}) Y_{\ell^{\prime}}^{m^{\prime}*}(\bf{\hat{r}_2}})\,,
    \label{eq:zeta_meas}
\end{equation}

\noindent where $\meas{\zeta}(\bm{r_1}, \bm{r_2}, \bm{x})$ is the anisotropic 3PCF evaluated at the spatial position  $\bm{x}$, $Y_{\ell}^{m}$ are orthonormal spherical harmonics, and $\meas{\zeta}_{\ell\; \ell^{\prime}}^{m\; m^{\prime}}$ are the coefficient of the expansion. These are computed as

\begin{equation}
\begin{split}
    \meas{\zeta}_{\ell\;\ell^{\prime}}^{m\; m^{\prime}}(r_1, r_2, \bm{x}) & = \delta(\bm{x})\int{d\Omega_1}\int{d\Omega_2\,\bar{\delta}(r_1,\bm{\hat{r}_1}, \bm{x}) \bar{\delta}(r_2,\bm{\hat{r}_2},\bm{x})\,Y_{\ell}^{m}(\bm{\hat{r}_1})Y_{\ell^{\prime}}^{m^{\prime}}(\bm{\hat{r}_2})} =\\
    &=\,\delta(\bm{x})\,a_{\ell}^m(\bm{\hat{r}_1}, \bm{x})\,a_{\ell^{\prime}}^{m^{\prime}}(\bm{\hat{r}_2}, \bm{x})\,
\end{split}
\label{eq:measured_mult}
\end{equation}

\noindent where, following the notation of \cite{Slepian:2017}, we implicitly defined

\begin{equation}
\begin{split}
    a_{\ell}^m(\bm{\hat{r}_i},\bm{x}) &=\, \int{d\Omega_i\,\bar{\delta}(r_i,\bm{\hat{r}_i}, \bm{x}) \,Y_{\ell}^{m}(\bm{\hat{r}_i})} = \\
    &=\, \int{d\Omega_i}\int{dr\,r^2 \phi(r,r_i,\bm{x})\,\delta(\bm{x + r})\,Y_{\ell}^{m}(\bm{\hat{r}_i})}\,,
\end{split}
\label{eq:alm}
\end{equation}

\noindent where $\phi(r, r_i,\bm{x})$ 
denotes a step function equal to unity within the $r_i$-bin and zero outside.
The symmetry around the azimuthal direction, together with the choice of a common line of sight, ensures that $m=m^{\prime}$.
The $\mathcal{O}(N^2)$ scaling is possible thanks to the  multipole expansion adopted in Eq.~(\ref{eq:zeta_meas}), that allows one to factor out the $\bm{\hat{r}}_1$ and $\bm{\hat{r}}_2$ dependencies in Eq.~(\ref{eq:measured_mult}). In this way, at any given position $\bm{x}$, it is not necessary to compute a multidimensional integral for each ($r_1$, $r_2$) pair. Instead one can pre-compute all the $a_{\ell}^{m}$ coefficients in each radial bin and combine them when needed.\\
Averaging the local multipole coefficients defined in Eq.~(\ref{eq:measured_mult}) over all positions $\bm{x}$ in the survey volume gives the global 3PCF moments $\meas{\zeta}_{\ell_1\; \ell_2}^m(r_1, r_2)$.
These multipoles are related to those used in the TripoSH expansion adopted in Eq.~(\ref{eq:TripoSH}) through  \citep{Sugiyama:2019}:

\begin{equation}
    \label{eq:slep2sugy}
    \meas{\zeta}_{\ell_1\, \ell_2\, L}(r_1, r_2) = (2L+1)\, H_{\ell_1\, \ell_2\, L} \sum_m \begin{pmatrix} \ell_1 & \ell_2 & L \\ m & -m & 0 \end{pmatrix} (-1)^m \meas{\zeta}_{\ell_1\; \ell_2}^m(r_1, r_2).
\end{equation}

\noindent We use Eq.~(\ref{eq:slep2sugy}) to compare  3PCF measurements with model predictions. \\
We measure the 3PCF for all triangles with sides spanning in the range $\left[30,\,150\right]\,\hMpc$ and set the bin size to $\Delta r^{\mathrm{3pt}} = 10 \hMpc$. The criteria used to select the triangle configurations for the analysis are discussed in Sec.~\ref{sec:3PCF-only}. As for the 2PCF, measurements in each bin are assigned to the corresponding bin center.
Finally, unless otherwise indicated, in this work we only focus on the first three isotropic ($\meas{\zeta}_{000}$, $\meas{\zeta}_{110}$, $\meas{\zeta}_{220}$) and the first three anisotropic ($\meas{\zeta}_{202}$, $\meas{\zeta}_{112}$, $\meas{\zeta}_{312}$) multipole moments of the 3PCF.

\subsection{Covariant errors} \label{sec:cov}
\begin{figure}
    \centering
    \includegraphics[width=1.\textwidth]{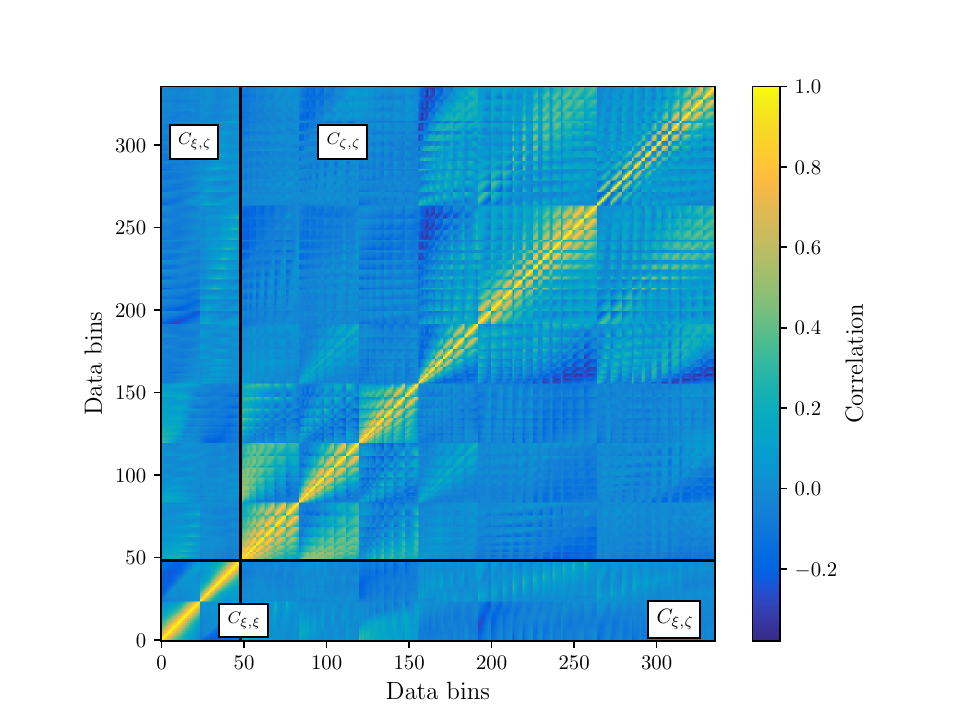}
    
    \caption{\small Example of a numerical covariance matrix used in one of the joint 2 and 3-point correlation analyses performed in this work. The data vector consists of the 2PCF monopole and quadrupole, measured in the interval $[32.5,\, 147.5]\,\hMpc$ in bins of $\Delta r^{\mathrm{2pt}} = 5\,\hMpc$, and all the six 3PCF multipole moments measured in Sec.(\ref{sec:measurements}), namely $\bm d= \left\{\meas{\zeta}_{000},\,\hat\zeta_{110},\,\meas{\zeta}_{220},\,\meas{\zeta}_{202},\,\meas{\zeta}_{112},\,\meas{\zeta}_{312} \right\}$. These latter are computed in the separation interval $[35,\, 145]\,\hMpc$ in bins of $\Delta r^{3pt} = 10\,\hMpc$. Only $\eta \ge 3$ configurations are considered.}
    \label{fig:covariance}
\end{figure}

We estimate the covariant errors  numerically,
from the 2PCF and 3PCF multipoles measured in the 3000 Pinocchio mock catalogs.
The covariance matrix is estimated as follows: 
\begin{equation}
    \hat{C}_{ij} = \frac{1}{N_m-1}\sum_{k=1}^{N_m}{\left(d_i^k-\bar{d}_i\right)\left(d_j^k-\bar{d}_j\right)} \, ,
    \label{eq:covariance}
\end{equation}
\noindent where $d_i^k$ represents the data vector and the indexes 
$i$ and $k$ identifies the bin and the mock catalog in the Pinocchio sample, respectively. The quantity $\bar{d}_i$ represents the average estimated from the $N_m=3000$ mocks.
The type and dimension of the data vector depend on the specific clustering analysis. For example, in the 3PCF-only analysis, $d_i^k$ specifies the multipoles of the anisotropic 3PCF. For the joint 2 and 3-point correlation analysis $d_i^k$ specifies the 2PCF and the 3PCF multipoles.
An illustrative example, referring to the joint 2 and 3-point correlation analysis is shown in Fig.~\ref{fig:covariance}. The 3PCF is computed for separations in the range $[35,\,145]\,\hMpc$ with bins of size $\Delta r^{3\mathrm{pt}} = 10\,\hMpc$. We impose $\eta = 3$, which requires $|r_1 - r_2| \ge \eta\,\Delta r^{3\mathrm{pt}}$, thereby selecting triangles with side-length differences of at least $30\,\hMpc$.
The resulting $336\times336$ matrix shown in the figure is divided into four quadrants. The bottom-left $48\times48$ sub-matrix $\hat{C}_{\xi, \xi}$ quantifies the covariant errors on the 2PCF multipoles. 
The upper right $288\times288$ sub-matrix $\hat{C}_{\zeta,\zeta}$
accounts for the  errors on the 3PCF multipoles. The two remaining  rectangular sub-matrices  $\hat{C}_{\xi,\zeta}$ and  $\hat{C}_{\zeta,\xi}$ are symmetric and account for the errors on the cross-correlation terms.\\
Because of the finite number of mock catalogs, the inverse of 
$\hat{C}_{ij}$ computed in Eq.~(\ref{eq:covariance}) is a biased estimate of the inverse covariance matrix. This bias can be corrected for through a multiplicative term \citep{Anderson:2003, Hartlap:2006}:

\begin{equation}
    C^{-1}=\left(\frac{N_m-N_d-2}{N_m-1}\right)\hat{C}^{-1} \, ,
    \label{eq:Hartlap}
\end{equation}
\noindent where $N_d$ is the size of the data vector $d_i^k$ (see also \citep{Sellentin:2015} for an alternative approach to the problem). The  
Eq.~(\ref{eq:Hartlap}) does not correct for errors in the covariance matrix, which propagate through to errors on parameters estimated from the posterior probability. To account for this effect, we follow \cite{Percival:2014} and apply the following correction factor on the parameters' variance:
\begin{equation}
    m_1 = \frac{1+B(N_d-N_p)}{1+A+B(N_p+1)} \, ,
    \label{eq:Perc1}
\end{equation}
where $N_p$ is the number of the inferred parameters and  

\begin{equation}
\begin{aligned}
    A &= \frac{2}{(N_m-N_d-1)(N_m-N_d-4)} \\[0.2cm]
    B &= \frac{N_m-N_d-2}{(N_m-N_d-1)(N_m-N_d-4)} \, .
\end{aligned}
\label{eq:Perc}
\end{equation}
\vspace{0.1cm}

\begin{figure}
    \centering
    \includegraphics[width=0.95\textwidth]{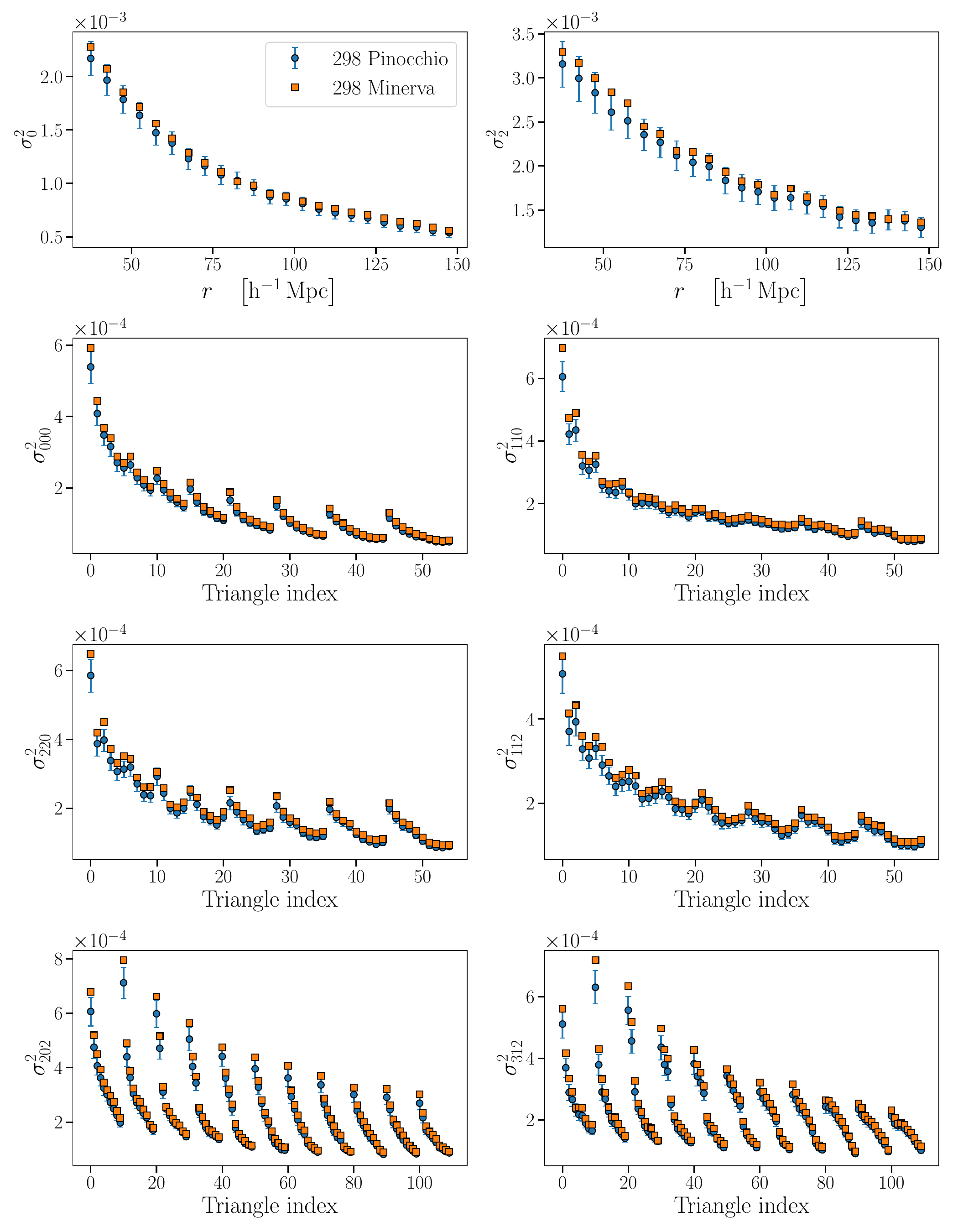}
    
    \caption{\small Comparison of variances between the entire Minerva simulation suite (orange squares) and the first 298 Pinocchio realizations (blue circles), which share the same initial conditions as Minerva. {\it Top row:} 2PCF monopole and quadrupole. {\it Remaining rows:} 3PCF multipoles with $r_{\mathrm{min}}^{3\mathrm{pt}}=45\,h^{-1}\mathrm{Mpc}$ and $\eta = 1$. Each triangle index $i$ denotes entries of the flattened $(r_1,r_2)$ grid and corresponds to a unique triangle configuration. Error-bars are estimated via bootstrap resampling and correspond to 95\% confidence intervals.}
    \label{fig:variance_2pcf_3pcf}
\end{figure}

\noindent The covariance matrix defined in Eq.~(\ref{eq:covariance}) quantifies the uncertainty associated with a single halo catalog of $\sim 3.375 \hGpc$.
In our analysis, it is used to estimate the likelihood of individual catalogs, which are then combined to determine the total likelihood, as described in the next section. 
The resulting best-fit parameters are therefore equivalent to those obtained from a clustering analysis over a survey volume of approximately $1000\,\hGpc$, with corresponding uncertainties. 
These reduced errors allow for a precise evaluation of the anisotropic 2PCF and 3PCF models and for testing potential biases in the inferred cosmological parameters. However, they remain significantly smaller than the uncertainties expected in ongoing cosmological surveys, which cover significantly smaller volumes. \\
We note that our covariance matrix is derived from mock halo catalogs constructed from the Pinocchio simulations, which use a mass threshold different from that of the Minerva halo catalogs employed in the analysis. This mismatch may bias the results when applying this covariance matrix to the clustering analysis of the halo 2PCF and 3PCF measured in the Minerva catalogs. 
To assess the potential impact, we carried out a series of tests comparing the covariance matrix from the 298 Minerva simulations with that derived from 298 Pinocchio mocks generated with matching initial conditions. \\
In Fig.~\ref{fig:variance_2pcf_3pcf}, we compare the variance of the 2PCF and 3PCF obtained from the Pinocchio mocks (blue circles) with that from the Minerva simulations (orange squares). The top two panels show the variances of the 2PCF monopole and quadrupole as a function of pair separation, while the remaining panels show the variances of the 3PCF tripolar multipoles analyzed in this work as a function of the triangle index. For the 3PCF, we set $r_{\mathrm{min}}^{3\mathrm{pt}}=40\,\hMpc$ and $\eta \ge 1$, with the triangle index $i$ labeling entries of the flattened $(r_1,r_2)$ grid, each corresponding to a unique triangle configuration. Error bars, estimated via bootstrap resampling, indicate the 95\% confidence interval.

\begin{figure}
    \centering
    \includegraphics[width=1\textwidth]{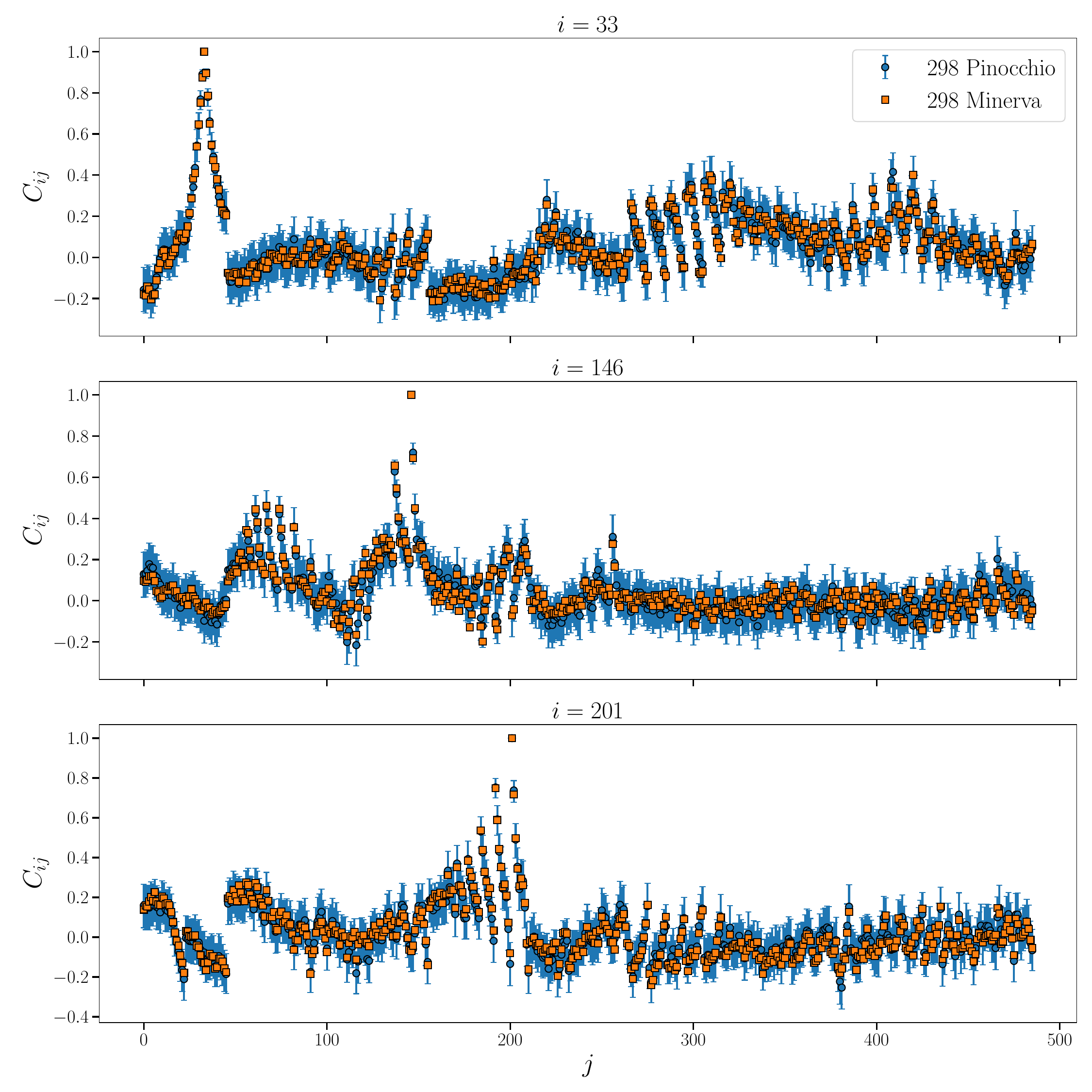}
    \caption{\small Comparison of the correlation indices measured from the full Minerva simulation suite (orange squares) with those extracted from the first 298 Pinocchio realizations (blue circles), which share the same initial conditions as Minerva. The three panels correspond to three different rows of the correlation matrix, as reported in the figure. Error-bars are estimated via bootstrap resampling and correspond to the 95\% confidence interval.} 
    \label{fig:correlation_2pcf_3pcf}
\end{figure}

\noindent The results show that the Pinocchio mocks systematically overestimate the variance of the Minerva mocks by 5–10 \% across the entire range of scales and triangle configurations considered. The magnitude of this bias always remains within the 95\% confidence interval, with the only exception of highly squeezed 3PCF configurations. This result is unsurprising since the third-order LPT approximation used in the Pinocchio simulations is not expected to fully capture nonlinear dynamics. Moreover, these findings are consistent with previous studies of \cite{Blot:2019, Colavincenzo:2019, Oddo:2020, Veropalumbo:2022, Rizzo:2023}.
Similarly, we assess the capability of the Pinocchio halo catalogs to reproduce the correlation structure underlying the 298 Minerva realizations. Results are shown in Fig.~\ref{fig:correlation_2pcf_3pcf}, where 3 different, randomly extracted lines of the Pinocchio correlation matrix (blue circles) are compared with their Minerva counterparts (orange squares). Again, uncertainties are computed via bootstrap resampling and represent the 95\% confidence interval.
We detect no systematic differences between the two cases and conclude that, despite the approximate nature of the Pinocchio mocks and the difference in mass cut, the structure of the covariance matrix from the Pinocchio mocks matches that of the Minerva catalogs. For these reasons we use them in our clustering analysis.

\section{Parameter inference}
\label{sec:parinf}
Our anisotropic 2PCF and 3PCF models are characterized by different sets of free parameters that we infer, in a Bayesian framework, from the comparison to the measurements in the mock catalogs. The parameter vector $\bm\theta$ is obtained by sampling the posterior probability $P(\bm\mu(\bm\theta)|\bm d)\propto \mathcal{L}(\bm d|\bm\mu)P(\bm\theta)$,
where $\mu(\bm\theta)$ is the $\bm\theta$-dependent model prediction and $\bm d \equiv d_i^k$ is the data vector. The quantity $ P(\bm\theta)$ represents the prior probability and $\mathcal{L}(\bm d|\bm\mu(\bm\theta))$ is the likelihood
function computed from the 298 independent halo 2PCF and 3PCF measurements obtained from the Minerva mock catalogs. 
Since these are all independent measurements we follow \cite{Oddo:2020,Oddo:2021,Veropalumbo:2022, Rizzo:2023} and estimate the total log-likelihood function as the sum of the log-likelihood of each single realization $\alpha$ which we assume to be Gaussian. As a result, modulo an additive normalization constant, we can write:

\begin{equation}
    \log{\mathcal{L}_{tot}}=\sum_\alpha{\log{\mathcal{L}_\alpha}}= -\frac12\sum_\alpha{\chi_\alpha^2} \, ,
    \label{eq:loglike}
\end{equation}

\noindent where the sum is over the 298 mock catalogs and, for each of them, the $\chi^2$ function is

\begin{equation}
    \chi_\alpha^2=\left(\bm d_\alpha-\bm \mu(\bm\theta)\right)^T\hat{C}^{-1}\left(\bm d_\alpha-\bm \mu(\bm\theta)\right).
\end{equation}

\noindent The datavector, the model, and consequently the parameter vector depend on the specific analysis being performed. In this work, we focus on two main cases.\\
The first analysis relies solely on the 3PCF and considers a data vector composed of the six multipole moments measured in Sec.~\ref{sec:measurements}, namely $\bm d= \left\{\meas{\zeta}_{000},\,\hat\zeta_{110},\,\meas{\zeta}_{220},\,\meas{\zeta}_{202},\,\meas{\zeta}_{112},\,\meas{\zeta}_{312} \right\}$. In this analysis, the free parameters are the growth rate of fluctuations, $f$, the isotropic dilation $\alpha$, the AP parameter $\varepsilon$, and the linear, quadratic, and tidal bias coefficients, that is $b_1$, $b_2$ and $b_{\mathcal{G}_2}$ respectively. Since in a 3PCF-only analysis the {\it rms} clustering amplitude is fully degenerate with $b_1$ and $f$, we fix its value to that of the Minerva simulations, i.e. $\sigma_8=0.828$.\\
In the second case we perform a joint  2PCF and 3PCF analysis. In this case the datavector includes, in addition to the previous 3PCF multipoles, the monopole and quadrupole moments of the halo 2PCF and we let all the model parameters, including $\sigma_8$, free to vary within their allowed ranges.
In Tab.~\ref{tab:priors} we list the model parameters used in these two analyses along with their priors. In all cases we assume flat priors, $\mathcal{U}(...)$, except for $\sigma_8$  which, in the case of 3PCF-only analysis, is fixed to its reference value.\\
To sample the posterior probability distributions we adopt the nested sampling Monte Carlo algorithm MLFriends \cite{Buchner:2014, Buchner:2019} as implemented in the
\texttt{UltraNest} package \citep{Buchner:2021}. We ensure the convergence of our chains adopting the \texttt{Ultranest} built-in stopping criteria, i.e. requiring the uncertainty in the log-evidence to be below 0.5, the Kullback-Leibler divergence to be less than 0.5, and the effective sample size to exceed 400.

\vspace{0.1cm}
\begin{table}[h!]
\centering
\begin{tabular}{c c c}
\hline
\textbf{Parameter} & \textbf{3PCF} & \textbf{2+3PCF} \\
\hline
$\sigma_8$ & $0.828$ & $\mathcal{U}(0.5, 1.5)$ \\
$f$ & $\mathcal{U}(0, 1.5)$ & $\mathcal{U}(0, 1.5)$ \\
$b_1$ & $\mathcal{U}(0.5, 5)$ & $\mathcal{U}(0.5, 5)$ \\
$b_2$ & $\mathcal{U}(-10, 10)$ & $\mathcal{U}(-10, 10)$ \\
$b_{\mathcal{G}_2}$ & $\mathcal{U}(-5, 5)$ & $\mathcal{U}(-5, 5)$ \\
$b_{\Gamma_3}$ & \textbackslash{} & $\mathcal{U}(-10, 10)$ \\
$c_0$ & \textbackslash{} & $\mathcal{U}(-100, 100)$ \\
$c_2$ & \textbackslash{} & $\mathcal{U}(-100, 100)$ \\
$c_\mathrm{nlo}$ & \textbackslash{} & $\mathcal{U}(-1000, 1000)$ \\
$\varepsilon$ & $\mathcal{U}(-0.025, 0.025)$ & $\mathcal{U}(-0.025, 0.025)$ \\
$\alpha$ & $\mathcal{U}(0.95, 1.05)$ & $\mathcal{U}(0.95, 1.05)$ \\
\hline
\end{tabular}
\caption{Prior ranges assigned to each parameter for the 3PCF-only and joint 2+3PCF analyses. Uniform priors are used in all cases, except for $\sigma_8$ in the 3PCF-only analysis, which was fixed to the value used in the Minerva simulations. Parameters marked with \textbackslash{} are not considered in the 3PCF-only analysis as they do not enter the 3PCF model at tree-level.}
\label{tab:priors}
\end{table}

\section{Results}
\label{sec:results}

In this section we present the results obtained for the clustering analyses anticipated in Sec.~\ref{sec:parinf}: the 3PCF-only analysis and the 2+3PCF ones.

\subsection{3PCF-only analysis}
\label{sec:3PCF-only}
\begin{figure}
    \centering
    \includegraphics[width=1.0\textwidth]{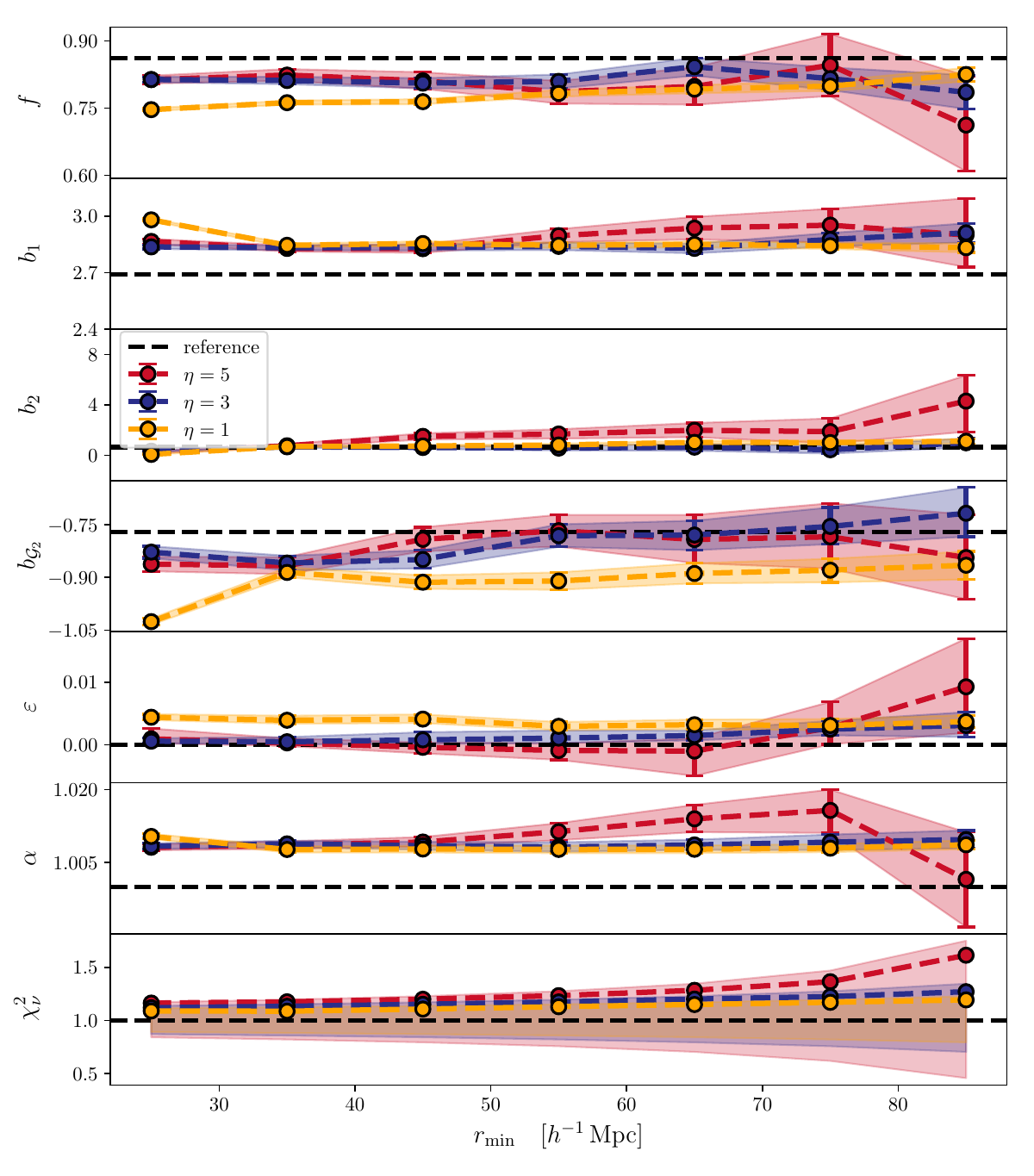}
    \caption{\small \emph{Top six panels}: best-fit values of the parameters $f$, $b_1$, $b_2$, $b_{\mathcal{G}_2}$, $\varepsilon$ and $\alpha$ as a function of the minimum triangle size $r_{\mathrm{min}}^{\mathrm{3pt}}$. Different marker styles and colors correspond to distinct triangle configurations defined by the $\eta$ parameter, as indicated in the legend.  Error bars represent the 68\% uncertainty interval derived from the 1D marginalized posterior distributions.
    The horizontal black dashed lines, instead, correspond to the reference values of the parameters estimated by \cite{Veropalumbo:2022} in real space.
    \emph{Bottom panel}: reduced $\chi^2$ value estimated from the 298 Minerva halo catalogs as a function of $r_{\mathrm{min}}^{\mathrm{3pt}}$. The colored bands  represent the 95\% confidence region associated to a $\chi^2$ distribution characterized by $N_{\mathrm{real}}N_{\mathrm{bin}} - 6$ {\it dof}. Different colors are used for the different values of $\eta$ considered in the analysis.}
    \label{fig:rmin_dep}
\end{figure}

The aim of the 3PCF-only analysis is twofold. On the one hand, it allows us to test the precision and accuracy of the anisotropic 3PCF model presented in Sec.~\ref{sec:model}. On the other, it provides a means to quantify how the inclusion of anisotropic multipoles modifies or enhances the constraints relative to the standard isotropic 3PCF analysis. \\
Since the 3PCF model adopted here relies on a tree-level expansion and does not fully capture nonlinear effects on small scales, it is necessary to determine the range of scales and triangle configurations where the it can be reliably applied. To this end, we carry out a series of tests varying the minimum triangle side length, $r_{\mathrm{min}}^{\mathrm{3pt}}$, and the shape parameter $\eta$. \\
The results are summarized in Fig.~\ref{fig:rmin_dep}. 
In the six upper panels, we plot the best-fit values of $f$, $b_1$, $b_2$, $b_{\mathcal{G}_2}$, $\varepsilon$ and $\alpha$ as a function of  $r_{\mathrm{min}}^{\mathrm{3pt}}$ for three different values of $\eta$, denoted by different  colors, as indicated in the legend. The error bars, together with their shaded bands, represent the statistical uncertainty of the 3PCF measurement for a volume equivalent to the combined 298 Minerva simulation boxes, approximately $1000\,\hGpc$. Error bars increase with  $r_{\mathrm{min}}^{\mathrm{3pt}}$  and with $\eta$ due to the corresponding reduction in the number of triangular configurations.
Since in the analysis we adopt the same cosmological model as in the Minerva simulations, the AP and isotropic dilation parameters are fixed to $\varepsilon = 0$ and $\alpha = 1$.
The reference values of the bias parameters, instead, were estimated by \cite{Veropalumbo:2022} from the clustering analyses of the Minerva halos in real space.

\begin{figure}
    \centering \includegraphics[width=1\textwidth]{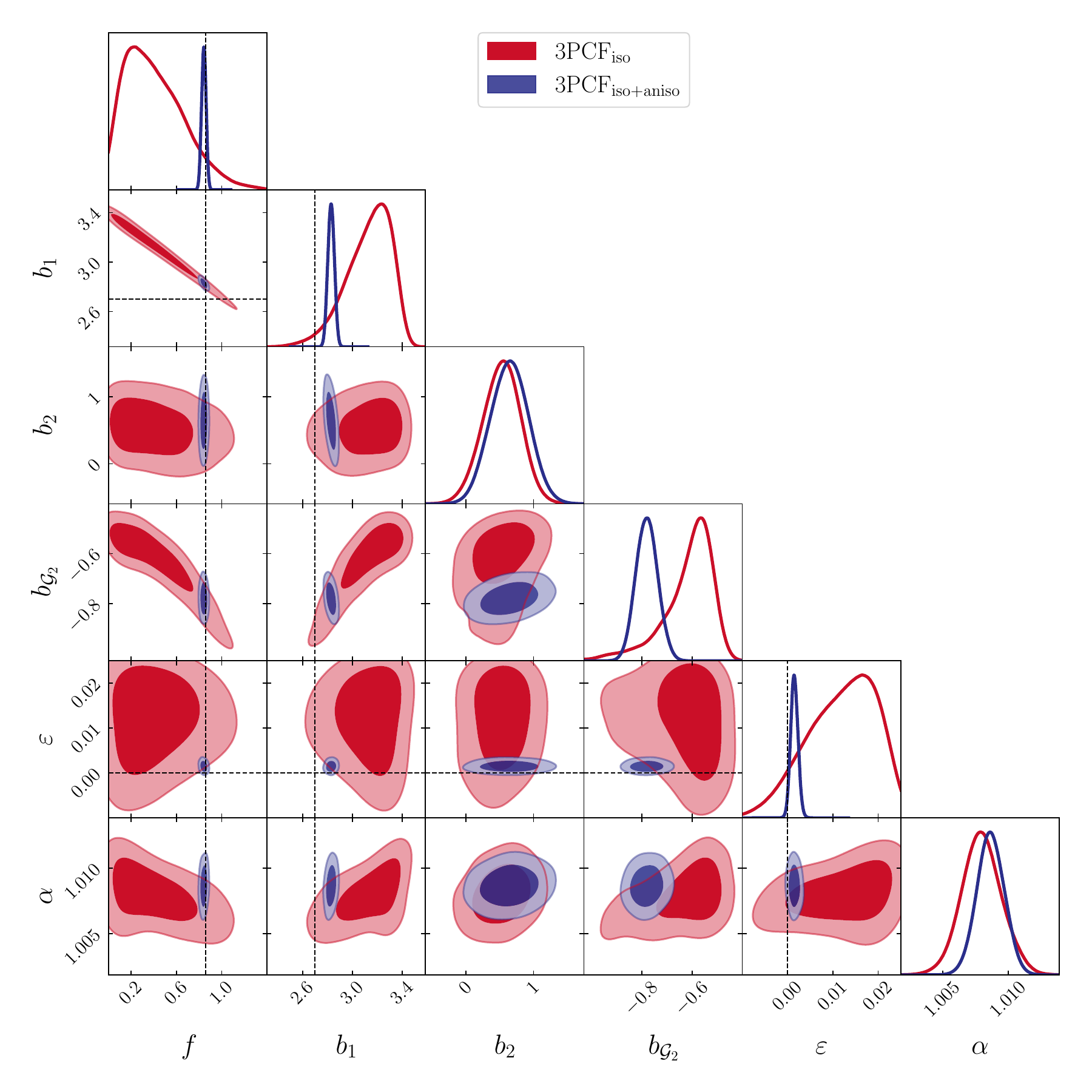}
    \caption{\small Marginalized 1D and 2D posterior probability contours for the parameters   $f$, $b_1$, $b_2$, $b_{\mathcal{G}_2}$, $\varepsilon$ and $\alpha$ estimated from a 3PCF-only analysis. The red contours consider only the isotropic multipole moments ($\meas{\zeta}_{000}$, $\meas{\zeta}_{110}$ and $\meas{\zeta}_{220}$ ), while the blue contours include both isotropic and anisotropic multipoles ($\meas{\zeta}_{000}$, $\meas{\zeta}_{110}$, $\meas{\zeta}_{220}$, $\meas{\zeta}_{202}$, $\meas{\zeta}_{112}$, and $\meas{\zeta}_{312}$). The black dashed lines represent the reference values of the main parameters. The probability contours refer to a survey volume of $1000\,\hGpc$, considering triangles with sizes above the reference value $r_{\min}^{\mathrm{3pt}}=65\,\hMpc$ and shape parameter $\eta=3$.}  \label{fig:iso_vs_aniso}
\end{figure}

\noindent The figure evidences distinct behaviors among the best-fit parameters. While $b_1$ and $\alpha$ remain essentially insensitive to variations in $r_{\mathrm{min}}^{3\mathrm{pt}}$, the growth rate $f$ shows a slight scale dependence, particularly for $\eta=1$, where it gradually approaches its reference value as the minimum separation increases. In this regime, $f$ is systematically underestimated, with deviations reaching up to 10\% at small $r_{\mathrm{min}}^{3\mathrm{pt}}$. Configurations with larger $\eta$ converge more rapidly, and for $r_{\min}^{3\mathrm{pt}} \gtrsim 65\,h^{-1}\mathrm{Mpc}$ the residual offset drops below the 2$\sigma$ level. Interestingly, this underestimate of $f$ is partially compensated by a $\sim 5\%$ overestimate of $b_1$.\\
The $\alpha$ parameter is systematically overestimated by roughly 0.8\%, a bias that is statistically significant given the reduced uncertainties associated with the large volume adopted in this analysis. The magnitude of this offset is largely independent of $r_{\mathrm{min}}^{3\mathrm{pt}}$, except for $\eta = 5$, where the bias grows with the scale until $r_{\min}^{\mathrm{3pt}}=75\,h^{-1}\hMpc$ and then suddenly drops.\\
Regarding the remaining parameters, $\varepsilon$ shows no significant shift for $\eta=3$ and $\eta=5$, whereas for $\eta=1$ it is systematically overestimated, reflecting the influence of highly nonlinear triangular configurations. Similar trends are observed for $b_{\mathcal{G}2}$: its best-fit value converges rapidly to the expected value for $r_{\min}^{3\mathrm{pt}} \gtrsim 45\,h^{-1}\mathrm{Mpc}$ when $\eta = 3$ and $5$, while for $\eta=1$ it remains systematically underestimated by roughly 15\% across the full range of minimum separations considered. The $b_2$ parameter exhibits a distinct trend: it is accurately recovered and largely insensitive to $r_{\min}^{3\mathrm{pt}}$ for $\eta=1$ and $3$, whereas for $\eta=5$ it shows a mild overestimate.\\
To assess the quality of the fits, the bottom panel of the figure shows the reduced $\chi^2$ associated with the total likelihood defined in Eq.~\ref{eq:loglike}. In particular, we define $\chi^2 \equiv -2\log \mathcal{L}_{\mathrm{tot}}\,$, where $\mathcal{L}_{\mathrm{tot}}$ is the joint likelihood over all Minerva realizations. In this setup, the effective number of degrees of freedom is given by the total number of data bins entering the combined data vector, i.e. $N_{\mathrm{dof}} = N_{\mathrm{real}}\,N_{\mathrm{bin}} - N_{\mathrm{par}}\,,$
where $N_{\mathrm{bin}}$ is the number of bins in the data vector of a single realization, $N_{\mathrm{real}}$ is the number of realizations, and $N_{\mathrm{par}}=6$ is the number of free parameters in the fit.
The results indicate that the model provides a good fit to the data across all the scales and triangle configurations explored in this analysis.\\
Based on these results, we restrict our 3PCF analysis on scales larger than $r_{\text{min}}^{\text{3pt}} = 65 \, h^{-1} \text{Mpc}$ and to triangle configurations identified by setting $\eta=3$. With this choice, the estimated value of the growth rate $f$, as well as that of the linear bias $b_1$, is expected to suffer from a small bias ($\sim 3\%)$, which remains comparable to the uncertainty of our analysis. 
This is not the case, however, for the isotropic dilation parameter $\alpha$, whose overestimate is strongly significant at all scales.\\ 
One possible avenue to extend the analysis below $65\,h^{-1}\mathrm{Mpc}$, and to possibly alleviate these biases, is to employ a more sophisticated 1-loop perturbation theory model for the 3PCF, such as that proposed by \cite{Guidi:2023}.
However, the magnitude of the shifts reported for $f$ and $b_1$ is modest (at the level of a few percent) and would therefore contribute only marginally to the overall error budget of any realistic cosmological survey conducted over plausible volumes. This is not the case for $\alpha$: a na\"{i}ve extrapolation of the statistical uncertainties of our analysis to a survey volume of $\sim 40\,\hGpc$, comparable to that expected for the Euclid mission upon completion, suggests that the observed systematic shift would be of the same order as the expected statistical error. \\
\noindent Having assessed the reliability of our model and quantified the theoretical systematics affecting the estimation of key cosmological parameters, we now gauge the impact of explicitly modeling and measuring the anisotropic component of the 3PCF.
To this end, we have performed a 3PCF analysis using only its isotropic component ($\meas{\zeta}_{000}$, $\meas{\zeta}_{110}$, and $\meas{\zeta}_{220}$) and compared the results with those obtained when both isotropic and anisotropic multipoles are included. The outcomes, restricted to configurations with $r_{\min}^{3\mathrm{pt}} = 65\,h^{-1}\mathrm{Mpc}$ and $\eta = 3$, are shown in Fig.~\ref{fig:iso_vs_aniso} where probability contours obtained in the anisotropic 3PCF analysis  (blue) are superimposed to those estimated in the isotropic case (red).

\noindent The results demonstrate that including the anisotropic multipoles successfully break the $f$-$b_1$ degeneracy evident in the isotropic analysis. This leads to a substantial improvement in the precision of the tidal bias parameter, $b_{\mathcal{G}_2}$, while the posterior of $b_2$ is only marginally tightened. Moreover, the anisotropic multipoles enable meaningful constraints on $\varepsilon$, which remains largely unconstrained in the isotropic-only case, and slightly enhance the precision of the $\alpha$ estimates.
These findings corroborate those obtained from a similar analysis  performed in Fourier space, i.e. using the halo bispectrum, by \cite{Rizzo:2023} \footnote{This analysis, however, did not account for the effect of AP geometric distortions.}, and reinforce the conclusion that the anisotropic multipoles of three-point statistics contain significant cosmological information and should be included in clustering analyses.

\subsection{Joint 2PCF and 3PCF analysis}
\label{sec:3PCF+2PCF}
\begin{figure}
    \centering \includegraphics[width=1.\textwidth]{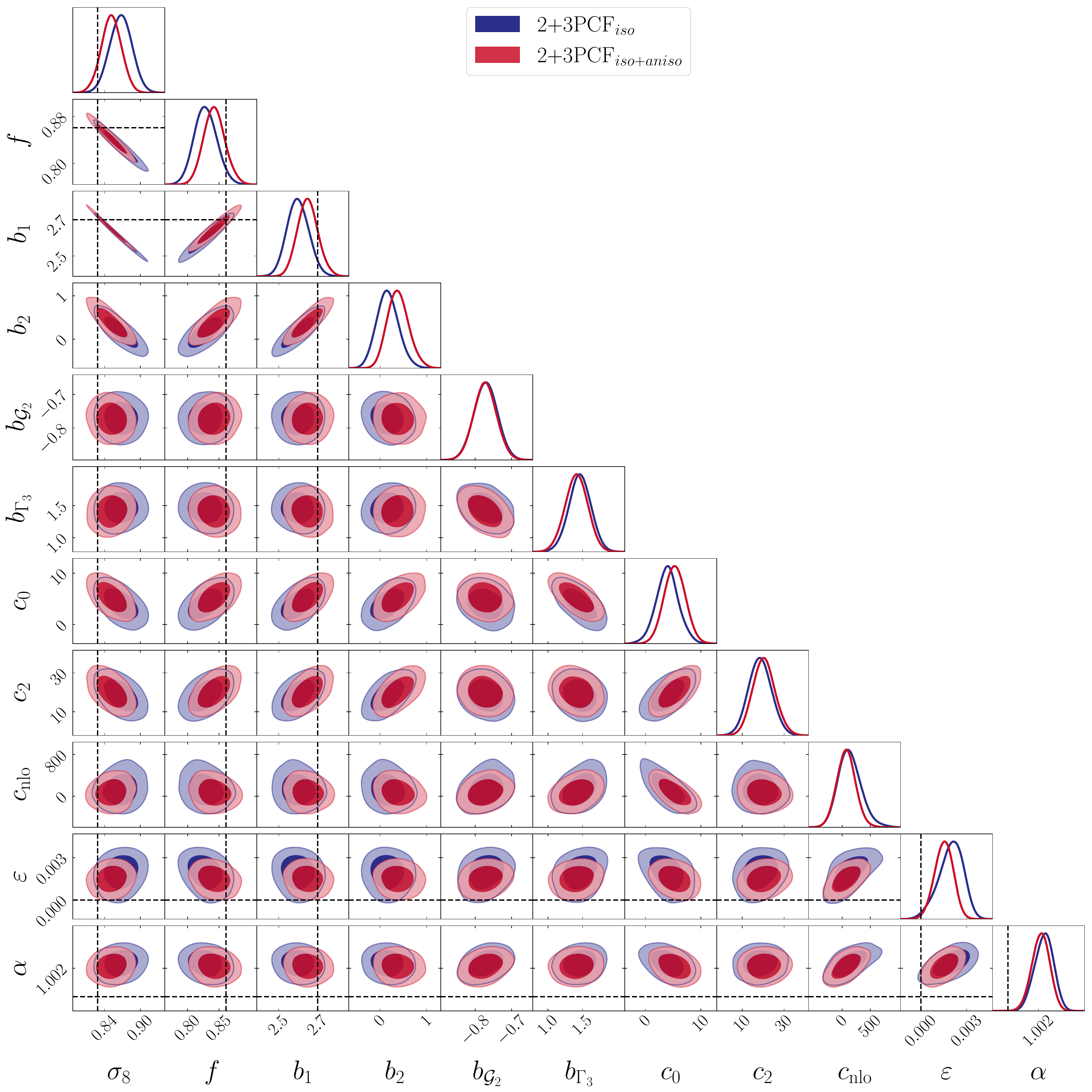}
    \caption{\small Marginalized 1 and 2D posterior probability contours for all the free parameters considered in this work. The results of the 2PCF-only analysis are depicted in red, while those coming from the joint  2+3PCF$_{\rm{iso+aniso}}$ analysis in blue. Reference values are indicated by horizontal and vertical black dashed lines when available.
    As in Fig.~\ref{fig:iso_vs_aniso} the results are obtained for survey volume of $1000\,\hGpc$ and triangles sizes above $r_{\min}^{\mathrm{3pt}}=65\,\hMpc$. }
\label{fig:joint_2pt}
\end{figure}

To combine information from the 2-point and 3-point statistics we compare
a composite datavector comprising 2PCF and 3PCF multipoles with the predictions of the anisotropic 1-loop 2PCF and tree-level 3PCF models described in Sec.~\ref{sec:model}.
As in the previous section, we consider two separate cases to appreciate the impact of the anisotropic multipoles. The first one considers all the 3PCF multipoles (3PCF$_{\rm{iso+aniso}}$: $\meas{\zeta}_{000}$, $\meas{\zeta}_{110}$, $\meas{\zeta}_{220}$, $\meas{\zeta}_{202}$, $\meas{\zeta}_{112}$, $\meas{\zeta}_{312}$). The second case considers only the isotropic ones (3PCF$_{\rm{iso}}$: $\meas{\zeta}_{000}$, $\meas{\zeta}_{110}$, $\meas{\zeta}_{220}$).\\
Since combining 2-point and 3-point statistics can potentially break the degeneracy among the parameters $\sigma_8$, $f$, and $b_1$ \cite{Veropalumbo:2021}, in this joint analysis, we allow $\sigma_8$ to vary freely.\\
The outcomes of the joint analysis are presented in the triangle plot of Fig.~\ref{fig:joint_2pt}, where the posterior probability contours from the 2+3PCF$_{\rm{iso}}$ case (blue) are compared with those from 2+3PCF$_{\rm{iso+aniso}}$ (red).
The black dashed lines indicate expected values.
As in the 3PCF-only analysis discussed in Sec.~\ref{sec:3PCF-only}, we have considered a total volume of about $1000\,h^{-3}\mathrm{Gpc^3}$, and have limited the 3PCF analysis to scales larger than $r_{\min}^{\mathrm{3pt}}=65\,\hMpc$ and to triangles characterized by $\eta = 3$. For the 2PCF, we restrict the analysis to separations larger than $r_{\min}^{\mathrm{2pt}}=37.5\, \hMpc$, for which the 1-loop 2PCF model provides an adequate description as shown in Appendix \ref{app:A1}.

\begin{figure}
    \centering \includegraphics[width=1.\textwidth]{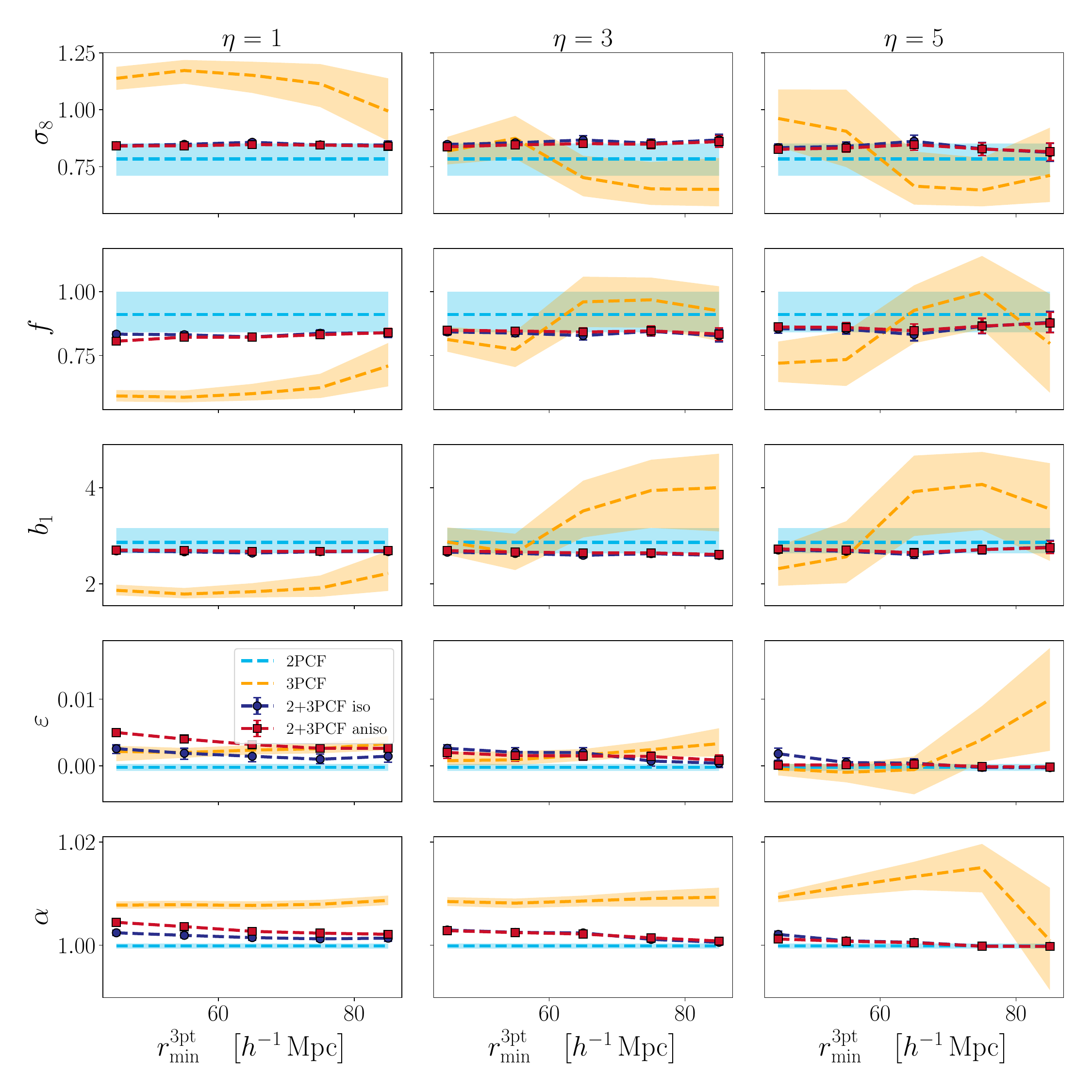}
    \caption{\small Best fit values of the parameters $f$, $\sigma_8$, $b_1$, $\alpha$ and $\varepsilon$ as a function of $r_{\min}^{\mathrm{3pt}}$ for three different values of the shape parameter $\eta$. Each row of the figure shows the best fit estimate of one of the aforementioned parameters as measured in a 2+3PCF$_{\mathrm{iso}}$ (blue), 2+3PCF$_{\mathrm{iso+aniso}}$ (red), 2PCF-only (cyan) and 3PCF-only (yellow) analysis, for a different value of $\eta$. In all the cases, error bars represent the 68\% uncertainty regions as computed from the relative 1D marginalized posterior distribution. }
\label{fig:params_1d_joint}
\end{figure}

\noindent The addition of the anisotropic 3PCF multipoles shifts the best-fit values of $\sigma_8$, $f$, $b_1$, and $\varepsilon$ closer to their expected values. These shifts are small but systematically in the right direction. However, they are not accompanied by a general reduction in the uncertainties, except in the case of $\varepsilon$, for which the posterior distribution contours become slightly tighter.
To further assess the impact of including the anisotropic 3PCF multipoles, we repeated the analysis while varying $r_{\min}^{\mathrm{3pt}}$ and $\eta$, keeping $r_{\min}^{\mathrm{2pt}} = 37.5\,\hMpc$ fixed.
The results are presented in Fig.~\ref{fig:params_1d_joint}, which shows the best-fit values of $\sigma_8$, $f$, $b_1$, $\alpha$, and $\varepsilon$ as a function of $r_{\min}^{3\mathrm{pt}}$ for three different values of $\eta$, indicated above the top panels. In each panel, blue circles and red squares represent the best-fit values obtained from the 2+3PCF$_{\mathrm{iso}}$ and 2+3PCF$_{\mathrm{iso+aniso}}$ analyses, respectively. The error bars correspond to the 68\% uncertainty intervals derived from the 1D marginalized posterior distributions. For comparison, the dashed cyan and yellow curves, with their respective shaded 68\% confidence regions, show the results from the 2PCF-only and anisotropic 3PCF-only analyses.

\begin{figure}
    \centering \includegraphics[width=1.\textwidth]{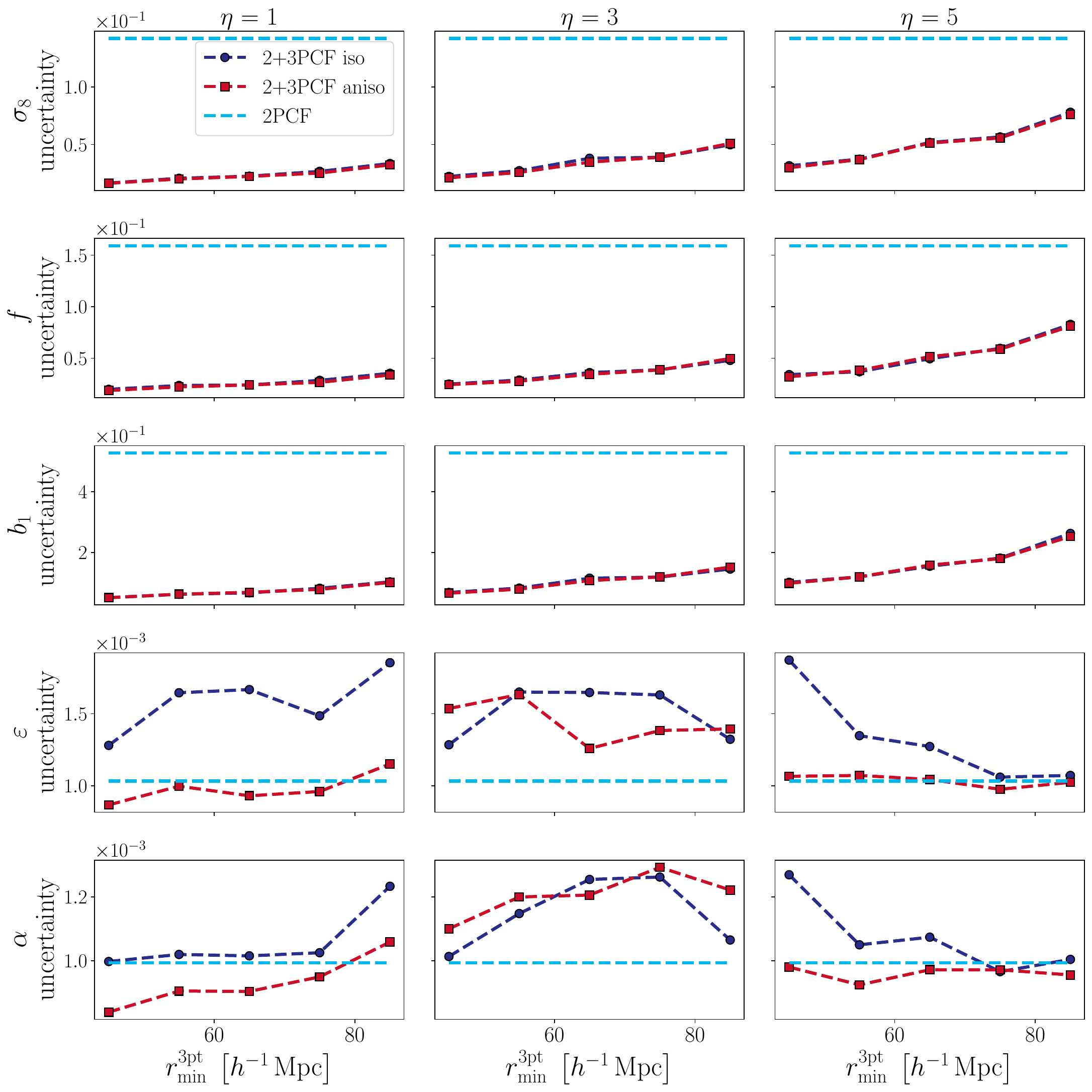}
    \caption{\small Amplitude of the 68\% uncertainty regions computed from the 1D marginalized posterior distributions of the parameters $f$, $\sigma_8$, $b_1$, $\alpha$ and $\varepsilon$. Each row of the figure shows the amplitude of the uncertainty region of one of the aforementioned parameters as a function of $r_{\min}^{\mathrm{3pt}}$ and for different values of the shape parameter $\eta$, indicated above each top panel. Results are reported for the 2+3PCF$_{\mathrm{iso}}$ (blue), 2+3PCF$_{\mathrm{iso+aniso}}$ (red) and 2PCF-only (cyan) case. }
\label{fig:errors_1d_joint_selected}
\end{figure}

\noindent These results demonstrate that combining the 2PCF and 3PCF statistics mitigates the $f$–$\sigma_8$–$b_1$ degeneracy present in the individual 2PCF and 3PCF analyses, thereby reducing the bias in their estimates. Remarkably, the best-fit values remain largely insensitive to the choice of the minimum scale $r_{\min}^{3\mathrm{pt}}$ and to the triangle configurations considered.
In contrast, the joint analysis reduces, but does not fully eliminate, the bias in the $\alpha$ and $\varepsilon$ estimates observed in the 3PCF-only analysis, as quantified by the dashed yellow curves in the panels. The magnitude of this positive bias is small but non-negligible, and it decreases with increasing $r_{\min}^{3\mathrm{pt}}$ and $\eta$. This trend indicates a gradual recovery of the true values (and convergence toward the 2PCF-only results) on scales where nonlinear effects become negligible and the 3PCF contributes little additional information beyond that already provided by the 2PCF.
This interpretation is supported by the results shown in Fig.~\ref{fig:errors_1d_joint_selected}, which displays the amplitude of the marginalized 68\% uncertainty intervals corresponding to the estimates of Fig.~\ref{fig:params_1d_joint}. The color scheme matches that of the previous figure: blue circles represent the joint isotropic analysis, red squares the isotropic+anisotropic case, and the cyan dashed line the 2PCF-only constraints.

\noindent These results confirm those shown in Fig.~\ref{fig:joint_2pt}, namely that the inclusion of the anisotropic 3PCF information substantially reduces the uncertainties on the AP and isotropic dilation parameters, particularly on $\varepsilon$, compared to the 2+3PCF$_{\mathrm{iso}}$ analysis, while leaving the other parameters essentially unaffected. It is worth noting, however, that the joint analysis does not improve on $\alpha$ and $\varepsilon$ relative to the 2PCF-only case across most of the scales and triangle configurations considered. 
This arises from the small, yet significant, tension between the estimates of these parameters obtained from the 2PCF and 3PCF analyses, as shown in Fig.~\ref{fig:params_1d_joint}. This discrepancy prevents the combination of the two statistics from producing a substantial reduction in the uncertainties on $\alpha$ and $\varepsilon$, and in some cases even increases their error bars relative to the 2PCF-only results.\\
Moreover, while the uncertainties on $f$, $\sigma_8$, and $b_1$ in the two joint analyses decrease steadily as $r_{\min}^{3\mathrm{pt}}$ is decreased, the uncertainties on $\alpha$ and $\varepsilon$ do not exhibit a clear trend with this scale.
For $\eta = 3$, the uncertainties increase as $r_{\min}^{3\mathrm{pt}}$ decreases.
A similar trend is observed for $\eta = 5$ in the 2+3PCF$_{\mathrm{iso}}$ case, whereas in the 2+3PCF$_{\mathrm{iso+aniso}}$ analysis, the uncertainties remain essentially independent of $r_{\min}^{3\mathrm{pt}}$. 
In contrast, for $\eta = 1$, the uncertainties in the AP parameters obtained from the 2+3PCF$_{\mathrm{iso+aniso}}$ analysis decrease as $r_{\min}^{3\mathrm{pt}}$ is reduced, resulting in tighter constraints than those from the 2PCF-only analysis. However, this occurs precisely in the regime where systematic biases in the AP parameters are most pronounced. This behavior is not observed in the 2+3PCF$_{\mathrm{iso}}$ case, where the uncertainties on $\alpha$, and especially on $\varepsilon$, remain larger than in the 2PCF-only analysis.\\
Finally, while combining the 2PCF with the 3PCF does not improve the constraints on the AP parameters relative to the 2PCF-only analysis, the joint fit effectively breaks the $f$–$b_1$–$\sigma_8$ degeneracy and substantially reduces their uncertainties. However, the addition of the 3PCF multipoles has a negligible impact on these parameters and does not improve the quality of the results.

\section{Conclusions}
\label{sec:conclusions}
In this work, we have presented a new implementation of both the anisotropic 3PCF model and its estimator. To validate these tools, we analyzed a large suite of simulated halo catalogs and used the results to assess whether, and to what extent, the inclusion of anisotropic multipoles enhances clustering analyses based on the 3PCF alone, or in combination with the more standard anisotropic 2PCF.
The anisotropic 3PCF model considered in this work is based on the leading-order Eulerian perturbation theory. It relies on the halo bispectrum model of 
\cite{Scoccimarro:1999}, which we expand on the TripoSH basis following \cite{Sugiyama:2019}. This approach reduces the dimensionality of the problem from a 5D function to a set of 2D multipoles, which are then mapped to configuration space via a 2D Hankel transform performed using the 2D FFT-Log algorithm 
\cite{Fang:2020}.
This procedure is computationally intensive due to the TripoSH transform. To speed it up, we have implemented a two-step procedure. First, we use \texttt{HEALPix} \cite{Gorski:2005} to efficiently perform standard spherical harmonics transforms. The resulting spherical harmonics coefficients are then integrated in 1D and summed over to compute the tripolar multipole moments \cite{Fang:2020}.
To assess the performance of our model, we compared its predictions with those obtained using the \texttt{HITOMI} package \cite{Sugiyama:2023}
and remain well within the statistical uncertainties expected for analyses of the same halo population over a volume of
$\sim 1000\,h^{-3}\mathrm{Gpc}^3$, corresponding to the total volume of the mock halo catalogs used in our study. Further details on these validation tests are provided in Appendix~\ref{app:B}.
This 3PCF model implementation is now publicly available in the form of a self-contained package dubbed \texttt{Mod3L}.\\
Our 3PCF estimator is based on the approach proposed by \cite{Slepian:2015, Slepian:2018} to efficiently measure the anisotropic modes of the 3PCF in large spectroscopic catalogs. This method allows one to reduce the computational cost in the triplet counting from an $\mathcal{O}(N^3)$ scaling to that of a  more manageable $\mathcal{O}(N^2)$ scaling.
The dramatic reduction in computation time is achieved utilizing the properties of spherical harmonics to express the local contribution to the 3PCF in terms of the direction of two triangle sides, without  requiring knowledge of their relative angle. As a result, all contributions can be pre-computed, thereby accelerating the calculation.
To validate our 3PCF estimator we compared our results with those obtained from the public \texttt{Triumvirate} package \cite{Wang:2023}. 
As shown in Appendix \ref{app:A}, when the two estimators are applied to the same set of 298 halo catalogs, spanning a total volume of $1000\,h^{-3}\mathrm{Gpc}^3$, their measurements agree almost everywhere within the $1\sigma$ statistical uncertainties.
The rare $2\sigma$ discrepancies can be attributed to the grid-based nature of \texttt{Triumvirate}, which makes its results more sensitive to the choice of mesh resolution.
This anisotropic 3PCF estimator, dubbed \texttt{MeasCorr}, is now publicly available. To the best of our knowledge, it constitutes the first public implementation of the pair-counting version of the 3PCF estimator proposed by \cite{Slepian:2018}, thereby providing the community with a practical tool that was previously unavailable. \\
To perform an absolute, rather than a relative, validation of  both the 3PCF model and its estimator and to evaluate random and systematic errors of the corresponding anisotropic 3PCF analysis we have used 298 halo catalogs extracted from the $z=1$ outputs of the Minerva N-body simulations \cite{Grieb:2016}. This choice and that of the minimum halo mass are motivated by the goal of comparing our results to those of \cite{Oddo:2020, Oddo:2021, Rizzo:2023}, who performed a similar analysis in Fourier space. Covariant errors were estimated from an independent set of 3000 halo catalogs extracted from the Pinocchio simulations \cite{Monaco:2013, Munari:2017}, which were generated assuming the same cosmological model as the Minerva simulations but adopting a slightly lower mass threshold. To assess the adequacy of the Pinocchio-based covariance matrix for the analysis of the Minerva mock catalogs, we compared two covariance matrices: one derived from 298 Minerva halo catalogs and the other from the corresponding 298 Pinocchio halo catalogs, obtained by evolving the same initial conditions as the Minerva simulations. The two matrices exhibit the same covariance structure, with the Pinocchio-based variance found to be 5–10 \% smaller, in agreement with the findings of \cite{Blot:2019,Rizzo:2023}. After applying the standard corrections of \cite{Hartlap:2006, Percival:2014} to account for the finite number of mocks, the Pinocchio-based covariance matrix is adequate for the analyses presented in this work.\\
We performed two different analyses.
In the first, we considered only the 3PCF, for which we measured the first three isotropic multipoles ($\meas{\zeta}_{000}$, $\meas{\zeta}_{110}$, and $\meas{\zeta}_{220}$) together with the first three anisotropic multipoles ($\meas{\zeta}_{202}$, $\meas{\zeta}_{112}$, and $\meas{\zeta}_{312}$).
To compare these measurements with the model predictions, we estimated the posterior probability distribution of the growth rate $f$, the clustering amplitude $\sigma_8$, the AP parameter $\varepsilon$, the isotropic dilation $\alpha$, the bias coefficients $b_1$, $b_2$, and $b_{\mathcal{G}2}$, as well as the EFT counterterms $c_1$, $c_2$, and $c_{\mathrm{nlo}}$. Instead, we fix the clustering amplitude $\sigma_8$ to the value adopted in the Minerva simulations, since within this 3PCF model it is fully degenerate with $b_1$ and $f$.
Our results show that the inclusion of anisotropic multipoles in the 3PCF analysis effectively breaks the $f$–$b_1$ degeneracy and reduces the uncertainties on all parameters, with the exception of $b_2$. This finding is consistent with the corresponding Fourier-space analysis of \cite{Rizzo:2023}. \\
The $\alpha$ parameter is systematically overestimated in both the isotropic and anisotropic analyses. This bias is robust to changes in the minimum scale and triangle configurations. While the magnitude of the effect is small, less than 1\%, it is nonetheless detectable owing to the large effective volume of this validation test.
No analogous bias is observed for $\varepsilon$.\\
These results suggest that the origin of this discrepancy, absent in the 2PCF-only analysis performed for comparison, lies in the limitations of the 3PCF model adopted here, which is based on a tree-level expansion and is expected to fail on small scales and for squeezed triangle configurations.\\
In the second analysis, we combined the 2PCF with the 3PCF data vectors, considering two versions of the latter: one including and one excluding the anisotropic multipoles. 
Moreover, in the joint analyses, we allowed all parameters to vary freely, including $\sigma_8$. The results show that a joint analysis performed using only the isotropic multipoles of the 3PCF significantly alleviates the $f$–$\sigma_8$–$b_1$ degeneracy. It also reduces the overestimation of $\alpha$, bringing its best-fit value to well below 1\% of the true value. It is worth emphasizing, however, that the magnitude of this residual bias is very small and would contribute negligibly to the error budget of a joint 2PCF+3PCF clustering analysis conducted on a survey of similar tracers covering a volume comparable to that of state-of-the-art redshift surveys.\\
Repeating the joint 2PCF+3PCF analysis with anisotropic multipoles ($\meas{\zeta}_{202}$, $\meas{\zeta}_{112}$, $\meas{\zeta}_{312}$) reduces the statistical uncertainties of the AP and isotropic dilation parameters, but fails to correct their systematic overestimation. The apparent gains stem mainly from small scales and squeezed configurations, precisely where the tree-level model is least reliable and biases are largest. Consequently, the joint analysis offers no real improvement over the 2PCF-only case across most scales and triangle shapes, and in some cases even worsens the constraints on $\alpha$ and $\varepsilon$.
The enhancement in precision for the remaining cosmological parameters proves to be modest, with the uncertainties on $\sigma_8$, $f$, and $b_1$ shrinking by only about a few percents.
This demonstrates that, unlike in the 3PCF-only analyses, the addition of anisotropic multipoles contributes little extra information in the joint analysis. \\
 To further investigate these seemingly surprising results, we carried out an analysis analogous to that of \cite{Gualdi:2020}, but employed a different model with a parametrization consistent with that adopted in this work.
Specifically, we generated synthetic data vectors from a tree-level anisotropic 3PCF model combined with a 1-loop 2PCF model and repeated the same joint analysis applied to the simulated data. As detailed in Appendix \ref{app:D}, in this case the inclusion of anisotropic 3PCF multipoles substantially reduces the statistical uncertainty on $\varepsilon$, while leaving the errors on the other parameters, including $\alpha$, essentially unchanged.
The gain in constraining power grows more pronounced when the 3PCF analysis is pushed to smaller scales, with the uncertainty on $\varepsilon$ decreasing by as much as $\sim 50$ \% for $r_{\min}^{3\mathrm{pt}} = 45\,\hMpc$ and $\eta = 1$.
These results confirm that the limited impact of the anisotropic multipoles in our joint analysis originates from the shortcomings of the tree-level 3PCF model which,
at the redshift and for the tracers considered in our analysis, fails on small scales and for isosceles triangle configurations. 
\\
As a natural next step, we plan to extend this analysis by employing an improved anisotropic 3PCF model based on the 1-loop framework we have previously developed \cite{Guidi:2023}, thereby enabling us to probe the nonlinear regime and exploit the anisotropic information encoded in the 3PCF.
The magnitude of this improvement is expected to depend on both the tracer population and the redshift of the sample, and should be carefully weighted against other potential sources of systematic error when assessing its impact on the analysis of real datasets. Ultimately, a definitive evaluation will require dedicated studies based on realistic simulated catalogs that incorporate survey-specific observational effects.

\noindent 

\acknowledgments{ 
\label{sec:ack}
The authors thank M.S. Wang and F. Beutler, N.S. Sugiyama, M. Moresco and H. Gil-Marín for useful discussions and comments on the results. Finally, we thank A.G. Sanchez and P. Monaco for providing us with the Minerva and Pinocchio simulations and for their useful comments on the manuscript.\\
This work was supported by the ASI/INAF agreement n. 2018-23-HH.0 “Scientific activity for Euclid mission, Phase D”, the research grant ‘From Darklight to DM: understanding the galaxy/matter connection to measure the Universe’, the INFN project “InDark”. MG ackowledges the financial contribution from the grant PRIN-MUR 2022 2022NY2ZRS 001 “Optimizing the extraction of cosmological information from Large Scale Structure analysis in view of the next large spectroscopic surveys” supported by Next Generation EU.}

\appendix
\section{Impact of the minimum scale cut on the 2PCF analysis.}
\label{app:A1}

\begin{figure}
    \centering \includegraphics[width=1.\textwidth]{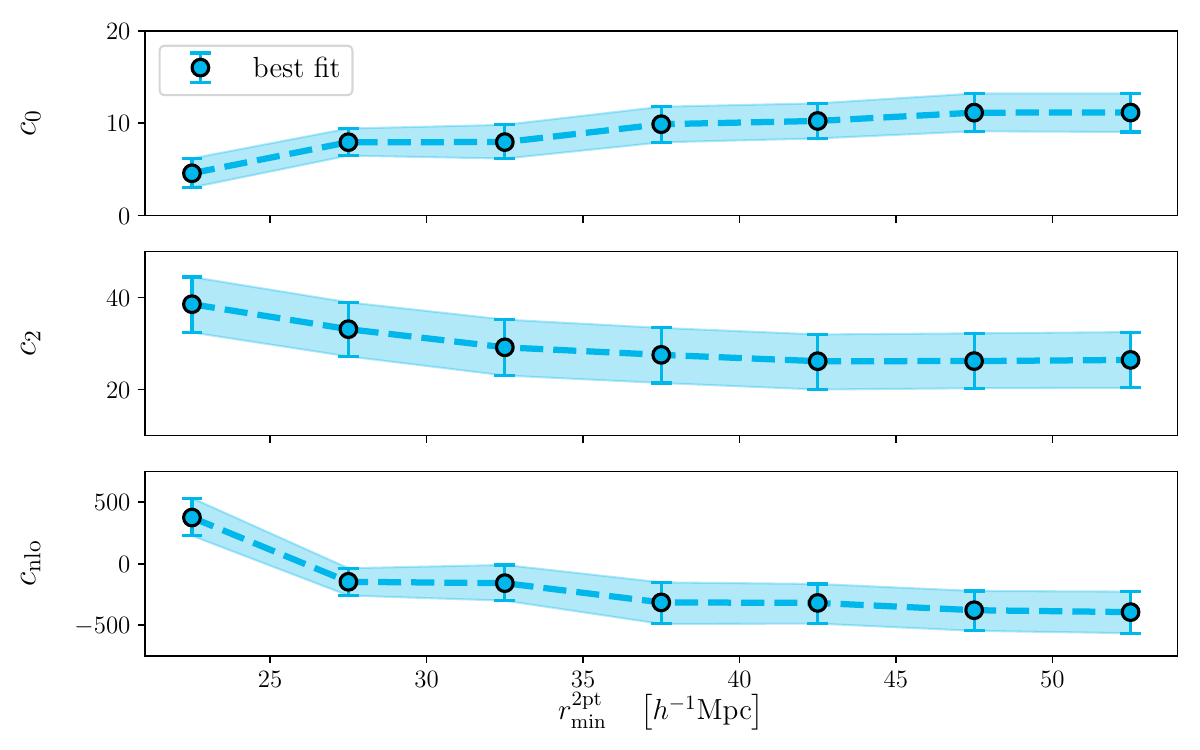}
    \caption{\small Best-fit values of the EFT counterterms $c_0$, $c_2$ and $c_{\mathrm{nlo}}$ as a function of the minimum separation $r_{\min}^{\mathrm{2pt}}$. The filled dots represent the value of the best fit parameters estimated from the 298 halo catalogs. Error bars and the shaded regions indicate their 68\% uncertainty interval.}
\label{fig:2pcf_ctrms_rmin_dep}
\end{figure}

In this appendix, we assess the sensitivity of the EFT-based 2PCF model presented in Sec.~\ref{sec:model} to the choice of the minimum pair separation, $r_{\min}^{\mathrm{2pt}}$, used in the analysis. To this end, we compare the monopole and quadrupole of the 2PCFs measured from 298 Minerva halo catalogs with the predictions of our model, estimating the best-fit values of the EFT counterterms as well as the
 $f$, $b_1$, $\varepsilon$ and $\alpha$ parameters. 
The value of $\sigma_8$  is kept fixed to its reference value in the simulations. Covariant errors are numerically estimated from the 3000 2PCFs measured in the Pinocchio mock catalogs.
Figure \ref{fig:2pcf_ctrms_rmin_dep} presents the best-fit estimates of $c_0$, $c_2$ and $c_{\mathrm{nlo}}$ as a function of $r_{\min}^{\mathrm{2pt}}$. 
The blue dots represent the best-fit values obtained from the 298 Minerva simulations, while the error bars indicate their 68\% credible intervals.

\begin{figure}
    \centering \includegraphics[width=0.95\textwidth]{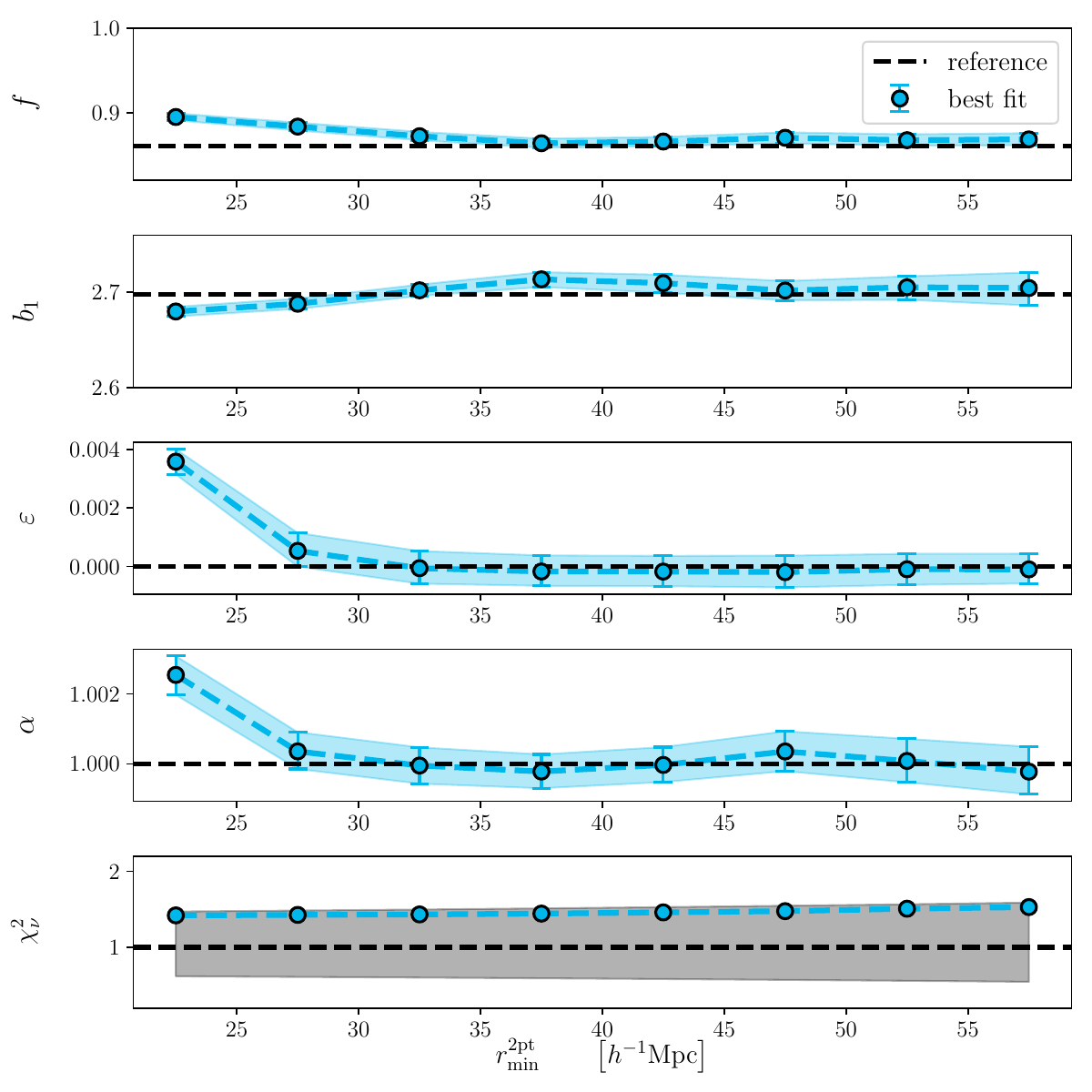}
    \caption{\small Best-fit values of the parameters $f$, $b_1$, $\alpha$ and $\varepsilon$ as a function of the minimum pair separation $r_{\min}^{\mathrm{2pt}}$ (top and mid panels). The filled dots represent the mean value of the best fit estimated from the 298 Minerva halo catalogs. Error bars and the shaded regions indicate their 68\% uncertainty intervals. In the bottom panel we show the reduced $\chi^2$ values averaged over the 298 Minerva simulation as a function of $r_{\min}^{\mathrm{2pt}}$ (blue dots) along with their 1$\sigma$ uncertainty (blue errorbar and band). The grey band represent the 95\% confidence region associated to a $\xi^2$ distribution characterized by $N_{bin} - 8$ d.o.f, where $N_{bin}$ is the number of bins composing the datavector and 8 the number of free parameters in the analysis. }
\label{fig:2pcf_fb_rmin_dep}
\end{figure}

\noindent The results indicate no significant running of the EFT counterterms down to minimum separations of $27.5\,\hMpc$. Above this scale, they do not provide significant evidence of a breakdown of the model. However, mild trends are visible for $r_{\min}^{\mathrm{2pt}}< 37.5\,\hMpc$ for all three counterterms.
The same test was performed for the parameters $f$, $b_1$, $\alpha$ and $\varepsilon$ with the results shown in Fig. \ref{fig:2pcf_fb_rmin_dep}, analogous to Fig. \ref{fig:rmin_dep} from the 3PCF analysis. Unlike the EFT counterterms, both $f$ and $b_1$ parameters exhibit a significant running for $r<37.5 \hMpc$, suggesting that the EFT model is inadequate for describing the 2PCF at separations smaller than $37.5 \hMpc$,  despite providing a good fit to the measured 2PCF, as indicated by the reduced $\chi^2$ shown in the bottom panel. The parameters $\alpha$ and $\varepsilon$ instead, demonstrate to be more robust to the choice of the minimum separation $r_{\min}^{\mathrm{2pt}}$, providing estimates consistent with expectation values down to $27.5\,\hMpc$. These findings motivate our choice to restrict the joint 2PCF+3PCF analysis to scales above $r_{\min}^{\mathrm{2pt}} = 37.5 \hMpc$.

\noindent A closer inspection of the first panel reveals that the best-fit value of the growth rate $f$ is consistently larger than expected. The magnitude of the mismatch is approximately 1\%, comparable to the reported error bars. However, its magnitude is significantly smaller than what is anticipated in the analysis of real data from upcoming spectroscopic surveys, which will probe considerably smaller volumes.

\section{Comparing 3PCF estimators: \texttt{MeasCORR} vs. \texttt{Triumvirate}}
\label{app:A}
In this appendix, we compare the multipole moments measured by the \texttt{MeasCORR} 3PCF estimator with those obtained using \texttt{Triumvirate} \cite{Wang:2023}. \texttt{Triumvirate} is a \texttt{Python} package implementing the FFT-based version of the Slepian-Eisenstein estimator \cite{Slepian:2015}. It builds upon an earlier implementation available in the \texttt{HITOMI} package \cite{Sugiyama:2023}.\\
The core idea behind \texttt{Triumvirate}'s approach is to evaluate the convolution integrals defined in Eq.(\ref{eq:alm}) using the FFTLog algorithm. This method, in principle, reduces the computational cost of the Slepian-Eisenstein estimator to an $\mathcal{O}(N\ln{N})$ scaling.
However, to achieve this, FFT-based estimators reconstruct the underlying density field by assigning each object to a cubic mesh grid, 
making them more susceptible to gridding and survey geometry effects compared to simpler pair-counting methods.
For a thorough discussion about pros and cons of FFT-based estimators we refer the reader to \cite{Slepian:2015a} and references therein.\\
To compare performances we have estimated the six 3PCF multipoles considered in the analysis on the entire set of 298 Minerva mock catalogs using \texttt{MeasCORR} and \texttt{Triumvirate}. \\
For \texttt{Triumvirate}, which relies on cubic grid partitioning, we considered two cases: a $256^3$ mesh grid and a  $320^3$ mesh grid. A Triangular Shaped Cloud mass assignment scheme was used in both cases.
For \texttt{MeasCORR} we used the same set up  as in Sec.~\ref{sec:threepoint_est}.
For both estimators the plane parallel approximation has been adopted.

\begin{figure}
    \centering \includegraphics[width=1.\textwidth]{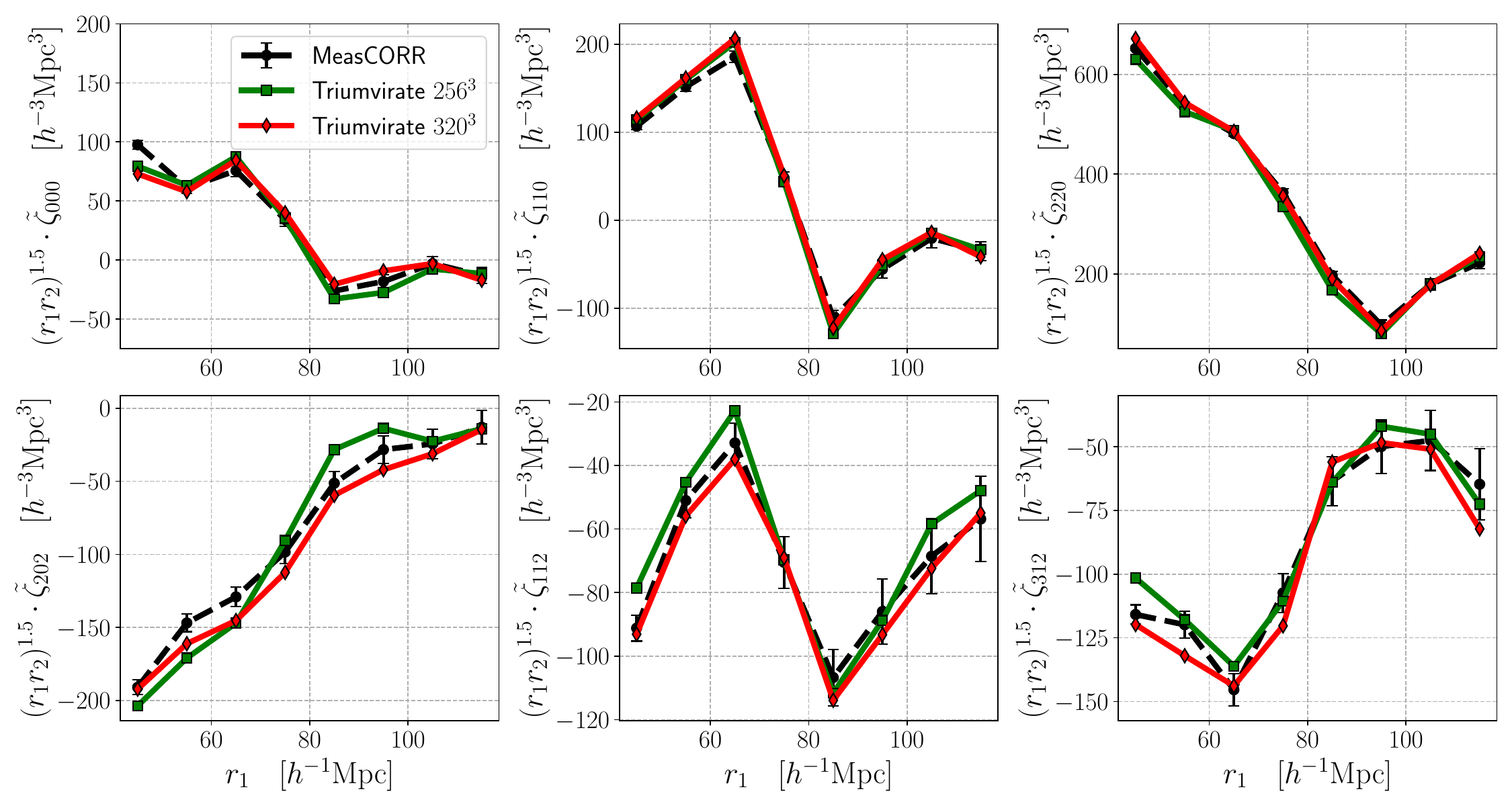}
    \caption{\small Average over the entire suite of 298 Minerva realizations of the six 3PCF multipole moments considered in our analysis as measured by \texttt{MeasCORR} (black dots and errorbars) and \texttt{Triumvirate}. Triangle configurations characterized by $r_2 = r_1+30\hMpc$ are displayed.
    The \texttt{Triumvirate} measurements have been obtained for two different partitionings of the box volume:  $256^3$ (green dots) and $320^3$ mesh grids (red dots). \texttt{MeasCORR} specifics are set as
    described in Sec.\ref{sec:threepoint_est} and the errorbars refer to the survey volume of $1000\hGpc$, corresponding to the aggregate volume of the entire suite of Minerva halo catalogs.}
\label{fig:triumvirate}
\end{figure}

\noindent 
The results are shown in Fig.~\ref{fig:triumvirate}. The values of the six multipoles were computed for a specific set of triangle configurations in which $r_2 = r_1 + 30\hMpc$ as a function of the triangle side $r_1$.
The \texttt{MeasCORR} measurements (black dots with errorbars) are compared with those obtained with 
\texttt{Triumvirate} using a $256^3$ (green squares) and a $320^3$ mesh grid (red diamonds). The errorbars associated to \texttt{MeasCorr} measurements represent their $rms$ scatter, i.e. the uncertainties expected for a dark matter halo catalog of about $1000\hGpc$. \\
The multipoles obtained from the two estimators agree within the error bars across most of the $r_1$ range considered with the only exception of the $\tilde{\zeta}_{202}$ multipole for which deviations of the order of $2\sigma$ are observed on all scales for both the $256^{3}$ and $320^{3}$ case.
We note that the difference between the \texttt{Triumvirate} results obtained using $320^3$ rather than $256^3$ grid points are of the same order as those between 
 \texttt{MeasCORR} and \texttt{Triumvirate}, with no apparent reduction of the mismatch with the increase of the number of gridpoints. This suggests that the results of the 
 \texttt{Triumvirate} estimator are sensitive to the choice fo the gridsize. In fact the mean halo separation in the Minerva catalogs
($\bar{l} \simeq 16.3 \,h^{-1}\,\mathrm{Mpc}$) is similar to the mesh size used by \texttt{Triumvirate}, resulting in a non negligible number of empty or scarcely populated cells, especially in the case of $320^3$ grid cells.
Conversely, adopting a coarser grid would introduce gridding effects that heavily bias the 3PCF estimates on the scales of our interest, potentially skewing the results.
In conclusion, while an FFT-based estimator may be beneficial for densely populated surveys or for evaluating the 3PCF of their window function (see e.g. \cite{Pardede:2022} and references therein), a pair counting estimator could be preferable for surveys characterized by a lower density.
We postpone a more detailed investigation of these effects to future studies.

\section{Comparing anisotropic 3PCF models: \texttt{Mod3l} vs.  \texttt{HITOMI}}
\label{app:B}
To validate our \texttt{Mod3l} package, we compared its 3PCF multipole predictions with those generated by the publicly available \texttt{HITOMI} package \citep{Sugiyama:2023}. \texttt{HITOMI} differs from \texttt{Mod3l} in several aspects.
The two main differences concern the strategy adopted to perform the TripoSH expansion of the bispectrum and the method used to map the bispectrum multipoles to their configuration space analogs. Specifically:

\begin{itemize}
    \item \texttt{HITOMI} uses the \texttt{Cuba} library \cite{Hahn:2005} to evaluate 
    the multidimensional integrals in Eq.~(\ref{eq:TripoSH}). \texttt{Mod3l}, on the other hand, approaches the calculation in two steps. The first one uses  \texttt{HEALPix} to perform a standard spherical harmonic decomposition. The second one computes the 1 dimensional integral defined Eq.(25) of \cite{Sugiyama:2019}.
    \item \texttt{HITOMI} does not estimate the 2-dimensional Hankel transform in Eq.~(\ref{eq:TriRelBtoZ}).
    Instead, it evaluates the integral through two nested calls of the 1-dimensional FFTLog algorithm.
\end{itemize}

\begin{figure}
    \centering \includegraphics[width=1.\textwidth]{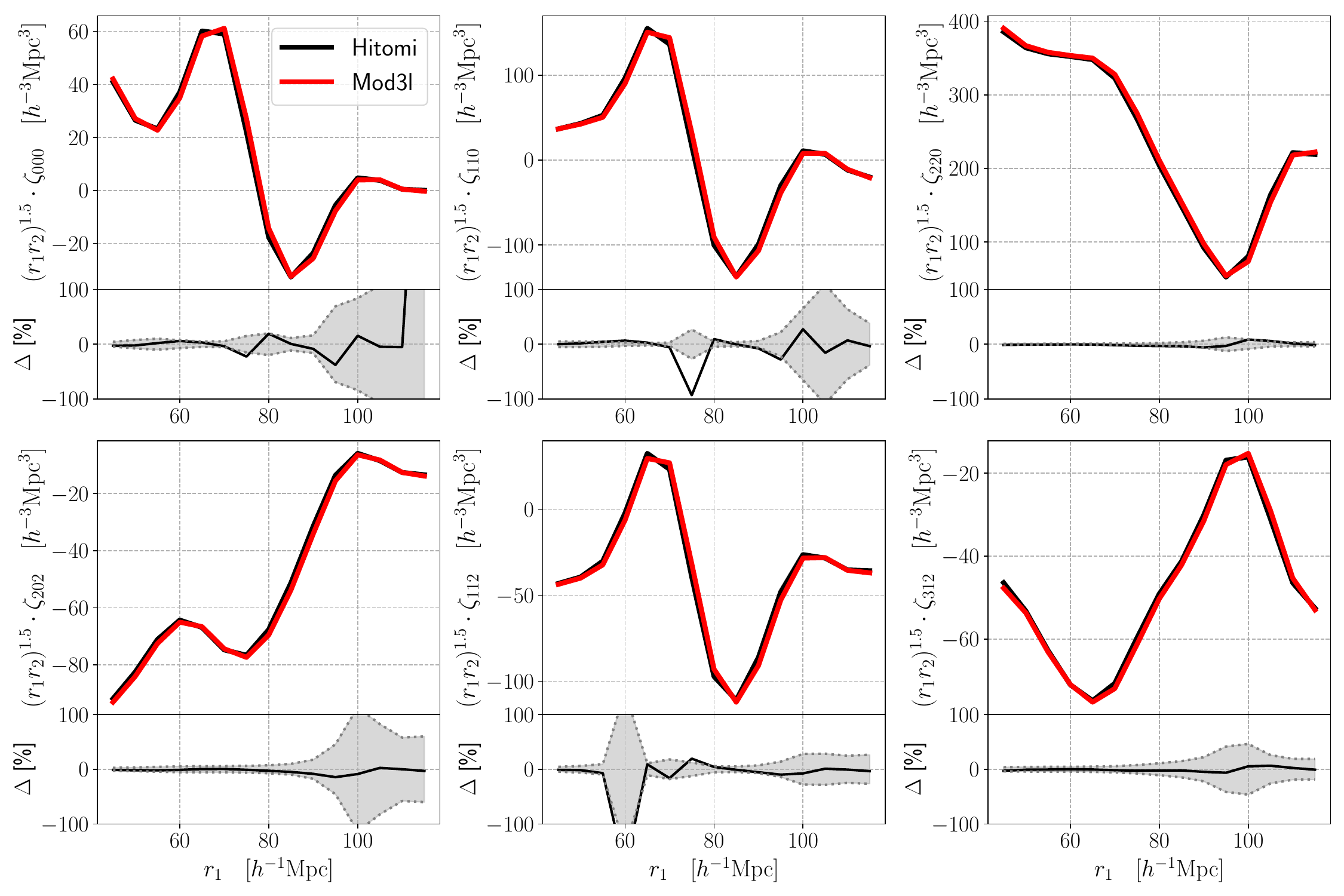}
    \caption{\small Isotropic (top panels) and anisotropic (bottom panels) 3PCF multipoles as a function of the length of triangle side $r_1$ for triangle configurations characterized by $r_2 = r_1 + 30\hMpc$. Each panel is divided in two sub-panels. In the upper ones, the 3PCF multipoles modeled by \texttt{Mod3l} (red line) are compared to those predicted by \texttt{HITOMI} (black). Both models have been obtained starting from the same matter power spectrum and RSD and nuisance parameters, namely $(f,\, b_1,\, b_2,\, b_{\mathcal{G}_2}) = (0.86,\, 2.7,\, 0.8,\, 0)$. The bottom sub-panels, on the other hand, display the percentage difference for each multipole between the two independent implementations of the model (black line), as well as the percentage deviation of the $1\sigma$ uncertainty, relative to \texttt{HITOMI}, expected for a survey of dark matter halos covering about 1000 $\hGpc$.}
\label{fig:hitomi}
\end{figure}

\noindent We compare the two sets of six multipoles obtained using the two methods in Fig.~\ref{fig:hitomi}. Both sets refer to the same choice of matter power spectrum and employ identical RSD and nuisance parameters, specifically $(f,\, b_1,\, b_2,\, b_{\mathcal{G}_2}) = (0.86,\, 2.7,\, 0.8,\, 0)$.
The multipoles shown in the upper part of the six panels are the same as in Fig.~\ref{fig:triumvirate} and refer to the same triangle configurations.
The black and the red curve show the \texttt{HITOMI} and the \texttt{Mod3l} predictions as a function of the triangle side $r_1$, respectively.

\noindent A simple visual inspection reveals that \texttt{HITOMI} and \texttt{Mod3l} are in good agreement. To quantify this match, we plot the per cent difference between the two models, $\Delta$.
The bottom part of each panel shows that the mismatch between the two models never exceeds
20\%, except at the zero crossing locations, and is of the same order as the 1$\sigma$ uncertainty expected from an aggregate volume of about 1000 $\hGpc$, indicating a very good agreement between the predictions of the two models.

\section{Joint analysis on synthetic noiseless datavector}
\label{app:D}
To further investigate the impact of the anisotropic 3PCF multipoles on the joint 2+3PCF analysis, particularly when incorporating information from nonlinear scales where the tree-level 3PCF model is known to break down, we repeated the analysis of Section~\ref{sec:3PCF+2PCF}, this time using noiseless synthetic data vectors instead of those derived from the mock catalogs. These vectors were constructed with a 1-loop 2PCF model, adopting $r_{\min}^{2\mathrm{pt}} = 37.5,h^{-1}\mathrm{Mpc}$, in combination with a tree-level 3PCF model, with $r_{\min}^{3\mathrm{pt}}$ and $\eta$ set to the same values as in the original analysis. The model parameters were set to resemble those characterizing the $z=1$ Minerva snapshots used in Sec.~\ref{sec:3PCF-only} and \ref{sec:3PCF+2PCF}. Their values are explicitly reported in Tab.~\ref{tab:mod_par}. 

\begin{table}[h!]
\centering
\begin{tabular}{c c c c c c c c c c c}
\hline
$\sigma_8$ & $f$ & $b_1$ & $b_2$ & $b_{\mathcal{G}_2}$ & $b_{\Gamma_3}$ & $c_0$ & $c_2$ & $c_\mathrm{nlo}$ & $\varepsilon$ & $\alpha$ \\
\hline
0.828 & 0.86 & 2.7 & 0.8 & -0.8 & 1.4 & 0 & 0 & 0 & 1 & 0 \\
\hline
\end{tabular}
\caption{Fixed model parameters used to generate the synthetic data vectors.}
\label{tab:mod_par}
\end{table}

\begin{figure}
    \centering \includegraphics[width=1.\textwidth]{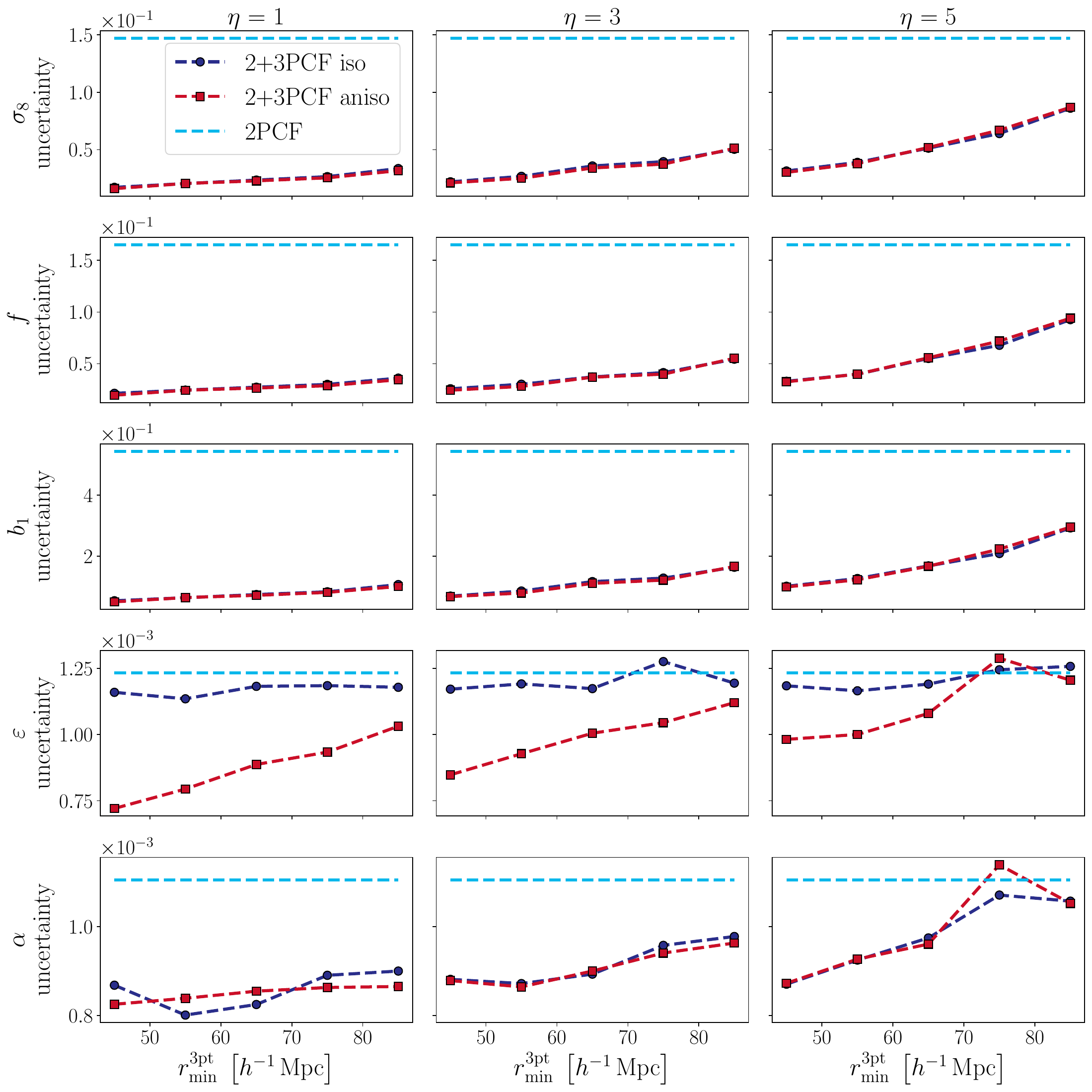}
    \caption{\small Amplitude of the 68\% uncertainty regions computed from the 1D marginalized posterior distributions of the parameters $f$, $\sigma_8$, $b_1$, $\alpha$ and $\varepsilon$ measured from synthetic noiseless datavectors. Each row of the figure shows the amplitude of the uncertainty region of one of the aforementioned parameters as a function of $r_{\min}^{\mathrm{3pt}}$ and for different values of the shape parameter $\eta$, indicated above each top panel. Results are reported for the 2+3PCF$_{\mathrm{iso}}$ (blue), 2+3PCF$_{\mathrm{iso+aniso}}$ (red) and 2PCF-only (cyan) case.}
\label{fig:errors_1d_joint_selected_sint2pt}
\end{figure}

\noindent The results are presented in Fig.~\ref{fig:errors_1d_joint_selected_sint2pt}, which, analogously to Fig.~\ref{fig:errors_1d_joint_selected}, shows the amplitude of the marginalized 68\% uncertainty intervals of the parameters $\sigma_8$, $f$, $b_1$, $\varepsilon$, and $\alpha$ as a function of $r_{\min}^{3\mathrm{pt}}$, for three different values of $\eta$ indicated above the top panels. The same color coding, symbols, and line styles as in the previous figure are adopted.

\noindent The results obtained from the synthetic data analysis show clear differences with respect to those from the mock catalogs. First, the uncertainties of all parameters, including $\alpha$ and $\varepsilon$, increase monotonically with $r_{\min}^{3\mathrm{pt}}$. Second, the inclusion of the 3PCF leads to a significant improvement compared to 2PCF-only case (cyan). This improvement is not limited to $\sigma_8$, $f$ and $b_1$ as in Fig.~\ref{fig:errors_1d_joint_selected}, but extends also to the AP {\bf and isotropic dilation} parameters. This reinforces our interpretation that the absence of gains relative to the 2PCF-analysis observed in Fig.~\ref{fig:errors_1d_joint_selected} originates from the tension between the 2PCF and 3PCF constraints on $\alpha$. Specifically, when only the monopole of the 3PCF is included, the improvement is confined to $\alpha$ and is up to $\sim35\%$ for the most extreme triangular configurations considered in the analysis. This result is not surprising since the monopole carries little information on the anisotropic warping parameter $\varepsilon$.
A substantial gain in this parameter is achieved once the anisotropic 3PCF multipoles are included, with its uncertainty reduced by up to $\sim 60\%$ relative to the 2PCF-only analysis and by $\sim 50\%$ relative to the 2+3PCF$_{\mathrm{iso}}$ case when the 3PCF analysis is extended down to small scales and squeezed triangle configurations.
These results further confirm that, for the massive halo sample at $z=1$ considered in our analysis, most of the constraining power of the anisotropic 3PCF is encoded in triangle configurations and scales that cannot be reliably captured by the tree-level model.\\
In agreement with the results of Sec.\ref{sec:3PCF+2PCF}, the inclusion of the anisotropic 3PCF multipoles does not lead to any noticeable improvement in the other parameters compared to the standard 2+3PCF$_{\mathrm{iso}}$ analysis. These findings are in qualitative agreement with those of \cite{Sugiyama:2021}, who likewise found that adding the anisotropic 3PCF multipoles leads to a significant reduction in the error on the anisotropic warping parameter $\varepsilon$, while yields negligible improvement for the dilation parameter $\alpha$.
In contrast, they appear to be in tension with the Fourier space analysis of \cite{Gualdi:2020}, who reported that the inclusion of anisotropic multipoles reduces the uncertainties on all key parameters, rather than solely $\varepsilon$.
However, this discrepancy may be only apparent and reflect methodological differences between the two analyses rather than a fundamental inconsistency.
In particular, \cite{Gualdi:2020} employed a different multipole expansion of the bispectrum, used a model characterized by an alternative set of parameters, including the local primordial non-Gaussianity parameter, $f_{NL}$, and focused on different tracers at a different redshift while probing larger scales than those covered by our mock analysis. 
These distinctions make a direct comparison between the two analysis
non-trivial, and likely explain the divergent conclusions regarding the constraining power of the 3-point statistics' quadrupole.

\newpage
\bibliographystyle{JHEP}
\bibliography{3PCF.bib}

@article{Mellier:2024,
       author = {{Euclid Collaboration} and {Mellier}, Y. and {Abdurro'uf} and {Acevedo Barroso}, J.~A. and {Ach{\'u}carro}, A. and {Adamek}, J. and {Adam}, R. and {Addison}, G.~E. and {Aghanim}, N. and {Aguena}, M. and {Ajani}, V. and {Akrami}, Y. and {Al-Bahlawan}, A. and {Alavi}, A. and {Albuquerque}, I.~S. and {Alestas}, G. and {Alguero}, G. and {Allaoui}, A. and {Allen}, S.~W. and {Allevato}, V. and {Alonso-Tetilla}, A.~V. and {Altieri}, B. and {Alvarez-Candal}, A. and {Amara}, A. and {Amendola}, L. and {Amiaux}, J. and {Andika}, I.~T. and {Andreon}, S. and {Andrews}, A. and {Angora}, G. and {Angulo}, R.~E. and {Annibali}, F. and {Anselmi}, A. and {Anselmi}, S. and {Arcari}, S. and {Archidiacono}, M. and {Aric{\`o}}, G. and {Arnaud}, M. and {Arnouts}, S. and {Asgari}, M. and {Asorey}, J. and {Atayde}, L. and {Atek}, H. and {Atrio-Barandela}, F. and {Aubert}, M. and {Aubourg}, E. and {Auphan}, T. and {Auricchio}, N. and {Aussel}, B. and {Aussel}, H. and {Avelino}, P.~P. and {Avgoustidis}, A. and {Avila}, S. and {Awan}, S. and {Azzollini}, R. and {Baccigalupi}, C. and {Bachelet}, E. and {Bacon}, D. and {Baes}, M. and {Bagley}, M.~B. and {Bahr-Kalus}, B. and {Balaguera-Antolinez}, A. and {Balbinot}, E. and {Balcells}, M. and {Baldi}, M. and {Baldry}, I. and {Balestra}, A. and {Ballardini}, M. and {Ballester}, O. and {Balogh}, M. and {Ba{\~n}ados}, E. and {Barbier}, R. and {Bardelli}, S. and {Barreiro}, T. and {Barriere}, J. -C. and {Barros}, B.~J. and {Barthelemy}, A. and {Bartolo}, N. and {Basset}, A. and {Battaglia}, P. and {Battisti}, A.~J. and {Baugh}, C.~M. and {Baumont}, L. and {Bazzanini}, L. and {Beaulieu}, J. -P. and {Beckmann}, V. and {Belikov}, A.~N. and {Bel}, J. and {Bellagamba}, F. and {Bella}, M. and {Bellini}, E. and {Benabed}, K. and {Bender}, R. and {Benevento}, G. and {Bennett}, C.~L. and {Benson}, K. and {Bergamini}, P. and {Bermejo-Climent}, J.~R. and {Bernardeau}, F. and {Bertacca}, D. and {Berthe}, M. and {Berthier}, J. and {Bethermin}, M. and {Beutler}, F. and {Bevillon}, C. and {Bhargava}, S. and {Bhatawdekar}, R. and {Bisigello}, L. and {Biviano}, A. and {Blake}, R.~P. and {Blanchard}, A. and {Blazek}, J. and {Blot}, L. and {Bosco}, A. and {Bodendorf}, C. and {Boenke}, T. and {B{\"o}hringer}, H. and {Bolzonella}, M. and {Bonchi}, A. and {Bonici}, M. and {Bonino}, D. and {Bonino}, L. and {Bonvin}, C. and {Bon}, W. and {Booth}, J.~T. and {Borgani}, S. and {Borlaff}, A.~S. and {Borsato}, E. and {Bosco}, A. and {Bose}, B. and {Botticella}, M.~T. and {Boucaud}, A. and {Bouche}, F. and {Boucher}, J.~S. and {Boutigny}, D. and {Bouvard}, T. and {Bouy}, H. and {Bowler}, R.~A.~A. and {Bozza}, V. and {Bozzo}, E. and {Branchini}, E. and {Brau-Nogue}, S. and {Brekke}, P. and {Bremer}, M.~N. and {Brescia}, M. and {Breton}, M. -A. and {Brinchmann}, J. and {Brinckmann}, T. and {Brockley-Blatt}, C. and {Brodwin}, M. and {Brouard}, L. and {Brown}, M.~L. and {Bruton}, S. and {Bucko}, J. and {Buddelmeijer}, H. and {Buenadicha}, G. and {Buitrago}, F. and {Burger}, P. and {Burigana}, C. and {Busillo}, V. and {Busonero}, D. and {Cabanac}, R. and {Cabayol-Garcia}, L. and {Cagliari}, M.~S. and {Caillat}, A. and {Caillat}, L. and {Calabrese}, M. and {Calabro}, A. and {Calderone}, G. and {Calura}, F. and {Camacho Quevedo}, B. and {Camera}, S. and {Campos}, L. and {Canas-Herrera}, G. and {Candini}, G.~P. and {Cantiello}, M. and {Capobianco}, V. and {Cappellaro}, E. and {Cappelluti}, N. and {Cappi}, A. and {Caputi}, K.~I. and {Cara}, C. and {Carbone}, C. and {Cardone}, V.~F. and {Carella}, E. and {Carlberg}, R.~G. and {Carle}, M. and {Carminati}, L. and {Caro}, F. and {Carrasco}, J.~M. and {Carretero}, J. and {Carrilho}, P. and {Carron Duque}, J. and {Carry}, B. and {Carvalho}, A. and {Carvalho}, C.~S. and {Casas}, R. and {Casas}, S. and {Casenove}, P. and {Casey}, C.~M. and {Cassata}, P. and {Castander}, F.~J. and {Castelao}, D. and {Castellano}, M. and {Castiblanco}, L. and {Castignani}, G. and {Castro}, T. and {Cavet}, C. and {Cavuoti}, S. and {Chabaud}, P. -Y. and {Chambers}, K.~C. and {Charles}, Y. and {Charlot}, S. and {Chartab}, N. and {Chary}, R. and {Chaumeil}, F. and {Cho}, H. and {Chon}, G. and {Ciancetta}, E. and {Ciliegi}, P. and {Cimatti}, A. and {Cimino}, M. and {Cioni}, M. -R.~L. and {Claydon}, R. and {Cleland}, C. and {Cl{\'e}ment}, B. and {Clements}, D.~L. and {Clerc}, N. and {Clesse}, S. and {Codis}, S. and {Cogato}, F. and {Colbert}, J. and {Cole}, R.~E. and {Coles}, P. and {Collett}, T.~E. and {Collins}, R.~S. and {Colodro-Conde}, C. and {Colombo}, C. and {Combes}, F. and {Conforti}, V. and {Congedo}, G. and {Conseil}, S. and {Conselice}, C.~J. and {Contarini}, S. and {Contini}, T. and {Conversi}, L. and {Cooray}, A.~R. and {Copin}, Y. and {Corasaniti}, P. -S. and {Corcho-Caballero}, P. and {Corcione}, L. and {Cordes}, O. and {Corpace}, O. and {Correnti}, M. and {Costanzi}, M. and {Costille}, A. and {Courbin}, F. and {Courcoult Mifsud}, L. and {Courtois}, H.~M. and {Cousinou}, M. -C. and {Covone}, G. and {Cowell}, T. and {Cragg}, C. and {Cresci}, G. and {Cristiani}, S. and {Crocce}, M. and {Cropper}, M. and {E Crouzet}, P. and {Csizi}, B. and {Cuby}, J. -G. and {Cucchetti}, E. and {Cucciati}, O. and {Cuillandre}, J. -C. and {Cunha}, P.~A.~C. and {Cuozzo}, V. and {Daddi}, E. and {D'Addona}, M. and {Dafonte}, C. and {Dagoneau}, N. and {Dalessandro}, E. and {Dalton}, G.~B. and {D'Amico}, G. and {Dannerbauer}, H. and {Danto}, P. and {Das}, I. and {Da Silva}, A. and {da Silva}, R. and {Daste}, G. and {Davies}, J.~E. and {Davini}, S. and {de Boer}, T. and {Decarli}, R. and {De Caro}, B. and {Degaudenzi}, H. and {Degni}, G. and {de Jong}, J.~T.~A. and {de la Bella}, L.~F. and {de la Torre}, S. and {Delhaise}, F. and {Delley}, D. and {Delucchi}, G. and {De Lucia}, G. and {Denniston}, J. and {De Paolis}, F. and {De Petris}, M. and {Derosa}, A. and {Desai}, S. and {Desjacques}, V. and {Despali}, G. and {Desprez}, G. and {De Vicente-Albendea}, J. and {Deville}, Y. and {Dias}, J.~D.~F. and {D{\'\i}az-S{\'a}nchez}, A. and {Diaz}, J.~J. and {Di Domizio}, S. and {Diego}, J.~M. and {Di Ferdinando}, D. and {Di Giorgio}, A.~M. and {Dimauro}, P. and {Dinis}, J. and {Dolag}, K. and {Dolding}, C. and {Dole}, H. and {Dom{\'\i}nguez S{\'a}nchez}, H. and {Dor{\'e}}, O. and {Dournac}, F. and {Douspis}, M. and {Dreihahn}, H. and {Droge}, B. and {Dryer}, B. and {Dubath}, F. and {Duc}, P. -A. and {Ducret}, F. and {Duffy}, C. and {Dufresne}, F. and {Duncan}, C.~A.~J. and {Dupac}, X. and {Duret}, V. and {Durrer}, R. and {Durret}, F. and {Dusini}, S. and {Ealet}, A. and {Eggemeier}, A. and {Eisenhardt}, P.~R.~M. and {Elbaz}, D. and {Elkhashab}, M.~Y. and {Ellien}, A. and {Endicott}, J. and {Enia}, A. and {Erben}, T. and {Escartin Vigo}, J.~A. and {Escoffier}, S. and {Escudero Sanz}, I. and {Essert}, J. and {Ettori}, S. and {Ezziati}, M. and {Fabbian}, G. and {Fabricius}, M. and {Fang}, Y. and {Farina}, A. and {Farina}, M. and {Farinelli}, R. and {Farrens}, S. and {Faustini}, F. and {Feltre}, A. and {Ferguson}, A.~M.~N. and {Ferrando}, P. and {Ferrari}, A.~G. and {Ferr{\'e}-Mateu}, A. and {Ferreira}, P.~G. and {Ferreras}, I. and {Ferrero}, I. and {Ferriol}, S. and {Ferruit}, P. and {Filleul}, D. and {Finelli}, F. and {Finkelstein}, S.~L. and {Finoguenov}, A. and {Fiorini}, B. and {Flentge}, F. and {Focardi}, P. and {Fonseca}, J. and {Fontana}, A. and {Fontanot}, F. and {Fornari}, F. and {Fosalba}, P. and {Fossati}, M. and {Fotopoulou}, S. and {Fouchez}, D. and {Fourmanoit}, N. and {Frailis}, M. and {Fraix-Burnet}, D. and {Franceschi}, E. and {Franco}, A. and {Franzetti}, P. and {Freihoefer}, J. and {Frittoli}, G. and {Frugier}, P. -A. and {Frusciante}, N. and {Fumagalli}, A. and {Fumagalli}, M. and {Fumana}, M. and {Fu}, Y. and {Gabarra}, L. and {Galeotta}, S. and {Galluccio}, L. and {Ganga}, K. and {Gao}, H. and {Garc{\'\i}a-Bellido}, J. and {Garcia}, K. and {Gardner}, J.~P. and {Garilli}, B. and {Gaspar-Venancio}, L. -M. and {Gasparetto}, T. and {Gautard}, V. and {Gavazzi}, R. and {Gaztanaga}, E. and {Genolet}, L. and {Genova Santos}, R. and {Gentile}, F. and {George}, K. and {Ghaffari}, Z. and {Giacomini}, F. and {Gianotti}, F. and {Gibb}, G.~P.~S. and {Gillard}, W. and {Gillis}, B. and {Ginolfi}, M. and {Giocoli}, C. and {Girardi}, M. and {Giri}, S.~K. and {Goh}, L.~W.~K. and {G{\'o}mez-Alvarez}, P. and {Gonzalez}, A.~H. and {Gonzalez}, E.~J. and {Gonzalez}, J.~C. and {Gouyou Beauchamps}, S. and {Gozaliasl}, G. and {Gracia-Carpio}, J. and {Grandis}, S. and {Granett}, B.~R. and {Granvik}, M. and {Grazian}, A. and {Gregorio}, A. and {Grenet}, C. and {Grillo}, C. and {Grupp}, F. and {Gruppioni}, C. and {Gruppuso}, A. and {Guerbuez}, C. and {Guerrini}, S. and {Guidi}, M. and {Guillard}, P. and {Gutierrez}, C.~M. and {Guttridge}, P. and {Guzzo}, L. and {Gwyn}, S. and {Haapala}, J. and {Haase}, J. and {Haddow}, C.~R. and {Hailey}, M. and {Hall}, A. and {Hall}, D. and {Hamaus}, N. and {Haridasu}, B.~S. and {Harnois-D{\'e}raps}, J. and {Harper}, C. and {Hartley}, W.~G. and {Hasinger}, G. and {Hassani}, F. and {Hatch}, N.~A. and {Haugan}, S.~V.~H. and {H{\"a}u{\ss}ler}, B. and {Heavens}, A. and {Heisenberg}, L. and {Helmi}, A. and {Helou}, G. and {Hemmati}, S. and {Henares}, K. and {Herent}, O. and {Hern{\'a}ndez-Monteagudo}, C. and {Heuberger}, T. and {Hewett}, P.~C. and {Heydenreich}, S. and {Hildebrandt}, H. and {Hirschmann}, M. and {Hjorth}, J. and {Hoar}, J. and {Hoekstra}, H. and {Holland}, A.~D. and {Holliman}, M.~S. and {Holmes}, W. and {Hook}, I. and {Horeau}, B. and {Hormuth}, F. and {Hornstrup}, A. and {Hosseini}, S. and {Hu}, D. and {Hudelot}, P. and {Hudson}, M.~J. and {Huertas-Company}, M. and {Huff}, E.~M. and {Hughes}, A.~C.~N. and {Humphrey}, A. and {Hunt}, L.~K. and {Huynh}, D.~D. and {Ibata}, R. and {Ichikawa}, K. and {Iglesias-Groth}, S. and {Ilbert}, O. and {Ili{\'c}}, S. and {Ingoglia}, L. and {Iodice}, E. and {Israel}, H. and {Israelsson}, U.~E. and {Izzo}, L. and {Jablonka}, P. and {Jackson}, N. and {Jacobson}, J. and {Jafariyazani}, M. and {Jahnke}, K. and {Jansen}, H. and {Jarvis}, M.~J. and {Jasche}, J. and {Jauzac}, M. and {Jeffrey}, N. and {Jhabvala}, M. and {Jimenez-Teja}, Y. and {Jimenez Mu{\~n}oz}, A. and {Joachimi}, B. and {Johansson}, P.~H. and {Joudaki}, S. and {Jullo}, E. and {Kajava}, J.~J.~E. and {Kang}, Y. and {Kannawadi}, A. and {Kansal}, V. and {Karagiannis}, D. and {K{\"a}rcher}, M. and {Kashlinsky}, A. and {Kazandjian}, M.~V. and {Keck}, F. and {Keih{\"a}nen}, E. and {Kerins}, E. and {Kermiche}, S. and {Khalil}, A. and {Kiessling}, A. and {Kiiveri}, K. and {Kilbinger}, M. and {Kim}, J. and {King}, R. and {Kirkpatrick}, C.~C. and {Kitching}, T. and {Kluge}, M. and {Knabenhans}, M. and {Knapen}, J.~H. and {Knebe}, A. and {Kneib}, J. -P. and {Kohley}, R. and {Koopmans}, L.~V.~E. and {Koskinen}, H. and {Koulouridis}, E. and {Kou}, R. and {Kov{\'a}cs}, A. and {Kova\{{\v{c}}\}i{\'c}}, I. and {Kowalczyk}, A. and {Koyama}, K. and {Kraljic}, K. and {Krause}, O. and {Kruk}, S. and {Kubik}, B. and {Kuchner}, U. and {Kuijken}, K. and {K{\"u}mmel}, M. and {Kunz}, M. and {Kurki-Suonio}, H. and {Lacasa}, F. and {Lacey}, C.~G. and {La Franca}, F. and {Lagarde}, N. and {Lahav}, O. and {Laigle}, C. and {La Marca}, A. and {La Marle}, O. and {Lamine}, B. and {Lam}, M.~C. and {Lan{\c{c}}on}, A. and {Landt}, H. and {Langer}, M. and {Lapi}, A. and {Larcheveque}, C. and {Larsen}, S.~S. and {Lattanzi}, M. and {Laudisio}, F. and {Laugier}, D. and {Laureijs}, R. and {Lavaux}, G. and {Lawrenson}, A. and {Lazanu}, A. and {Lazeyras}, T. and {Le Boulc'h}, Q. and {Le Brun}, A.~M.~C. and {Le Brun}, V. and {Leclercq}, F. and {Lee}, S. and {Le Graet}, J. and {Legrand}, L. and {Leirvik}, K.~N. and {Le Jeune}, M. and {Lembo}, M. and {Le Mignant}, D. and {Lepinzan}, M.~D. and {Lepori}, F. and {Lesci}, G.~F. and {Lesgourgues}, J. and {Leuzzi}, L. and {Levi}, M.~E. and {Liaudat}, T.~I. and {Libet}, G. and {Liebing}, P. and {Ligori}, S. and {Lilje}, P.~B. and {Lin}, C. -C. and {Linde}, D. and {Linder}, E. and {Lindholm}, V. and {Linke}, L. and {Li}, S. -S. and {Liu}, S.~J. and {Lloro}, I. and {Lobo}, F.~S.~N. and {Lodieu}, N. and {Lombardi}, M. and {Lombriser}, L. and {Lonare}, P. and {Longo}, G. and {L{\'o}pez-Caniego}, M. and {Lopez Lopez}, X. and {Alvarez}, J. Lorenzo and {Loureiro}, A. and {Loveday}, J. and {Lusso}, E. and {Macias-Perez}, J. and {Maciaszek}, T. and {Magliocchetti}, M. and {Magnard}, F. and {Magnier}, E.~A. and {Magro}, A. and {Mahler}, G. and {Mainetti}, G. and {Maino}, D. and {Maiorano}, E. and {Maiorano}, E. and {Malavasi}, N. and {Mamon}, G.~A. and {Mancini}, C. and {Mandelbaum}, R. and {Manera}, M. and {Manj{\'o}n-Garc{\'\i}a}, A. and {Mannucci}, F. and {Mansutti}, O. and {Manteiga Outeiro}, M. and {Maoli}, R. and {Maraston}, C. and {Marcin}, S. and {Marcos-Arenal}, P. and {Margalef-Bentabol}, B. and {Marggraf}, O. and {Marinucci}, D. and {Marinucci}, M. and {Markovic}, K. and {Marleau}, F.~R. and {Marpaud}, J. and {Martignac}, J. and {Mart{\'\i}n-Fleitas}, J. and {Martin-Moruno}, P. and {Martin}, E.~L. and {Martinelli}, M. and {Martinet}, N. and {Martin}, H. and {Martins}, C.~J.~A.~P. and {Marulli}, F. and {Massari}, D. and {Massey}, R. and {Masters}, D.~C. and {Matarrese}, S. and {Matsuoka}, Y. and {Matthew}, S. and {Maughan}, B.~J. and {Mauri}, N. and {Maurin}, L. and {Maurogordato}, S. and {McCarthy}, K. and {McConnachie}, A.~W. and {McCracken}, H.~J. and {McDonald}, I. and {McEwen}, J.~D. and {McPartland}, C.~J.~R. and {Medinaceli}, E. and {Mehta}, V. and {Mei}, S. and {Melchior}, M. and {Melin}, J. -B. and {M{\'e}nard}, B. and {Mendes}, J. and {Mendez-Abreu}, J. and {Meneghetti}, M. and {Mercurio}, A. and {Merlin}, E. and {Metcalf}, R.~B. and {Meylan}, G. and {Migliaccio}, M. and {Mignoli}, M. and {Miller}, L. and {Miluzio}, M. and {Milvang-Jensen}, B. and {Mimoso}, J.~P. and {Miquel}, R. and {Miyatake}, H. and {Mobasher}, B. and {Mohr}, J.~J. and {Monaco}, P. and {Mongui{\'o}}, M. and {Montoro}, A. and {Mora}, A. and {Moradinezhad Dizgah}, A. and {Moresco}, M. and {Moretti}, C. and {Morgante}, G. and {Morisset}, N. and {Moriya}, T.~J. and {Morris}, P.~W. and {Mortlock}, D.~J. and {Moscardini}, L. and {Mota}, D.~F. and {Moustakas}, L.~A. and {Moutard}, T. and {M{\"u}ller}, T. and {Munari}, E. and {Murphree}, G. and {Murray}, C. and {Murray}, N. and {Musi}, P. and {Nadathur}, S. and {Nagam}, B.~C. and {Nagao}, T. and {Naidoo}, K. and {Nakajima}, R. and {Nally}, C. and {Natoli}, P. and {Navarro-Alsina}, A. and {Navarro Girones}, D. and {Neissner}, C. and {Nersesian}, A. and {Nesseris}, S. and {Nguyen-Kim}, H.~N. and {Nicastro}, L. and {Nichol}, R.~C. and {Nielbock}, M. and {Niemi}, S. -M. and {Nieto}, S. and {Nilsson}, K. and {Noller}, J. and {Norberg}, P. and {Nourizonoz}, A. and {Ntelis}, P. and {Nucita}, A.~A. and {Nugent}, P. and {Nunes}, N.~J. and {Nutma}, T. and {Ocampo}, I. and {Odier}, J. and {Oesch}, P.~A. and {Oguri}, M. and {Magalhaes Oliveira}, D. and {Onoue}, M. and {Oosterbroek}, T. and {Oppizzi}, F. and {Ordenovic}, C. and {Osato}, K. and {Pacaud}, F. and {Pace}, F. and {Padilla}, C. and {Paech}, K. and {Pagano}, L. and {Page}, M.~J. and {Palazzi}, E. and {Paltani}, S. and {Pamuk}, S. and {Pandolfi}, S. and {Paoletti}, D. and {Paolillo}, M. and {Papaderos}, P. and {Pardede}, K. and {Parimbelli}, G. and {Parmar}, A. and {Partmann}, C. and {Pasian}, F. and {Passalacqua}, F. and {Paterson}, K. and {Patrizii}, L. and {Pattison}, C. and {Paulino-Afonso}, A. and {Paviot}, R. and {Peacock}, J.~A. and {Pearce}, F.~R. and {Pedersen}, K. and {Peel}, A. and {Peletier}, R.~F. and {Pellejero Ibanez}, M. and {Pello}, R. and {Penny}, M.~T. and {Percival}, W.~J. and {Perez-Garrido}, A. and {Perotto}, L. and {Pettorino}, V. and {Pezzotta}, A. and {Pezzuto}, S. and {Philippon}, A. and {Piersanti}, O. and {Pietroni}, M. and {Piga}, L. and {Pilo}, L. and {Pires}, S. and {Pisani}, A. and {Pizzella}, A. and {Pizzuti}, L. and {Plana}, C. and {Polenta}, G. and {Pollack}, J.~E. and {Poncet}, M. and {P{\"o}ntinen}, M. and {Pool}, P. and {Popa}, L.~A. and {Popa}, V. and {Popp}, J. and {Porciani}, C. and {Porth}, L. and {Potter}, D. and {Poulain}, M. and {Pourtsidou}, A. and {Pozzetti}, L. and {Prandoni}, I. and {Pratt}, G.~W. and {Prezelus}, S. and {Prieto}, E. and {Pugno}, A. and {Quai}, S. and {Quilley}, L. and {Racca}, G.~D. and {Raccanelli}, A. and {R{\'a}cz}, G. and {Radinovi{\'c}}, S. and {Radovich}, M. and {Ragagnin}, A. and {Ragnit}, U. and {Raison}, F. and {Ramos-Chernenko}, N. and {Ranc}, C. and {Raylet}, N. and {Rebolo}, R. and {Refregier}, A. and {Reimberg}, P. and {Reiprich}, T.~H. and {Renk}, F. and {Renzi}, A. and {Retre}, J. and {Revaz}, Y. and {Reyl{\'e}}, C. and {Reynolds}, L. and {Rhodes}, J. and {Ricci}, F. and {Ricci}, M. and {Riccio}, G. and {Ricken}, S.~O. and {Rissanen}, S. and {Risso}, I. and {Rix}, H. -W. and {Robin}, A.~C. and {Rocca-Volmerange}, B. and {Rocci}, P. -F. and {Rodenhuis}, M. and {Rodighiero}, G. and {Rodriguez Monroy}, M. and {Rollins}, R.~P. and {Romanello}, M. and {Roman}, J. and {Romelli}, E. and {Romero-Gomez}, M. and {Roncarelli}, M. and {Rosati}, P. and {Rosset}, C. and {Rossetti}, E. and {Roster}, W. and {Rottgering}, H.~J.~A. and {Rozas-Fern{\'a}ndez}, A. and {Ruane}, K. and {Rubino-Martin}, J.~A. and {Rudolph}, A. and {Ruppin}, F. and {Rusholme}, B. and {Sacquegna}, S. and {S{\'a}ez-Casares}, I. and {Saga}, S. and {Saglia}, R. and {Sahl{\'e}n}, M. and {Saifollahi}, T. and {Sakr}, Z. and {Salvalaggio}, J. and {Salvaterra}, R. and {Salvati}, L. and {Salvato}, M. and {Salvignol}, J. -C. and {S{\'a}nchez}, A.~G. and {Sanchez}, E. and {Sanders}, D.~B. and {Sapone}, D. and {Saponara}, M. and {Sarpa}, E. and {Sarron}, F. and {Sartori}, S. and {Sassolas}, B. and {Sauniere}, L. and {Sauvage}, M. and {Sawicki}, M. and {Scaramella}, R. and {Scarlata}, C. and {Scharr{\'e}}, L. and {Schaye}, J. and {Schewtschenko}, J.~A. and {Schindler}, J. -T. and {Schinnerer}, E. and {Schirmer}, M. and {Schmidt}, F. and {Schmidt}, F. and {Schmidt}, M. and {Schneider}, A. and {Schneider}, M. and {Schneider}, P. and {Sch{\"o}neberg}, N. and {Schrabback}, T. and {Schultheis}, M. and {Schulz}, S. and {Schwartz}, J. and {Sciotti}, D. and {Scodeggio}, M. and {Scognamiglio}, D. and {Scott}, D. and {Scottez}, V. and {Secroun}, A. and {Sefusatti}, E. and {Seidel}, G. and {Seiffert}, M. and {Sellentin}, E. and {Selwood}, M. and {Semboloni}, E. and {Sereno}, M. and {Serjeant}, S. and {Serrano}, S. and {Shankar}, F. and {Sharples}, R.~M. and {Short}, A. and {Shulevski}, A. and {Shuntov}, M. and {Sias}, M. and {Sikkema}, G. and {Silvestri}, A. and {Simon}, P. and {Sirignano}, C. and {Sirri}, G. and {Skottfelt}, J. and {Slezak}, E. and {Sluse}, D. and {Smith}, G.~P. and {Smith}, L.~C. and {Smith}, R.~E. and {Smit}, S.~J.~A. and {Soldano}, F. and {Solheim}, B.~G.~B. and {Sorce}, J.~G. and {Sorrenti}, F. and {Soubrie}, E. and {Spinoglio}, L. and {Spurio Mancini}, A. and {Stadel}, J. and {Stagnaro}, L. and {Stanco}, L. and {Stanford}, S.~A. and {Starck}, J. -L. and {Stassi}, P. and {Steinwagner}, J. and {Stern}, D. and {Stone}, C. and {Strada}, P. and {Strafella}, F. and {Stramaccioni}, D. and {Surace}, C. and {Sureau}, F. and {Suyu}, S.~H. and {Swindells}, I. and {Szafraniec}, M. and {Szapudi}, I. and {Taamoli}, S. and {Talia}, M. and {Tallada-Cresp{\'\i}}, P. and {Tanidis}, K. and {Tao}, C. and {Tarr{\'\i}o}, P. and {Tavagnacco}, D. and {Taylor}, A.~N. and {Taylor}, J.~E. and {Taylor}, P.~L. and {Teixeira}, E.~M. and {Tenti}, M. and {Teodoro Idiago}, P. and {Teplitz}, H.~I. and {Tereno}, I. and {Tessore}, N. and {Testa}, V. and {Testera}, G. and {Tewes}, M. and {Teyssier}, R. and {Theret}, N. and {Thizy}, C. and {Thomas}, P.~D. and {Toba}, Y. and {Toft}, S. and {Toledo-Moreo}, R. and {Tolstoy}, E. and {Tommasi}, E. and {Torbaniuk}, O. and {Torradeflot}, F. and {Tortora}, C. and {Tosi}, S. and {Tosti}, S. and {Trifoglio}, M. and {Troja}, A. and {Trombetti}, T. and {Tronconi}, A. and {Tsedrik}, M. and {Tsyganov}, A. and {Tucci}, M. and {Tutusaus}, I. and {Uhlemann}, C. and {Ulivi}, L. and {Urbano}, M. and {Vacher}, L. and {Vaillon}, L. and {Valdes}, I. and {Valentijn}, E.~A. and {Valenziano}, L. and {Valieri}, C. and {Valiviita}, J. and {Van den Broeck}, M. and {Vassallo}, T. and {Vavrek}, R. and {Venemans}, B. and {Venhola}, A. and {Ventura}, S. and {Verdoes Kleijn}, G. and {Vergani}, D. and {Verma}, A. and {Vernizzi}, F. and {Veropalumbo}, A. and {Verza}, G. and {Vescovi}, C. and {Vibert}, D. and {Viel}, M. and {Vielzeuf}, P. and {Viglione}, C. and {Viitanen}, A. and {Villaescusa-Navarro}, F. and {Vinciguerra}, S. and {Visticot}, F. and {Voggel}, K. and {von Wietersheim-Kramsta}, M. and {Vriend}, W.~J. and {Wachter}, S. and {Walmsley}, M. and {Walth}, G. and {Walton}, D.~M. and {Walton}, N.~A. and {Wander}, M. and {Wang}, L. and {Wang}, Y. and {Weaver}, J.~R. and {Weller}, J. and {Whalen}, D.~J. and {Wiesmann}, M. and {Wilde}, J. and {Williams}, O.~R. and {Winther}, H. -A. and {Wittje}, A. and {Wong}, J.~H.~W. and {Wright}, A.~H. and {Yankelevich}, V. and {Yeung}, H.~W. and {Youles}, S. and {Yung}, L.~Y.~A. and {Zacchei}, A. and {Zalesky}, L. and {Zamorani}, G. and {Zamorano Vitorelli}, A. and {Zanoni Marc}, M. and {Zennaro}, M. and {Zerbi}, F.~M. and {Zinchenko}, I.~A. and {Zoubian}, J. and {Zucca}, E. and {Zumalacarregui}, M.},
        title = "{Euclid. I. Overview of the Euclid mission}",
      journal = {arXiv e-prints},
         year = 2024,
        month = may,
          eid = {arXiv:2405.13491},
        pages = {arXiv:2405.13491},
          doi = {10.48550/arXiv.2405.13491},
archivePrefix = {arXiv},
       eprint = {2405.13491},
 primaryClass = {astro-ph.CO},
       adsurl = {https://ui.adsabs.harvard.edu/abs/2024arXiv240513491E}
}

@article{Wang:2023,
       author = {{Wang}, Mike and {Beutler}, Florian and {Sugiyama}, Naonori},
        title = "{Triumvirate: A Python/C++ package for three-point clustering measurements}",
      journal = {The Journal of Open Source Software},
         year = 2023,
        month = nov,
       volume = {8},
       number = {91},
          eid = {5571},
        pages = {5571},
          doi = {10.21105/joss.05571},
archivePrefix = {arXiv},
       eprint = {2304.03643},
 primaryClass = {astro-ph.IM},
       adsurl = {https://ui.adsabs.harvard.edu/abs/2023JOSS....8.5571W}
}

@article{Sugiyama:2023c,
       author = {{Sugiyama}, Naonori S. and {Yamauchi}, Daisuke and {Kobayashi}, Tsutomu and {Fujita}, Tomohiro and {Arai}, Shun and {Hirano}, Shin'ichi and {Saito}, Shun and {Beutler}, Florian and {Seo}, Hee-Jong},
        title = "{First test of the consistency relation for the large-scale structure using the anisotropic three-point correlation function of BOSS DR12 galaxies}",
      journal = {MNRAS},
         year = 2023,
        month = sep,
       volume = {524},
       number = {2},
        pages = {1651-1667},
          doi = {10.1093/mnras/stad1935},
archivePrefix = {arXiv},
       eprint = {2305.01142},
 primaryClass = {astro-ph.CO},
       adsurl = {https://ui.adsabs.harvard.edu/abs/2023MNRAS.524.1651S}
}

@article{Sugiyama:2023,
       author = {{Sugiyama}, Naonori S. and {Yamauchi}, Daisuke and {Kobayashi}, Tsutomu and {Fujita}, Tomohiro and {Arai}, Shun and {Hirano}, Shin'ichi and {Saito}, Shun and {Beutler}, Florian and {Seo}, Hee-Jong},
        title = "{New constraints on cosmological modified gravity theories from anisotropic three-point correlation functions of BOSS DR12 galaxies}",
      journal = {MNRAS},
         year = 2023,
        month = aug,
       volume = {523},
       number = {2},
        pages = {3133-3191},
          doi = {10.1093/mnras/stad1505},
archivePrefix = {arXiv},
       eprint = {2302.06808},
 primaryClass = {astro-ph.CO},
       adsurl = {https://ui.adsabs.harvard.edu/abs/2023MNRAS.523.3133S}
}

@article{Guidi:2023,
       author = {{Guidi}, M. and {Veropalumbo}, A. and {Branchini}, E. and {Eggemeier}, A. and {Carbone}, C.},
        title = "{Modelling the next-to-leading order matter three-point correlation function using FFTLog}",
      journal = {JCAP},
         year = 2023,
        month = aug,
       volume = {2023},
       number = {8},
          eid = {066},
        pages = {066},
          doi = {10.1088/1475-7516/2023/08/066},
archivePrefix = {arXiv},
       eprint = {2212.07382},
 primaryClass = {astro-ph.CO},
       adsurl = {https://ui.adsabs.harvard.edu/abs/2023JCAP...08..066G}
}

@article{Rizzo:2023,
       author = {{Rizzo}, Federico and {Moretti}, Chiara and {Pardede}, Kevin and {Eggemeier}, Alexander and {Oddo}, Andrea and {Sefusatti}, Emiliano and {Porciani}, Cristiano and {Monaco}, Pierluigi},
        title = "{The halo bispectrum multipoles in redshift space}",
      journal = {JCAP},
         year = 2023,
        month = jan,
       volume = {2023},
       number = {1},
          eid = {031},
        pages = {031},
          doi = {10.1088/1475-7516/2023/01/031},
archivePrefix = {arXiv},
       eprint = {2204.13628},
 primaryClass = {astro-ph.CO},
       adsurl = {https://ui.adsabs.harvard.edu/abs/2023JCAP...01..031R}
}

@article{Pardede:2022,
       author = {{Pardede}, Kevin and {Rizzo}, Federico and {Biagetti}, Matteo and {Castorina}, Emanuele and {Sefusatti}, Emiliano and {Monaco}, Pierluigi},
        title = "{Bispectrum-window convolution via Hankel transform}",
      journal = {JCAP},
         year = 2022,
        month = oct,
       volume = {2022},
       number = {10},
          eid = {066},
        pages = {066},
          doi = {10.1088/1475-7516/2022/10/066},
archivePrefix = {arXiv},
       eprint = {2203.04174},
 primaryClass = {astro-ph.CO},
       adsurl = {https://ui.adsabs.harvard.edu/abs/2022JCAP...10..066P}
}

@article{Veropalumbo:2022,
       author = {{Veropalumbo}, A. and {Binetti}, A. and {Branchini}, E. and {Moresco}, M. and {Monaco}, P. and {Oddo}, A. and {S{\'a}nchez}, A.~G. and {Sefusatti}, E.},
        title = "{The halo 3-point correlation function: a methodological analysis}",
      journal = {JCAP},
         year = 2022,
        month = sep,
       volume = {2022},
       number = {9},
          eid = {033},
        pages = {033},
          doi = {10.1088/1475-7516/2022/09/033},
archivePrefix = {arXiv},
       eprint = {2206.00672},
 primaryClass = {astro-ph.CO},
       adsurl = {https://ui.adsabs.harvard.edu/abs/2022JCAP...09..033V}
}

@article{Philcox:2022B,
       author = {{Philcox}, Oliver H.~E. and {Ivanov}, Mikhail M. and {Cabass}, Giovanni and {Simonovi{\'c}}, Marko and {Zaldarriaga}, Matias and {Nishimichi}, Takahiro},
        title = "{Cosmology with the redshift-space galaxy bispectrum monopole at one-loop order}",
      journal = {Phys. Rev. D},
         year = 2022,
        month = aug,
       volume = {106},
       number = {4},
          eid = {043530},
        pages = {043530},
          doi = {10.1103/PhysRevD.106.043530},
archivePrefix = {arXiv},
       eprint = {2206.02800},
 primaryClass = {astro-ph.CO},
       adsurl = {https://ui.adsabs.harvard.edu/abs/2022PhRvD.106d3530P}
}

@article{Philcox:2021,
       author = {{Philcox}, Oliver H.~E.},
        title = "{Cosmology without window functions. II. Cubic estimators for the galaxy bispectrum}",
      journal = {Phys. Rev. D},
         year = 2021,
        month = dec,
       volume = {104},
       number = {12},
          eid = {123529},
        pages = {123529},
          doi = {10.1103/PhysRevD.104.123529},
archivePrefix = {arXiv},
       eprint = {2107.06287},
 primaryClass = {astro-ph.CO},
       adsurl = {https://ui.adsabs.harvard.edu/abs/2021PhRvD.104l3529P}
}

@article{Oddo:2021,
       author = {{Oddo}, Andrea and {Rizzo}, Federico and {Sefusatti}, Emiliano and {Porciani}, Cristiano and {Monaco}, Pierluigi},
        title = "{Cosmological parameters from the likelihood analysis of the galaxy power spectrum and bispectrum in real space}",
      journal = {JCAP},
         year = 2021,
        month = nov,
       volume = {2021},
       number = {11},
          eid = {038},
        pages = {038},
          doi = {10.1088/1475-7516/2021/11/038},
archivePrefix = {arXiv},
       eprint = {2108.03204},
 primaryClass = {astro-ph.CO},
       adsurl = {https://ui.adsabs.harvard.edu/abs/2021JCAP...11..038O}
}

@article{Veropalumbo:2021,
       author = {{Veropalumbo}, Alfonso and {S{\'a}ez Casares}, I{\~n}igo and {Branchini}, Enzo and {Granett}, Benjamin R. and {Guzzo}, Luigi and {Marulli}, Federico and {Moresco}, Michele and {Moscardini}, Lauro and {Pezzotta}, Andrea and {de la Torre}, Sylvain},
        title = "{A joint 2- and 3-point clustering analysis of the VIPERS PDR2 catalogue at $z = 1$: breaking the degeneracy of cosmological parameters}",
      journal = {MNRAS},
         year = 2021,
        month = oct,
       volume = {507},
       number = {1},
        pages = {1184-1201},
          doi = {10.1093/mnras/stab2205},
archivePrefix = {arXiv},
       eprint = {2106.12581},
 primaryClass = {astro-ph.CO},
       adsurl = {https://ui.adsabs.harvard.edu/abs/2021MNRAS.507.1184V}
}

@article{Karamanis:2021,
       author = {{Karamanis}, Minas and {Beutler}, Florian},
        title = "{hankl: A lightweight Python implementation of the FFTLog algorithm for Cosmology}",
      journal = {arXiv e-prints},
         year = 2021,
        month = jun,
          eid = {arXiv:2106.06331},
        pages = {arXiv:2106.06331},
          doi = {10.48550/arXiv.2106.06331},
archivePrefix = {arXiv},
       eprint = {2106.06331},
 primaryClass = {astro-ph.IM},
       adsurl = {https://ui.adsabs.harvard.edu/abs/2021arXiv210606331K}
}

@article{Umeh:2021,
       author = {{Umeh}, Obinna},
        title = "{Optimal computation of anisotropic galaxy three point correlation function multipoles using 2DFFTLOG formalism}",
      journal = {JCAP},
         year = 2021,
        month = may,
       volume = {2021},
       number = {5},
          eid = {035},
        pages = {035},
          doi = {10.1088/1475-7516/2021/05/035},
archivePrefix = {arXiv},
       eprint = {2011.05889},
 primaryClass = {astro-ph.CO},
       adsurl = {https://ui.adsabs.harvard.edu/abs/2021JCAP...05..035U}
}

@article{Sugiyama:2021,
       author = {{Sugiyama}, Naonori S. and {Saito}, Shun and {Beutler}, Florian and {Seo}, Hee-Jong},
        title = "{Towards a self-consistent analysis of the anisotropic galaxy two- and three-point correlation functions on large scales: application to mock galaxy catalogues}",
      journal = {MNRAS},
         year = 2021,
        month = feb,
       volume = {501},
       number = {2},
        pages = {2862-2896},
          doi = {10.1093/mnras/staa3725},
archivePrefix = {arXiv},
       eprint = {2010.06179},
 primaryClass = {astro-ph.CO},
       adsurl = {https://ui.adsabs.harvard.edu/abs/2021MNRAS.501.2862S}
}

@article{Fang:2020,
       author = {{Fang}, Xiao and {Eifler}, Tim and {Krause}, Elisabeth},
        title = "{2D-FFTLog: efficient computation of real-space covariance matrices for galaxy clustering and weak lensing}",
      journal = {MNRAS},
         year = 2020,
        month = sep,
       volume = {497},
       number = {3},
        pages = {2699-2714},
          doi = {10.1093/mnras/staa1726},
archivePrefix = {arXiv},
       eprint = {2004.04833},
 primaryClass = {astro-ph.CO},
       adsurl = {https://ui.adsabs.harvard.edu/abs/2020MNRAS.497.2699F}
}

@article{Chudaykin:2020,
       author = {{Chudaykin}, Anton and {Ivanov}, Mikhail M. and {Philcox}, Oliver H.~E. and {Simonovi{\'c}}, Marko},
        title = "{Nonlinear perturbation theory extension of the Boltzmann code CLASS}",
      journal = {Phys. Rev. D},
         year = 2020,
        month = sep,
       volume = {102},
       number = {6},
          eid = {063533},
        pages = {063533},
          doi = {10.1103/PhysRevD.102.063533},
archivePrefix = {arXiv},
       eprint = {2004.10607},
 primaryClass = {astro-ph.CO},
       adsurl = {https://ui.adsabs.harvard.edu/abs/2020PhRvD.102f3533C}
}

@article{Akrami:2020,
       author = {{Planck Collaboration} and {Akrami}, Y. and {Arroja}, F. and {Ashdown}, M. and {Aumont}, J. and {Baccigalupi}, C. and {Ballardini}, M. and {Banday}, A.~J. and {Barreiro}, R.~B. and {Bartolo}, N. and {Basak}, S. and {Benabed}, K. and {Bernard}, J. -P. and {Bersanelli}, M. and {Bielewicz}, P. and {Bock}, J.~J. and {Bond}, J.~R. and {Borrill}, J. and {Bouchet}, F.~R. and {Boulanger}, F. and {Bucher}, M. and {Burigana}, C. and {Butler}, R.~C. and {Calabrese}, E. and {Cardoso}, J. -F. and {Carron}, J. and {Challinor}, A. and {Chiang}, H.~C. and {Colombo}, L.~P.~L. and {Combet}, C. and {Contreras}, D. and {Crill}, B.~P. and {Cuttaia}, F. and {de Bernardis}, P. and {de Zotti}, G. and {Delabrouille}, J. and {Delouis}, J. -M. and {Di Valentino}, E. and {Diego}, J.~M. and {Donzelli}, S. and {Dor{\'e}}, O. and {Douspis}, M. and {Ducout}, A. and {Dupac}, X. and {Dusini}, S. and {Efstathiou}, G. and {Elsner}, F. and {En{\ss}lin}, T.~A. and {Eriksen}, H.~K. and {Fantaye}, Y. and {Fergusson}, J. and {Fernandez-Cobos}, R. and {Finelli}, F. and {Forastieri}, F. and {Frailis}, M. and {Franceschi}, E. and {Frolov}, A. and {Galeotta}, S. and {Galli}, S. and {Ganga}, K. and {Gauthier}, C. and {G{\'e}nova-Santos}, R.~T. and {Gerbino}, M. and {Ghosh}, T. and {Gonz{\'a}lez-Nuevo}, J. and {G{\'o}rski}, K.~M. and {Gratton}, S. and {Gruppuso}, A. and {Gudmundsson}, J.~E. and {Hamann}, J. and {Handley}, W. and {Hansen}, F.~K. and {Herranz}, D. and {Hivon}, E. and {Hooper}, D.~C. and {Huang}, Z. and {Jaffe}, A.~H. and {Jones}, W.~C. and {Keih{\"a}nen}, E. and {Keskitalo}, R. and {Kiiveri}, K. and {Kim}, J. and {Kisner}, T.~S. and {Krachmalnicoff}, N. and {Kunz}, M. and {Kurki-Suonio}, H. and {Lagache}, G. and {Lamarre}, J. -M. and {Lasenby}, A. and {Lattanzi}, M. and {Lawrence}, C.~R. and {Le Jeune}, M. and {Lesgourgues}, J. and {Levrier}, F. and {Lewis}, A. and {Liguori}, M. and {Lilje}, P.~B. and {Lindholm}, V. and {L{\'o}pez-Caniego}, M. and {Lubin}, P.~M. and {Ma}, Y. -Z. and {Mac{\'\i}as-P{\'e}rez}, J.~F. and {Maggio}, G. and {Maino}, D. and {Mandolesi}, N. and {Mangilli}, A. and {Marcos-Caballero}, A. and {Maris}, M. and {Martin}, P.~G. and {Mart{\'\i}nez-Gonz{\'a}lez}, E. and {Matarrese}, S. and {Mauri}, N. and {McEwen}, J.~D. and {Meerburg}, P.~D. and {Meinhold}, P.~R. and {Melchiorri}, A. and {Mennella}, A. and {Migliaccio}, M. and {Mitra}, S. and {Miville-Desch{\^e}nes}, M. -A. and {Molinari}, D. and {Moneti}, A. and {Montier}, L. and {Morgante}, G. and {Moss}, A. and {M{\"u}nchmeyer}, M. and {Natoli}, P. and {N{\o}rgaard-Nielsen}, H.~U. and {Pagano}, L. and {Paoletti}, D. and {Partridge}, B. and {Patanchon}, G. and {Peiris}, H.~V. and {Perrotta}, F. and {Pettorino}, V. and {Piacentini}, F. and {Polastri}, L. and {Polenta}, G. and {Puget}, J. -L. and {Rachen}, J.~P. and {Reinecke}, M. and {Remazeilles}, M. and {Renzi}, A. and {Rocha}, G. and {Rosset}, C. and {Roudier}, G. and {Rubi{\~n}o-Mart{\'\i}n}, J.~A. and {Ruiz-Granados}, B. and {Salvati}, L. and {Sandri}, M. and {Savelainen}, M. and {Scott}, D. and {Shellard}, E.~P.~S. and {Shiraishi}, M. and {Sirignano}, C. and {Sirri}, G. and {Spencer}, L.~D. and {Sunyaev}, R. and {Suur-Uski}, A. -S. and {Tauber}, J.~A. and {Tavagnacco}, D. and {Tenti}, M. and {Toffolatti}, L. and {Tomasi}, M. and {Trombetti}, T. and {Valiviita}, J. and {Van Tent}, B. and {Vielva}, P. and {Villa}, F. and {Vittorio}, N. and {Wandelt}, B.~D. and {Wehus}, I.~K. and {White}, S.~D.~M. and {Zacchei}, A. and {Zibin}, J.~P. and {Zonca}, A.},
        title = "{Planck 2018 results. X. Constraints on inflation}",
      journal = {A\&A},
         year = 2020,
        month = sep,
       volume = {641},
          eid = {A10},
        pages = {A10},
          doi = {10.1051/0004-6361/201833887},
archivePrefix = {arXiv},
       eprint = {1807.06211},
 primaryClass = {astro-ph.CO},
       adsurl = {https://ui.adsabs.harvard.edu/abs/2020A&A...641A..10P}
}

@article{Oddo:2020,
       author = {{Oddo}, Andrea and {Sefusatti}, Emiliano and {Porciani}, Cristiano and {Monaco}, Pierluigi and {S{\'a}nchez}, Ariel G.},
        title = "{Toward a robust inference method for the galaxy bispectrum: likelihood function and model selection}",
      journal = {JCAP},
         year = 2020,
        month = mar,
       volume = {2020},
       number = {3},
          eid = {056},
        pages = {056},
          doi = {10.1088/1475-7516/2020/03/056},
archivePrefix = {arXiv},
       eprint = {1908.01774},
 primaryClass = {astro-ph.CO},
       adsurl = {https://ui.adsabs.harvard.edu/abs/2020JCAP...03..056O}
}

@article{Sugiyama:2019,
       author = {{Sugiyama}, Naonori S. and {Saito}, Shun and {Beutler}, Florian and {Seo}, Hee-Jong},
        title = "{A complete FFT-based decomposition formalism for the redshift-space bispectrum}",
      journal = {MNRAS},
         year = 2019,
        month = mar,
       volume = {484},
       number = {1},
        pages = {364-384},
          doi = {10.1093/mnras/sty3249},
archivePrefix = {arXiv},
       eprint = {1803.02132},
 primaryClass = {astro-ph.CO},
       adsurl = {https://ui.adsabs.harvard.edu/abs/2019MNRAS.484..364S}
}

@article{Zonca:2019,
       author = {{Zonca}, Andrea and {Singer}, Leo and {Lenz}, Daniel and {Reinecke}, Martin and {Rosset}, Cyrille and {Hivon}, Eric and {Gorski}, Krzysztof},
        title = "{healpy: equal area pixelization and spherical harmonics transforms for data on the sphere in Python}",
      journal = {The Journal of Open Source Software},
         year = 2019,
        month = mar,
       volume = {4},
       number = {35},
          eid = {1298},
        pages = {1298},
          doi = {10.21105/joss.01298},
       adsurl = {https://ui.adsabs.harvard.edu/abs/2019JOSS....4.1298Z}
}

@article{Yankelevich:2019,
       author = {{Yankelevich}, Victoria and {Porciani}, Cristiano},
        title = "{Cosmological information in the redshift-space bispectrum}",
      journal = {MNRAS},
         year = 2019,
        month = feb,
       volume = {483},
       number = {2},
        pages = {2078-2099},
          doi = {10.1093/mnras/sty3143},
archivePrefix = {arXiv},
       eprint = {1807.07076},
 primaryClass = {astro-ph.CO},
       adsurl = {https://ui.adsabs.harvard.edu/abs/2019MNRAS.483.2078Y}
}

@article{Akeson:2019,
       author = {{Akeson}, Rachel and {Armus}, Lee and {Bachelet}, Etienne and {Bailey}, Vanessa and {Bartusek}, Lisa and {Bellini}, Andrea and {Benford}, Dominic and {Bennett}, David and {Bhattacharya}, Aparna and {Bohlin}, Ralph and {Boyer}, Martha and {Bozza}, Valerio and {Bryden}, Geoffrey and {Calchi Novati}, Sebastiano and {Carpenter}, Kenneth and {Casertano}, Stefano and {Choi}, Ami and {Content}, David and {Dayal}, Pratika and {Dressler}, Alan and {Dor{\'e}}, Olivier and {Fall}, S. Michael and {Fan}, Xiaohui and {Fang}, Xiao and {Filippenko}, Alexei and {Finkelstein}, Steven and {Foley}, Ryan and {Furlanetto}, Steven and {Kalirai}, Jason and {Gaudi}, B. Scott and {Gilbert}, Karoline and {Girard}, Julien and {Grady}, Kevin and {Greene}, Jenny and {Guhathakurta}, Puragra and {Heinrich}, Chen and {Hemmati}, Shoubaneh and {Hendel}, David and {Henderson}, Calen and {Henning}, Thomas and {Hirata}, Christopher and {Ho}, Shirley and {Huff}, Eric and {Hutter}, Anne and {Jansen}, Rolf and {Jha}, Saurabh and {Johnson}, Samson and {Jones}, David and {Kasdin}, Jeremy and {Kelly}, Patrick and {Kirshner}, Robert and {Koekemoer}, Anton and {Kruk}, Jeffrey and {Lewis}, Nikole and {Macintosh}, Bruce and {Madau}, Piero and {Malhotra}, Sangeeta and {Mandel}, Kaisey and {Massara}, Elena and {Masters}, Daniel and {McEnery}, Julie and {McQuinn}, Kristen and {Melchior}, Peter and {Melton}, Mark and {Mennesson}, Bertrand and {Peeples}, Molly and {Penny}, Matthew and {Perlmutter}, Saul and {Pisani}, Alice and {Plazas}, Andr{\'e}s and {Poleski}, Radek and {Postman}, Marc and {Ranc}, Cl{\'e}ment and {Rauscher}, Bernard and {Rest}, Armin and {Roberge}, Aki and {Robertson}, Brant and {Rodney}, Steven and {Rhoads}, James and {Rhodes}, Jason and {Ryan}, Russell, Jr. and {Sahu}, Kailash and {Sand}, David and {Scolnic}, Dan and {Seth}, Anil and {Shvartzvald}, Yossi and {Siellez}, Karelle and {Smith}, Arfon and {Spergel}, David and {Stassun}, Keivan and {Street}, Rachel and {Strolger}, Louis-Gregory and {Szalay}, Alexander and {Trauger}, John and {Troxel}, M.~A. and {Turnbull}, Margaret and {van der Marel}, Roeland and {von der Linden}, Anja and {Wang}, Yun and {Weinberg}, David and {Williams}, Benjamin and {Windhorst}, Rogier and {Wollack}, Edward and {Wu}, Hao-Yi and {Yee}, Jennifer and {Zimmerman}, Neil},
        title = "{The Wide Field Infrared Survey Telescope: 100 Hubbles for the 2020s}",
      journal = {arXiv e-prints},
         year = 2019,
        month = feb,
          eid = {arXiv:1902.05569},
        pages = {arXiv:1902.05569},
          doi = {10.48550/arXiv.1902.05569},
archivePrefix = {arXiv},
       eprint = {1902.05569},
 primaryClass = {astro-ph.IM},
       adsurl = {https://ui.adsabs.harvard.edu/abs/2019arXiv190205569A}
}

@article{Slepian:2018,
       author = {{Slepian}, Zachary and {Eisenstein}, Daniel J.},
        title = "{A practical computational method for the anisotropic redshift-space three-point correlation function}",
      journal = {MNRAS},
         year = 2018,
        month = aug,
       volume = {478},
       number = {2},
        pages = {1468-1483},
          doi = {10.1093/mnras/sty1063},
archivePrefix = {arXiv},
       eprint = {1709.10150},
 primaryClass = {astro-ph.CO},
       adsurl = {https://ui.adsabs.harvard.edu/abs/2018MNRAS.478.1468S}
}

@article{Ivanov:2018,
       author = {{Ivanov}, Mikhail M. and {Sibiryakov}, Sergey},
        title = "{Infrared resummation for biased tracers in redshift space}",
      journal = {JCAP},
         year = 2018,
        month = jul,
       volume = {2018},
       number = {7},
          eid = {053},
        pages = {053},
          doi = {10.1088/1475-7516/2018/07/053},
archivePrefix = {arXiv},
       eprint = {1804.05080},
 primaryClass = {astro-ph.CO},
       adsurl = {https://ui.adsabs.harvard.edu/abs/2018JCAP...07..053I}
}

@article{Simonovic:2018,
       author = {{Simonovi{\'c}}, Marko and {Baldauf}, Tobias and {Zaldarriaga}, Matias and {Carrasco}, John Joseph and {Kollmeier}, Juna A.},
        title = "{Cosmological perturbation theory using the FFTLog: formalism and connection to QFT loop integrals}",
      journal = {JCAP},
         year = 2018,
        month = apr,
       volume = {2018},
       number = {4},
          eid = {030},
        pages = {030},
          doi = {10.1088/1475-7516/2018/04/030},
archivePrefix = {arXiv},
       eprint = {1708.08130},
 primaryClass = {astro-ph.CO},
       adsurl = {https://ui.adsabs.harvard.edu/abs/2018JCAP...04..030S}
}

@article{Lewandowski:2018,
       author = {{Lewandowski}, Matthew and {Senatore}, Leonardo and {Prada}, Francisco and {Zhao}, Cheng and {Chuang}, Chia-Hsun},
        title = "{EFT of large scale structures in redshift space}",
      journal = {Phys. Rev. D},
         year = 2018,
        month = mar,
       volume = {97},
       number = {6},
          eid = {063526},
        pages = {063526},
          doi = {10.1103/PhysRevD.97.063526},
archivePrefix = {arXiv},
       eprint = {1512.06831},
 primaryClass = {astro-ph.CO},
       adsurl = {https://ui.adsabs.harvard.edu/abs/2018PhRvD..97f3526L}
}

@article{Desjacques:2018,
       author = {{Desjacques}, Vincent and {Jeong}, Donghui and {Schmidt}, Fabian},
        title = "{Large-scale galaxy bias}",
      journal = {Phys. Rep.},
         year = 2018,
        month = feb,
       volume = {733},
        pages = {1-193},
          doi = {10.1016/j.physrep.2017.12.002},
archivePrefix = {arXiv},
       eprint = {1611.09787},
 primaryClass = {astro-ph.CO},
       adsurl = {https://ui.adsabs.harvard.edu/abs/2018PhR...733....1D}
}

@article{Slepian:2017,
       author = {{Slepian}, Zachary and {Eisenstein}, Daniel J.},
        title = "{Modelling the large-scale redshift-space 3-point correlation function of galaxies}",
      journal = {MNRAS},
         year = 2017,
        month = aug,
       volume = {469},
       number = {2},
        pages = {2059-2076},
          doi = {10.1093/mnras/stx490},
archivePrefix = {arXiv},
       eprint = {1607.03109},
 primaryClass = {astro-ph.CO},
       adsurl = {https://ui.adsabs.harvard.edu/abs/2017MNRAS.469.2059S}
}

@article{Slepian:2017B,
       author = {{Slepian}, Zachary and {Eisenstein}, Daniel J. and {Brownstein}, Joel R. and {Chuang}, Chia-Hsun and {Gil-Mar{\'\i}n}, H{\'e}ctor and {Ho}, Shirley and {Kitaura}, Francisco-Shu and {Percival}, Will J. and {Ross}, Ashley J. and {Rossi}, Graziano and {Seo}, Hee-Jong and {Slosar}, An{\v{z}}e and {Vargas-Maga{\~n}a}, Mariana},
        title = "{Detection of baryon acoustic oscillation features in the large-scale three-point correlation function of SDSS BOSS DR12 CMASS galaxies}",
      journal = {MNRAS},
         year = 2017,
        month = aug,
       volume = {469},
       number = {2},
        pages = {1738-1751},
          doi = {10.1093/mnras/stx488},
archivePrefix = {arXiv},
       eprint = {1607.06097},
 primaryClass = {astro-ph.CO},
       adsurl = {https://ui.adsabs.harvard.edu/abs/2017MNRAS.469.1738S}
}

@article{Slepian:2017L,
       author = {{Slepian}, Zachary and {Eisenstein}, Daniel J. and {Beutler}, Florian and {Chuang}, Chia-Hsun and {Cuesta}, Antonio J. and {Ge}, Jian and {Gil-Mar{\'\i}n}, H{\'e}ctor and {Ho}, Shirley and {Kitaura}, Francisco-Shu and {McBride}, Cameron K. and {Nichol}, Robert C. and {Percival}, Will J. and {Rodr{\'\i}guez-Torres}, Sergio and {Ross}, Ashley J. and {Scoccimarro}, Rom{\'a}n and {Seo}, Hee-Jong and {Tinker}, Jeremy and {Tojeiro}, Rita and {Vargas-Maga{\~n}a}, Mariana},
        title = "{The large-scale three-point correlation function of the SDSS BOSS DR12 CMASS galaxies}",
      journal = {MNRAS},
         year = 2017,
        month = jun,
       volume = {468},
       number = {1},
        pages = {1070-1083},
          doi = {10.1093/mnras/stw3234},
archivePrefix = {arXiv},
       eprint = {1512.02231},
 primaryClass = {astro-ph.CO},
       adsurl = {https://ui.adsabs.harvard.edu/abs/2017MNRAS.468.1070S}
}

@article{Munari:2017,
       author = {{Munari}, Emiliano and {Monaco}, Pierluigi and {Sefusatti}, Emiliano and {Castorina}, Emanuele and {Mohammad}, Faizan G. and {Anselmi}, Stefano and {Borgani}, Stefano},
        title = "{Improving fast generation of halo catalogues with higher order Lagrangian perturbation theory}",
      journal = {MNRAS},
         year = 2017,
        month = mar,
       volume = {465},
       number = {4},
        pages = {4658-4677},
          doi = {10.1093/mnras/stw3085},
archivePrefix = {arXiv},
       eprint = {1605.04788},
 primaryClass = {astro-ph.CO},
       adsurl = {https://ui.adsabs.harvard.edu/abs/2017MNRAS.465.4658M}
}

@article{Sanchez:2017,
       author = {{S{\'a}nchez}, Ariel G. and {Scoccimarro}, Rom{\'a}n and {Crocce}, Mart{\'\i}n and {Grieb}, Jan Niklas and {Salazar-Albornoz}, Salvador and {Dalla Vecchia}, Claudio and {Lippich}, Martha and {Beutler}, Florian and {Brownstein}, Joel R. and {Chuang}, Chia-Hsun and {Eisenstein}, Daniel J. and {Kitaura}, Francisco-Shu and {Olmstead}, Matthew D. and {Percival}, Will J. and {Prada}, Francisco and {Rodr{\'\i}guez-Torres}, Sergio and {Ross}, Ashley J. and {Samushia}, Lado and {Seo}, Hee-Jong and {Tinker}, Jeremy and {Tojeiro}, Rita and {Vargas-Maga{\~n}a}, Mariana and {Wang}, Yuting and {Zhao}, Gong-Bo},
        title = "{The clustering of galaxies in the completed SDSS-III Baryon Oscillation Spectroscopic Survey: Cosmological implications of the configuration-space clustering wedges}",
      journal = {MNRAS},
         year = 2017,
        month = jan,
       volume = {464},
       number = {2},
        pages = {1640-1658},
          doi = {10.1093/mnras/stw2443},
archivePrefix = {arXiv},
       eprint = {1607.03147},
 primaryClass = {astro-ph.CO},
       adsurl = {https://ui.adsabs.harvard.edu/abs/2017MNRAS.464.1640S}
}

@article{Perko:2016,
       author = {{Perko}, Ashley and {Senatore}, Leonardo and {Jennings}, Elise and {Wechsler}, Risa H.},
        title = "{Biased Tracers in Redshift Space in the EFT of Large-Scale Structure}",
      journal = {arXiv e-prints},
         year = 2016,
        month = oct,
          eid = {arXiv:1610.09321},
        pages = {arXiv:1610.09321},
          doi = {10.48550/arXiv.1610.09321},
archivePrefix = {arXiv},
       eprint = {1610.09321},
 primaryClass = {astro-ph.CO},
       adsurl = {https://ui.adsabs.harvard.edu/abs/2016arXiv161009321P}
}

@article{DESI:2016,
       author = {{DESI Collaboration} and {Aghamousa}, Amir and {Aguilar}, Jessica and {Ahlen}, Steve and {Alam}, Shadab and {Allen}, Lori E. and {Allende Prieto}, Carlos and {Annis}, James and {Bailey}, Stephen and {Balland}, Christophe and {Ballester}, Otger and {Baltay}, Charles and {Beaufore}, Lucas and {Bebek}, Chris and {Beers}, Timothy C. and {Bell}, Eric F. and {Bernal}, Jos{\'e} Luis and {Besuner}, Robert and {Beutler}, Florian and {Blake}, Chris and {Bleuler}, Hannes and {Blomqvist}, Michael and {Blum}, Robert and {Bolton}, Adam S. and {Briceno}, Cesar and {Brooks}, David and {Brownstein}, Joel R. and {Buckley-Geer}, Elizabeth and {Burden}, Angela and {Burtin}, Etienne and {Busca}, Nicolas G. and {Cahn}, Robert N. and {Cai}, Yan-Chuan and {Cardiel-Sas}, Laia and {Carlberg}, Raymond G. and {Carton}, Pierre-Henri and {Casas}, Ricard and {Castander}, Francisco J. and {Cervantes-Cota}, Jorge L. and {Claybaugh}, Todd M. and {Close}, Madeline and {Coker}, Carl T. and {Cole}, Shaun and {Comparat}, Johan and {Cooper}, Andrew P. and {Cousinou}, M. -C. and {Crocce}, Martin and {Cuby}, Jean-Gabriel and {Cunningham}, Daniel P. and {Davis}, Tamara M. and {Dawson}, Kyle S. and {de la Macorra}, Axel and {De Vicente}, Juan and {Delubac}, Timoth{\'e}e and {Derwent}, Mark and {Dey}, Arjun and {Dhungana}, Govinda and {Ding}, Zhejie and {Doel}, Peter and {Duan}, Yutong T. and {Ealet}, Anne and {Edelstein}, Jerry and {Eftekharzadeh}, Sarah and {Eisenstein}, Daniel J. and {Elliott}, Ann and {Escoffier}, St{\'e}phanie and {Evatt}, Matthew and {Fagrelius}, Parker and {Fan}, Xiaohui and {Fanning}, Kevin and {Farahi}, Arya and {Farihi}, Jay and {Favole}, Ginevra and {Feng}, Yu and {Fernandez}, Enrique and {Findlay}, Joseph R. and {Finkbeiner}, Douglas P. and {Fitzpatrick}, Michael J. and {Flaugher}, Brenna and {Flender}, Samuel and {Font-Ribera}, Andreu and {Forero-Romero}, Jaime E. and {Fosalba}, Pablo and {Frenk}, Carlos S. and {Fumagalli}, Michele and {Gaensicke}, Boris T. and {Gallo}, Giuseppe and {Garcia-Bellido}, Juan and {Gaztanaga}, Enrique and {Pietro Gentile Fusillo}, Nicola and {Gerard}, Terry and {Gershkovich}, Irena and {Giannantonio}, Tommaso and {Gillet}, Denis and {Gonzalez-de-Rivera}, Guillermo and {Gonzalez-Perez}, Violeta and {Gott}, Shelby and {Graur}, Or and {Gutierrez}, Gaston and {Guy}, Julien and {Habib}, Salman and {Heetderks}, Henry and {Heetderks}, Ian and {Heitmann}, Katrin and {Hellwing}, Wojciech A. and {Herrera}, David A. and {Ho}, Shirley and {Holland}, Stephen and {Honscheid}, Klaus and {Huff}, Eric and {Hutchinson}, Timothy A. and {Huterer}, Dragan and {Hwang}, Ho Seong and {Illa Laguna}, Joseph Maria and {Ishikawa}, Yuzo and {Jacobs}, Dianna and {Jeffrey}, Niall and {Jelinsky}, Patrick and {Jennings}, Elise and {Jiang}, Linhua and {Jimenez}, Jorge and {Johnson}, Jennifer and {Joyce}, Richard and {Jullo}, Eric and {Juneau}, St{\'e}phanie and {Kama}, Sami and {Karcher}, Armin and {Karkar}, Sonia and {Kehoe}, Robert and {Kennamer}, Noble and {Kent}, Stephen and {Kilbinger}, Martin and {Kim}, Alex G. and {Kirkby}, David and {Kisner}, Theodore and {Kitanidis}, Ellie and {Kneib}, Jean-Paul and {Koposov}, Sergey and {Kovacs}, Eve and {Koyama}, Kazuya and {Kremin}, Anthony and {Kron}, Richard and {Kronig}, Luzius and {Kueter-Young}, Andrea and {Lacey}, Cedric G. and {Lafever}, Robin and {Lahav}, Ofer and {Lambert}, Andrew and {Lampton}, Michael and {Landriau}, Martin and {Lang}, Dustin and {Lauer}, Tod R. and {Le Goff}, Jean-Marc and {Le Guillou}, Laurent and {Le Van Suu}, Auguste and {Lee}, Jae Hyeon and {Lee}, Su-Jeong and {Leitner}, Daniela and {Lesser}, Michael and {Levi}, Michael E. and {L'Huillier}, Benjamin and {Li}, Baojiu and {Liang}, Ming and {Lin}, Huan and {Linder}, Eric and {Loebman}, Sarah R. and {Luki{\'c}}, Zarija and {Ma}, Jun and {MacCrann}, Niall and {Magneville}, Christophe and {Makarem}, Laleh and {Manera}, Marc and {Manser}, Christopher J. and {Marshall}, Robert and {Martini}, Paul and {Massey}, Richard and {Matheson}, Thomas and {McCauley}, Jeremy and {McDonald}, Patrick and {McGreer}, Ian D. and {Meisner}, Aaron and {Metcalfe}, Nigel and {Miller}, Timothy N. and {Miquel}, Ramon and {Moustakas}, John and {Myers}, Adam and {Naik}, Milind and {Newman}, Jeffrey A. and {Nichol}, Robert C. and {Nicola}, Andrina and {Nicolati da Costa}, Luiz and {Nie}, Jundan and {Niz}, Gustavo and {Norberg}, Peder and {Nord}, Brian and {Norman}, Dara and {Nugent}, Peter and {O'Brien}, Thomas and {Oh}, Minji and {Olsen}, Knut A.~G. and {Padilla}, Cristobal and {Padmanabhan}, Hamsa and {Padmanabhan}, Nikhil and {Palanque-Delabrouille}, Nathalie and {Palmese}, Antonella and {Pappalardo}, Daniel and {P{\^a}ris}, Isabelle and {Park}, Changbom and {Patej}, Anna and {Peacock}, John A. and {Peiris}, Hiranya V. and {Peng}, Xiyan and {Percival}, Will J. and {Perruchot}, Sandrine and {Pieri}, Matthew M. and {Pogge}, Richard and {Pollack}, Jennifer E. and {Poppett}, Claire and {Prada}, Francisco and {Prakash}, Abhishek and {Probst}, Ronald G. and {Rabinowitz}, David and {Raichoor}, Anand and {Ree}, Chang Hee and {Refregier}, Alexandre and {Regal}, Xavier and {Reid}, Beth and {Reil}, Kevin and {Rezaie}, Mehdi and {Rockosi}, Constance M. and {Roe}, Natalie and {Ronayette}, Samuel and {Roodman}, Aaron and {Ross}, Ashley J. and {Ross}, Nicholas P. and {Rossi}, Graziano and {Rozo}, Eduardo and {Ruhlmann-Kleider}, Vanina and {Rykoff}, Eli S. and {Sabiu}, Cristiano and {Samushia}, Lado and {Sanchez}, Eusebio and {Sanchez}, Javier and {Schlegel}, David J. and {Schneider}, Michael and {Schubnell}, Michael and {Secroun}, Aur{\'e}lia and {Seljak}, Uros and {Seo}, Hee-Jong and {Serrano}, Santiago and {Shafieloo}, Arman and {Shan}, Huanyuan and {Sharples}, Ray and {Sholl}, Michael J. and {Shourt}, William V. and {Silber}, Joseph H. and {Silva}, David R. and {Sirk}, Martin M. and {Slosar}, Anze and {Smith}, Alex and {Smoot}, George F. and {Som}, Debopam and {Song}, Yong-Seon and {Sprayberry}, David and {Staten}, Ryan and {Stefanik}, Andy and {Tarle}, Gregory and {Sien Tie}, Suk and {Tinker}, Jeremy L. and {Tojeiro}, Rita and {Valdes}, Francisco and {Valenzuela}, Octavio and {Valluri}, Monica and {Vargas-Magana}, Mariana and {Verde}, Licia and {Walker}, Alistair R. and {Wang}, Jiali and {Wang}, Yuting and {Weaver}, Benjamin A. and {Weaverdyck}, Curtis and {Wechsler}, Risa H. and {Weinberg}, David H. and {White}, Martin and {Yang}, Qian and {Yeche}, Christophe and {Zhang}, Tianmeng and {Zhao}, Gong-Bo and {Zheng}, Yi and {Zhou}, Xu and {Zhou}, Zhimin and {Zhu}, Yaling and {Zou}, Hu and {Zu}, Ying},
        title = "{The DESI Experiment Part II: Instrument Design}",
      journal = {arXiv e-prints},
         year = 2016,
        month = oct,
          eid = {arXiv:1611.00037},
        pages = {arXiv:1611.00037},
          doi = {10.48550/arXiv.1611.00037},
archivePrefix = {arXiv},
       eprint = {1611.00037},
 primaryClass = {astro-ph.IM},
       adsurl = {https://ui.adsabs.harvard.edu/abs/2016arXiv161100037D}
}

@article{DESI:2016b,
       author = {{DESI Collaboration} and {Aghamousa}, Amir and {Aguilar}, Jessica and {Ahlen}, Steve and {Alam}, Shadab and {Allen}, Lori E. and {Allende Prieto}, Carlos and {Annis}, James and {Bailey}, Stephen and {Balland}, Christophe and {Ballester}, Otger and {Baltay}, Charles and {Beaufore}, Lucas and {Bebek}, Chris and {Beers}, Timothy C. and {Bell}, Eric F. and {Bernal}, Jos{\'e} Luis and {Besuner}, Robert and {Beutler}, Florian and {Blake}, Chris and {Bleuler}, Hannes and {Blomqvist}, Michael and {Blum}, Robert and {Bolton}, Adam S. and {Briceno}, Cesar and {Brooks}, David and {Brownstein}, Joel R. and {Buckley-Geer}, Elizabeth and {Burden}, Angela and {Burtin}, Etienne and {Busca}, Nicolas G. and {Cahn}, Robert N. and {Cai}, Yan-Chuan and {Cardiel-Sas}, Laia and {Carlberg}, Raymond G. and {Carton}, Pierre-Henri and {Casas}, Ricard and {Castander}, Francisco J. and {Cervantes-Cota}, Jorge L. and {Claybaugh}, Todd M. and {Close}, Madeline and {Coker}, Carl T. and {Cole}, Shaun and {Comparat}, Johan and {Cooper}, Andrew P. and {Cousinou}, M. -C. and {Crocce}, Martin and {Cuby}, Jean-Gabriel and {Cunningham}, Daniel P. and {Davis}, Tamara M. and {Dawson}, Kyle S. and {de la Macorra}, Axel and {De Vicente}, Juan and {Delubac}, Timoth{\'e}e and {Derwent}, Mark and {Dey}, Arjun and {Dhungana}, Govinda and {Ding}, Zhejie and {Doel}, Peter and {Duan}, Yutong T. and {Ealet}, Anne and {Edelstein}, Jerry and {Eftekharzadeh}, Sarah and {Eisenstein}, Daniel J. and {Elliott}, Ann and {Escoffier}, St{\'e}phanie and {Evatt}, Matthew and {Fagrelius}, Parker and {Fan}, Xiaohui and {Fanning}, Kevin and {Farahi}, Arya and {Farihi}, Jay and {Favole}, Ginevra and {Feng}, Yu and {Fernandez}, Enrique and {Findlay}, Joseph R. and {Finkbeiner}, Douglas P. and {Fitzpatrick}, Michael J. and {Flaugher}, Brenna and {Flender}, Samuel and {Font-Ribera}, Andreu and {Forero-Romero}, Jaime E. and {Fosalba}, Pablo and {Frenk}, Carlos S. and {Fumagalli}, Michele and {Gaensicke}, Boris T. and {Gallo}, Giuseppe and {Garcia-Bellido}, Juan and {Gaztanaga}, Enrique and {Pietro Gentile Fusillo}, Nicola and {Gerard}, Terry and {Gershkovich}, Irena and {Giannantonio}, Tommaso and {Gillet}, Denis and {Gonzalez-de-Rivera}, Guillermo and {Gonzalez-Perez}, Violeta and {Gott}, Shelby and {Graur}, Or and {Gutierrez}, Gaston and {Guy}, Julien and {Habib}, Salman and {Heetderks}, Henry and {Heetderks}, Ian and {Heitmann}, Katrin and {Hellwing}, Wojciech A. and {Herrera}, David A. and {Ho}, Shirley and {Holland}, Stephen and {Honscheid}, Klaus and {Huff}, Eric and {Hutchinson}, Timothy A. and {Huterer}, Dragan and {Hwang}, Ho Seong and {Illa Laguna}, Joseph Maria and {Ishikawa}, Yuzo and {Jacobs}, Dianna and {Jeffrey}, Niall and {Jelinsky}, Patrick and {Jennings}, Elise and {Jiang}, Linhua and {Jimenez}, Jorge and {Johnson}, Jennifer and {Joyce}, Richard and {Jullo}, Eric and {Juneau}, St{\'e}phanie and {Kama}, Sami and {Karcher}, Armin and {Karkar}, Sonia and {Kehoe}, Robert and {Kennamer}, Noble and {Kent}, Stephen and {Kilbinger}, Martin and {Kim}, Alex G. and {Kirkby}, David and {Kisner}, Theodore and {Kitanidis}, Ellie and {Kneib}, Jean-Paul and {Koposov}, Sergey and {Kovacs}, Eve and {Koyama}, Kazuya and {Kremin}, Anthony and {Kron}, Richard and {Kronig}, Luzius and {Kueter-Young}, Andrea and {Lacey}, Cedric G. and {Lafever}, Robin and {Lahav}, Ofer and {Lambert}, Andrew and {Lampton}, Michael and {Landriau}, Martin and {Lang}, Dustin and {Lauer}, Tod R. and {Le Goff}, Jean-Marc and {Le Guillou}, Laurent and {Le Van Suu}, Auguste and {Lee}, Jae Hyeon and {Lee}, Su-Jeong and {Leitner}, Daniela and {Lesser}, Michael and {Levi}, Michael E. and {L'Huillier}, Benjamin and {Li}, Baojiu and {Liang}, Ming and {Lin}, Huan and {Linder}, Eric and {Loebman}, Sarah R. and {Luki{\'c}}, Zarija and {Ma}, Jun and {MacCrann}, Niall and {Magneville}, Christophe and {Makarem}, Laleh and {Manera}, Marc and {Manser}, Christopher J. and {Marshall}, Robert and {Martini}, Paul and {Massey}, Richard and {Matheson}, Thomas and {McCauley}, Jeremy and {McDonald}, Patrick and {McGreer}, Ian D. and {Meisner}, Aaron and {Metcalfe}, Nigel and {Miller}, Timothy N. and {Miquel}, Ramon and {Moustakas}, John and {Myers}, Adam and {Naik}, Milind and {Newman}, Jeffrey A. and {Nichol}, Robert C. and {Nicola}, Andrina and {Nicolati da Costa}, Luiz and {Nie}, Jundan and {Niz}, Gustavo and {Norberg}, Peder and {Nord}, Brian and {Norman}, Dara and {Nugent}, Peter and {O'Brien}, Thomas and {Oh}, Minji and {Olsen}, Knut A.~G. and {Padilla}, Cristobal and {Padmanabhan}, Hamsa and {Padmanabhan}, Nikhil and {Palanque-Delabrouille}, Nathalie and {Palmese}, Antonella and {Pappalardo}, Daniel and {P{\^a}ris}, Isabelle and {Park}, Changbom and {Patej}, Anna and {Peacock}, John A. and {Peiris}, Hiranya V. and {Peng}, Xiyan and {Percival}, Will J. and {Perruchot}, Sandrine and {Pieri}, Matthew M. and {Pogge}, Richard and {Pollack}, Jennifer E. and {Poppett}, Claire and {Prada}, Francisco and {Prakash}, Abhishek and {Probst}, Ronald G. and {Rabinowitz}, David and {Raichoor}, Anand and {Ree}, Chang Hee and {Refregier}, Alexandre and {Regal}, Xavier and {Reid}, Beth and {Reil}, Kevin and {Rezaie}, Mehdi and {Rockosi}, Constance M. and {Roe}, Natalie and {Ronayette}, Samuel and {Roodman}, Aaron and {Ross}, Ashley J. and {Ross}, Nicholas P. and {Rossi}, Graziano and {Rozo}, Eduardo and {Ruhlmann-Kleider}, Vanina and {Rykoff}, Eli S. and {Sabiu}, Cristiano and {Samushia}, Lado and {Sanchez}, Eusebio and {Sanchez}, Javier and {Schlegel}, David J. and {Schneider}, Michael and {Schubnell}, Michael and {Secroun}, Aur{\'e}lia and {Seljak}, Uros and {Seo}, Hee-Jong and {Serrano}, Santiago and {Shafieloo}, Arman and {Shan}, Huanyuan and {Sharples}, Ray and {Sholl}, Michael J. and {Shourt}, William V. and {Silber}, Joseph H. and {Silva}, David R. and {Sirk}, Martin M. and {Slosar}, Anze and {Smith}, Alex and {Smoot}, George F. and {Som}, Debopam and {Song}, Yong-Seon and {Sprayberry}, David and {Staten}, Ryan and {Stefanik}, Andy and {Tarle}, Gregory and {Sien Tie}, Suk and {Tinker}, Jeremy L. and {Tojeiro}, Rita and {Valdes}, Francisco and {Valenzuela}, Octavio and {Valluri}, Monica and {Vargas-Magana}, Mariana and {Verde}, Licia and {Walker}, Alistair R. and {Wang}, Jiali and {Wang}, Yuting and {Weaver}, Benjamin A. and {Weaverdyck}, Curtis and {Wechsler}, Risa H. and {Weinberg}, David H. and {White}, Martin and {Yang}, Qian and {Yeche}, Christophe and {Zhang}, Tianmeng and {Zhao}, Gong-Bo and {Zheng}, Yi and {Zhou}, Xu and {Zhou}, Zhimin and {Zhu}, Yaling and {Zou}, Hu and {Zu}, Ying},
        title = "{The DESI Experiment Part I: Science,Targeting, and Survey Design}",
      journal = {arXiv e-prints},
         year = 2016,
        month = oct,
          eid = {arXiv:1611.00036},
        pages = {arXiv:1611.00036},
          doi = {10.48550/arXiv.1611.00036},
archivePrefix = {arXiv},
       eprint = {1611.00036},
 primaryClass = {astro-ph.IM},
       adsurl = {https://ui.adsabs.harvard.edu/abs/2016arXiv161100036D}
}

@article{Blas:2016,
       author = {{Blas}, Diego and {Garny}, Mathias and {Ivanov}, Mikhail M. and {Sibiryakov}, Sergey},
        title = "{Time-sliced perturbation theory II: baryon acoustic oscillations and infrared resummation}",
      journal = {JCAP},
         year = 2016,
        month = jul,
       volume = {2016},
       number = {7},
          eid = {028},
        pages = {028},
          doi = {10.1088/1475-7516/2016/07/028},
archivePrefix = {arXiv},
       eprint = {1605.02149},
 primaryClass = {astro-ph.CO},
       adsurl = {https://ui.adsabs.harvard.edu/abs/2016JCAP...07..028B}
}

@article{Grieb:2016,
       author = {{Grieb}, Jan Niklas and {S{\'a}nchez}, Ariel G. and {Salazar-Albornoz}, Salvador and {Dalla Vecchia}, Claudio},
        title = "{Gaussian covariance matrices for anisotropic galaxy clustering measurements}",
      journal = {MNRAS},
         year = 2016,
        month = apr,
       volume = {457},
       number = {2},
        pages = {1577-1592},
          doi = {10.1093/mnras/stw065},
archivePrefix = {arXiv},
       eprint = {1509.04293},
 primaryClass = {astro-ph.CO},
       adsurl = {https://ui.adsabs.harvard.edu/abs/2016MNRAS.457.1577G}
}

@article{Sellentin:2015,
       author = {{Sellentin}, Elena and {Heavens}, Alan F.},
        title = "{Parameter inference with estimated covariance matrices}",
      journal = {MNRAS},
         year = 2016,
        month = feb,
       volume = {456},
       number = {1},
        pages = {L132-L136},
          doi = {10.1093/mnrasl/slv190},
archivePrefix = {arXiv},
       eprint = {1511.05969},
 primaryClass = {astro-ph.CO},
       adsurl = {https://ui.adsabs.harvard.edu/abs/2016MNRAS.456L.132S}
}

@article{Slepian:2015a,
       author = {{Slepian}, Zachary and {Eisenstein}, Daniel J.},
        title = "{Accelerating the two-point and three-point galaxy correlation functions using Fourier transforms}",
      journal = {MNRAS},
         year = 2016,
        month = jan,
       volume = {455},
       number = {1},
        pages = {L31-L35},
          doi = {10.1093/mnrasl/slv133},
archivePrefix = {arXiv},
       eprint = {1506.04746},
 primaryClass = {astro-ph.CO},
       adsurl = {https://ui.adsabs.harvard.edu/abs/2016MNRAS.455L..31S}
}

@article{Slepian:2015,
       author = {{Slepian}, Zachary and {Eisenstein}, Daniel J.},
        title = "{Computing the three-point correlation function of galaxies in $O(N^2)$ time}",
      journal = {MNRAS},
         year = 2015,
        month = dec,
       volume = {454},
       number = {4},
        pages = {4142-4158},
          doi = {10.1093/mnras/stv2119},
archivePrefix = {arXiv},
       eprint = {1506.02040},
 primaryClass = {astro-ph.CO},
       adsurl = {https://ui.adsabs.harvard.edu/abs/2015MNRAS.454.4142S}
}

@article{Senatore:2015,
       author = {{Senatore}, Leonardo},
        title = "{Bias in the effective field theory of large scale structures}",
      journal = {JCAP},
         year = 2015,
        month = nov,
       volume = {2015},
       number = {11},
        pages = {007-007},
          doi = {10.1088/1475-7516/2015/11/007},
archivePrefix = {arXiv},
       eprint = {1406.7843},
 primaryClass = {astro-ph.CO},
       adsurl = {https://ui.adsabs.harvard.edu/abs/2015JCAP...11..007S}
}

@article{Angulo:2015,
       author = {{Angulo}, Raul E. and {Foreman}, Simon and {Schmittfull}, Marcel and {Senatore}, Leonardo},
        title = "{The one-loop matter bispectrum in the Effective Field Theory of Large Scale Structures}",
      journal = {JCAP},
         year = 2015,
        month = oct,
       volume = {2015},
       number = {10},
        pages = {039-039},
          doi = {10.1088/1475-7516/2015/10/039},
archivePrefix = {arXiv},
       eprint = {1406.4143},
 primaryClass = {astro-ph.CO},
       adsurl = {https://ui.adsabs.harvard.edu/abs/2015JCAP...10..039A}
}

@article{Mirbabayi:2015,
       author = {{Mirbabayi}, Mehrdad and {Schmidt}, Fabian and {Zaldarriaga}, Matias},
        title = "{Biased tracers and time evolution}",
      journal = {JCAP},
         year = 2015,
        month = jul,
       volume = {2015},
       number = {7},
        pages = {030-030},
          doi = {10.1088/1475-7516/2015/07/030},
archivePrefix = {arXiv},
       eprint = {1412.5169},
 primaryClass = {astro-ph.CO},
       adsurl = {https://ui.adsabs.harvard.edu/abs/2015JCAP...07..030M}
}

@article{Slepian:2015v,
       author = {{Slepian}, Zachary and {Eisenstein}, Daniel J.},
        title = "{On the signature of the baryon-dark matter relative velocity in the two- and three-point galaxy correlation functions}",
      journal = {MNRAS},
         year = 2015,
        month = mar,
       volume = {448},
       number = {1},
        pages = {9-26},
          doi = {10.1093/mnras/stu2627},
archivePrefix = {arXiv},
       eprint = {1411.4052},
 primaryClass = {astro-ph.CO},
       adsurl = {https://ui.adsabs.harvard.edu/abs/2015MNRAS.448....9S}
}

@article{Senatore:2014,
       author = {{Senatore}, Leonardo and {Zaldarriaga}, Matias},
        title = "{Redshift Space Distortions in the Effective Field Theory of Large Scale Structures}",
      journal = {arXiv e-prints},
         year = 2014,
        month = sep,
          eid = {arXiv:1409.1225},
        pages = {arXiv:1409.1225},
          doi = {10.48550/arXiv.1409.1225},
archivePrefix = {arXiv},
       eprint = {1409.1225},
 primaryClass = {astro-ph.CO},
       adsurl = {https://ui.adsabs.harvard.edu/abs/2014arXiv1409.1225S}
}

@article{Assassi:2014,
       author = {{Assassi}, Valentin and {Baumann}, Daniel and {Green}, Daniel and {Zaldarriaga}, Matias},
        title = "{Renormalized halo bias}",
      journal = {JCAP},
         year = 2014,
        month = aug,
       volume = {2014},
       number = {8},
        pages = {056-056},
          doi = {10.1088/1475-7516/2014/08/056},
archivePrefix = {arXiv},
       eprint = {1402.5916},
 primaryClass = {astro-ph.CO},
       adsurl = {https://ui.adsabs.harvard.edu/abs/2014JCAP...08..056A}
}

@article{Porto:2014,
       author = {{Porto}, Rafael A. and {Senatore}, Leonardo and {Zaldarriaga}, Matias},
        title = "{The Lagrangian-space Effective Field Theory of large scale structures}",
      journal = {JCAP},
         year = 2014,
        month = may,
       volume = {2014},
       number = {5},
          eid = {022},
        pages = {022},
          doi = {10.1088/1475-7516/2014/05/022},
archivePrefix = {arXiv},
       eprint = {1311.2168},
 primaryClass = {astro-ph.CO},
       adsurl = {https://ui.adsabs.harvard.edu/abs/2014JCAP...05..022P}
}

@article{Percival:2014,
       author = {{Percival}, Will J. and {Ross}, Ashley J. and {S{\'a}nchez}, Ariel G. and {Samushia}, Lado and {Burden}, Angela and {Crittenden}, Robert and {Cuesta}, Antonio J. and {Magana}, Mariana Vargas and {Manera}, Marc and {Beutler}, Florian and {Chuang}, Chia-Hsun and {Eisenstein}, Daniel J. and {Ho}, Shirley and {McBride}, Cameron K. and {Montesano}, Francesco and {Padmanabhan}, Nikhil and {Reid}, Beth and {Saito}, Shun and {Schneider}, Donald P. and {Seo}, Hee-Jong and {Tojeiro}, Rita and {Weaver}, Benjamin A.},
        title = "{The clustering of Galaxies in the SDSS-III Baryon Oscillation Spectroscopic Survey: including covariance matrix errors}",
      journal = {MNRAS},
         year = 2014,
        month = apr,
       volume = {439},
       number = {3},
        pages = {2531-2541},
          doi = {10.1093/mnras/stu112},
archivePrefix = {arXiv},
       eprint = {1312.4841},
 primaryClass = {astro-ph.CO},
       adsurl = {https://ui.adsabs.harvard.edu/abs/2014MNRAS.439.2531P}
}

@article{Wang:2014,
       author = {{Wang}, Lile and {Reid}, Beth and {White}, Martin},
        title = "{An analytic model for redshift-space distortions}",
      journal = {MNRAS},
         year = 2014,
        month = jan,
       volume = {437},
       number = {1},
        pages = {588-599},
          doi = {10.1093/mnras/stt1916},
archivePrefix = {arXiv},
       eprint = {1306.1804},
 primaryClass = {astro-ph.CO},
       adsurl = {https://ui.adsabs.harvard.edu/abs/2014MNRAS.437..588W}
}

@article{Monaco:2013,
       author = {{Monaco}, P. and {Sefusatti}, E. and {Borgani}, S. and {Crocce}, M. and {Fosalba}, P. and {Sheth}, R.~K. and {Theuns}, T.},
        title = "{An accurate tool for the fast generation of dark matter halo catalogues}",
      journal = {MNRAS},
         year = 2013,
        month = aug,
       volume = {433},
       number = {3},
        pages = {2389-2402},
          doi = {10.1093/mnras/stt907},
archivePrefix = {arXiv},
       eprint = {1305.1505},
 primaryClass = {astro-ph.CO},
       adsurl = {https://ui.adsabs.harvard.edu/abs/2013MNRAS.433.2389M}
}

@article{Monaco:2013a,
       author = {{Monaco}, Pierluigi and {Theuns}, Tom and {Taffoni}, Giuliano},
        title = "{The pinocchio algorithm: pinpointing orbit-crossing collapsed hierarchical objects in a linear density field}",
      journal = {MNRAS},
         year = 2002,
        month = apr,
       volume = {331},
       number = {3},
        pages = {587-608},
          doi = {10.1046/j.1365-8711.2002.05162.x},
archivePrefix = {arXiv},
       eprint = {astro-ph/0109323},
 primaryClass = {astro-ph},
       adsurl = {https://ui.adsabs.harvard.edu/abs/2002MNRAS.331..587M}
}

@article{Carlson:2013,
       author = {{Carlson}, Jordan and {Reid}, Beth and {White}, Martin},
        title = "{Convolution Lagrangian perturbation theory for biased tracers}",
      journal = {MNRAS},
         year = 2013,
        month = feb,
       volume = {429},
       number = {2},
        pages = {1674-1685},
          doi = {10.1093/mnras/sts457},
archivePrefix = {arXiv},
       eprint = {1209.0780},
 primaryClass = {astro-ph.CO},
       adsurl = {https://ui.adsabs.harvard.edu/abs/2013MNRAS.429.1674C}
}

@article{Baldauf:2012,
       author = {{Baldauf}, Tobias and {Seljak}, Uro{\v{s}} and {Desjacques}, Vincent and {McDonald}, Patrick},
        title = "{Evidence for quadratic tidal tensor bias from the halo bispectrum}",
      journal = {Phys. Rev. D},
         year = 2012,
        month = oct,
       volume = {86},
       number = {8},
          eid = {083540},
        pages = {083540},
          doi = {10.1103/PhysRevD.86.083540},
archivePrefix = {arXiv},
       eprint = {1201.4827},
 primaryClass = {astro-ph.CO},
       adsurl = {https://ui.adsabs.harvard.edu/abs/2012PhRvD..86h3540B}
}

@article{Carrasco:2012,
       author = {{Carrasco}, John Joseph M. and {Hertzberg}, Mark P. and {Senatore}, Leonardo},
        title = "{The effective field theory of cosmological large scale structures}",
      journal = {Journal of High Energy Physics},
         year = 2012,
        month = sep,
       volume = {2012},
          eid = {82},
        pages = {82},
          doi = {10.1007/JHEP09(2012)082},
archivePrefix = {arXiv},
       eprint = {1206.2926},
 primaryClass = {astro-ph.CO},
       adsurl = {https://ui.adsabs.harvard.edu/abs/2012JHEP...09..082C}
}

@article{Baumann:2012,
       author = {{Baumann}, Daniel and {Nicolis}, Alberto and {Senatore}, Leonardo and {Zaldarriaga}, Matias},
        title = "{Cosmological non-linearities as an effective fluid}",
      journal = {JCAP},
         year = 2012,
        month = jul,
       volume = {2012},
       number = {7},
          eid = {051},
        pages = {051},
          doi = {10.1088/1475-7516/2012/07/051},
archivePrefix = {arXiv},
       eprint = {1004.2488},
 primaryClass = {astro-ph.CO},
       adsurl = {https://ui.adsabs.harvard.edu/abs/2012JCAP...07..051B}
}

@article{Bernardeau:2012,
       author = {{Bernardeau}, Francis and {Crocce}, Mart{\'\i}n and {Scoccimarro}, Rom{\'a}n},
        title = "{Constructing regularized cosmic propagators}",
      journal = {Phys. Rev. D},
         year = 2012,
        month = jun,
       volume = {85},
       number = {12},
          eid = {123519},
        pages = {123519},
          doi = {10.1103/PhysRevD.85.123519},
archivePrefix = {arXiv},
       eprint = {1112.3895},
 primaryClass = {astro-ph.CO},
       adsurl = {https://ui.adsabs.harvard.edu/abs/2012PhRvD..85l3519B}
}

@article{Laureijs:2011,
       author = {{Laureijs}, R. and {Amiaux}, J. and {Arduini}, S. and {Augu{\`e}res}, J. -L. and {Brinchmann}, J. and {Cole}, R. and {Cropper}, M. and {Dabin}, C. and {Duvet}, L. and {Ealet}, A. and {Garilli}, B. and {Gondoin}, P. and {Guzzo}, L. and {Hoar}, J. and {Hoekstra}, H. and {Holmes}, R. and {Kitching}, T. and {Maciaszek}, T. and {Mellier}, Y. and {Pasian}, F. and {Percival}, W. and {Rhodes}, J. and {Saavedra Criado}, G. and {Sauvage}, M. and {Scaramella}, R. and {Valenziano}, L. and {Warren}, S. and {Bender}, R. and {Castander}, F. and {Cimatti}, A. and {Le F{\`e}vre}, O. and {Kurki-Suonio}, H. and {Levi}, M. and {Lilje}, P. and {Meylan}, G. and {Nichol}, R. and {Pedersen}, K. and {Popa}, V. and {Rebolo Lopez}, R. and {Rix}, H. -W. and {Rottgering}, H. and {Zeilinger}, W. and {Grupp}, F. and {Hudelot}, P. and {Massey}, R. and {Meneghetti}, M. and {Miller}, L. and {Paltani}, S. and {Paulin-Henriksson}, S. and {Pires}, S. and {Saxton}, C. and {Schrabback}, T. and {Seidel}, G. and {Walsh}, J. and {Aghanim}, N. and {Amendola}, L. and {Bartlett}, J. and {Baccigalupi}, C. and {Beaulieu}, J. -P. and {Benabed}, K. and {Cuby}, J. -G. and {Elbaz}, D. and {Fosalba}, P. and {Gavazzi}, G. and {Helmi}, A. and {Hook}, I. and {Irwin}, M. and {Kneib}, J. -P. and {Kunz}, M. and {Mannucci}, F. and {Moscardini}, L. and {Tao}, C. and {Teyssier}, R. and {Weller}, J. and {Zamorani}, G. and {Zapatero Osorio}, M.~R. and {Boulade}, O. and {Foumond}, J.~J. and {Di Giorgio}, A. and {Guttridge}, P. and {James}, A. and {Kemp}, M. and {Martignac}, J. and {Spencer}, A. and {Walton}, D. and {Bl{\"u}mchen}, T. and {Bonoli}, C. and {Bortoletto}, F. and {Cerna}, C. and {Corcione}, L. and {Fabron}, C. and {Jahnke}, K. and {Ligori}, S. and {Madrid}, F. and {Martin}, L. and {Morgante}, G. and {Pamplona}, T. and {Prieto}, E. and {Riva}, M. and {Toledo}, R. and {Trifoglio}, M. and {Zerbi}, F. and {Abdalla}, F. and {Douspis}, M. and {Grenet}, C. and {Borgani}, S. and {Bouwens}, R. and {Courbin}, F. and {Delouis}, J. -M. and {Dubath}, P. and {Fontana}, A. and {Frailis}, M. and {Grazian}, A. and {Koppenh{\"o}fer}, J. and {Mansutti}, O. and {Melchior}, M. and {Mignoli}, M. and {Mohr}, J. and {Neissner}, C. and {Noddle}, K. and {Poncet}, M. and {Scodeggio}, M. and {Serrano}, S. and {Shane}, N. and {Starck}, J. -L. and {Surace}, C. and {Taylor}, A. and {Verdoes-Kleijn}, G. and {Vuerli}, C. and {Williams}, O.~R. and {Zacchei}, A. and {Altieri}, B. and {Escudero Sanz}, I. and {Kohley}, R. and {Oosterbroek}, T. and {Astier}, P. and {Bacon}, D. and {Bardelli}, S. and {Baugh}, C. and {Bellagamba}, F. and {Benoist}, C. and {Bianchi}, D. and {Biviano}, A. and {Branchini}, E. and {Carbone}, C. and {Cardone}, V. and {Clements}, D. and {Colombi}, S. and {Conselice}, C. and {Cresci}, G. and {Deacon}, N. and {Dunlop}, J. and {Fedeli}, C. and {Fontanot}, F. and {Franzetti}, P. and {Giocoli}, C. and {Garcia-Bellido}, J. and {Gow}, J. and {Heavens}, A. and {Hewett}, P. and {Heymans}, C. and {Holland}, A. and {Huang}, Z. and {Ilbert}, O. and {Joachimi}, B. and {Jennins}, E. and {Kerins}, E. and {Kiessling}, A. and {Kirk}, D. and {Kotak}, R. and {Krause}, O. and {Lahav}, O. and {van Leeuwen}, F. and {Lesgourgues}, J. and {Lombardi}, M. and {Magliocchetti}, M. and {Maguire}, K. and {Majerotto}, E. and {Maoli}, R. and {Marulli}, F. and {Maurogordato}, S. and {McCracken}, H. and {McLure}, R. and {Melchiorri}, A. and {Merson}, A. and {Moresco}, M. and {Nonino}, M. and {Norberg}, P. and {Peacock}, J. and {Pello}, R. and {Penny}, M. and {Pettorino}, V. and {Di Porto}, C. and {Pozzetti}, L. and {Quercellini}, C. and {Radovich}, M. and {Rassat}, A. and {Roche}, N. and {Ronayette}, S. and {Rossetti}, E. and {Sartoris}, B. and {Schneider}, P. and {Semboloni}, E. and {Serjeant}, S. and {Simpson}, F. and {Skordis}, C. and {Smadja}, G. and {Smartt}, S. and {Spano}, P. and {Spiro}, S. and {Sullivan}, M. and {Tilquin}, A. and {Trotta}, R. and {Verde}, L. and {Wang}, Y. and {Williger}, G. and {Zhao}, G. and {Zoubian}, J. and {Zucca}, E.},
        title = "{Euclid Definition Study Report}",
      journal = {arXiv e-prints},
         year = 2011,
        month = oct,
          eid = {arXiv:1110.3193},
        pages = {arXiv:1110.3193},
          doi = {10.48550/arXiv.1110.3193},
archivePrefix = {arXiv},
       eprint = {1110.3193},
 primaryClass = {astro-ph.CO},
       adsurl = {https://ui.adsabs.harvard.edu/abs/2011arXiv1110.3193L}
}

@article{McDonald:2009,
       author = {{McDonald}, Patrick and {Roy}, Arabindo},
        title = "{Clustering of dark matter tracers: generalizing bias for the coming era of precision LSS}",
      journal = {JCAP},
         year = 2009,
        month = aug,
       volume = {2009},
       number = {8},
          eid = {020},
        pages = {020},
          doi = {10.1088/1475-7516/2009/08/020},
archivePrefix = {arXiv},
       eprint = {0902.0991},
 primaryClass = {astro-ph.CO},
       adsurl = {https://ui.adsabs.harvard.edu/abs/2009JCAP...08..020M}
}

@article{Bernardeau:2008,
       author = {{Bernardeau}, Francis and {Crocce}, Mart{\'\i}n and {Scoccimarro}, Rom{\'a}n},
        title = "{Multipoint propagators in cosmological gravitational instability}",
      journal = {Phys. Rev. D},
         year = 2008,
        month = nov,
       volume = {78},
       number = {10},
          eid = {103521},
        pages = {103521},
          doi = {10.1103/PhysRevD.78.103521},
archivePrefix = {arXiv},
       eprint = {0806.2334},
 primaryClass = {astro-ph},
       adsurl = {https://ui.adsabs.harvard.edu/abs/2008PhRvD..78j3521B}
}

@article{Matsubara:2008b,
       author = {{Matsubara}, Takahiko},
        title = "{Nonlinear perturbation theory with halo bias and redshift-space distortions via the Lagrangian picture}",
      journal = {Phys. Rev. D},
         year = 2008,
        month = oct,
       volume = {78},
       number = {8},
          eid = {083519},
        pages = {083519},
          doi = {10.1103/PhysRevD.78.083519},
archivePrefix = {arXiv},
       eprint = {0807.1733},
 primaryClass = {astro-ph},
       adsurl = {https://ui.adsabs.harvard.edu/abs/2008PhRvD..78h3519M}
}

@article{Matsubara:2008a,
       author = {{Matsubara}, Takahiko},
        title = "{Resumming cosmological perturbations via the Lagrangian picture: One-loop results in real space and in redshift space}",
      journal = {Phys. Rev. D},
         year = 2008,
        month = mar,
       volume = {77},
       number = {6},
          eid = {063530},
        pages = {063530},
          doi = {10.1103/PhysRevD.77.063530},
archivePrefix = {arXiv},
       eprint = {0711.2521},
 primaryClass = {astro-ph},
       adsurl = {https://ui.adsabs.harvard.edu/abs/2008PhRvD..77f3530M}
}

@article{Guzzo:2008,
       author = {{Guzzo}, L. and {Pierleoni}, M. and {Meneux}, B. and {Branchini}, E. and {Le F{\`e}vre}, O. and {Marinoni}, C. and {Garilli}, B. and {Blaizot}, J. and {De Lucia}, G. and {Pollo}, A. and {McCracken}, H.~J. and {Bottini}, D. and {Le Brun}, V. and {Maccagni}, D. and {Picat}, J.~P. and {Scaramella}, R. and {Scodeggio}, M. and {Tresse}, L. and {Vettolani}, G. and {Zanichelli}, A. and {Adami}, C. and {Arnouts}, S. and {Bardelli}, S. and {Bolzonella}, M. and {Bongiorno}, A. and {Cappi}, A. and {Charlot}, S. and {Ciliegi}, P. and {Contini}, T. and {Cucciati}, O. and {de la Torre}, S. and {Dolag}, K. and {Foucaud}, S. and {Franzetti}, P. and {Gavignaud}, I. and {Ilbert}, O. and {Iovino}, A. and {Lamareille}, F. and {Marano}, B. and {Mazure}, A. and {Memeo}, P. and {Merighi}, R. and {Moscardini}, L. and {Paltani}, S. and {Pell{\`o}}, R. and {Perez-Montero}, E. and {Pozzetti}, L. and {Radovich}, M. and {Vergani}, D. and {Zamorani}, G. and {Zucca}, E.},
        title = "{A test of the nature of cosmic acceleration using galaxy redshift distortions}",
      journal = {Nature},
         year = 2008,
        month = jan,
       volume = {451},
       number = {7178},
        pages = {541-544},
          doi = {10.1038/nature06555},
archivePrefix = {arXiv},
       eprint = {0802.1944},
 primaryClass = {astro-ph},
       adsurl = {https://ui.adsabs.harvard.edu/abs/2008Natur.451..541G}
}

@article{Crocce:2008,
       author = {{Crocce}, Mart{\'\i}n and {Scoccimarro}, Rom{\'a}n},
        title = "{Nonlinear evolution of baryon acoustic oscillations}",
      journal = {Phys. Rev. D},
         year = 2008,
        month = jan,
       volume = {77},
       number = {2},
          eid = {023533},
        pages = {023533},
          doi = {10.1103/PhysRevD.77.023533},
archivePrefix = {arXiv},
       eprint = {0704.2783},
 primaryClass = {astro-ph},
       adsurl = {https://ui.adsabs.harvard.edu/abs/2008PhRvD..77b3533C}
}

@article{Eisenstein:2007,
       author = {{Eisenstein}, Daniel J. and {Seo}, Hee-Jong and {White}, Martin},
        title = "{On the Robustness of the Acoustic Scale in the Low-Redshift Clustering of Matter}",
      journal = {ApJ},
         year = 2007,
        month = aug,
       volume = {664},
       number = {2},
        pages = {660-674},
          doi = {10.1086/518755},
archivePrefix = {arXiv},
       eprint = {astro-ph/0604361},
 primaryClass = {astro-ph},
       adsurl = {https://ui.adsabs.harvard.edu/abs/2007ApJ...664..660E}
}

@article{Hartlap:2006,
       author = {{Hartlap}, J. and {Simon}, P. and {Schneider}, P.},
        title = "{Why your model parameter confidences might be too optimistic. Unbiased estimation of the inverse covariance matrix}",
      journal = {A\&A},
         year = 2007,
        month = mar,
       volume = {464},
       number = {1},
        pages = {399-404},
          doi = {10.1051/0004-6361:20066170},
archivePrefix = {arXiv},
       eprint = {astro-ph/0608064},
 primaryClass = {astro-ph},
       adsurl = {https://ui.adsabs.harvard.edu/abs/2007A&A...464..399H}
}

@article{Sefusatti:2006,
       author = {{Sefusatti}, Emiliano and {Crocce}, Mart{\'\i}n and {Pueblas}, Sebasti{\'a}n and {Scoccimarro}, Rom{\'a}n},
        title = "{Cosmology and the bispectrum}",
      journal = {Phys. Rev. D},
         year = 2006,
        month = jul,
       volume = {74},
       number = {2},
          eid = {023522},
        pages = {023522},
          doi = {10.1103/PhysRevD.74.023522},
archivePrefix = {arXiv},
       eprint = {astro-ph/0604505},
 primaryClass = {astro-ph},
       adsurl = {https://ui.adsabs.harvard.edu/abs/2006PhRvD..74b3522S}
}

@article{Crocce:2006,
       author = {{Crocce}, Mart{\'\i}n and {Scoccimarro}, Rom{\'a}n},
        title = "{Renormalized cosmological perturbation theory}",
      journal = {Phys. Rev. D},
         year = 2006,
        month = mar,
       volume = {73},
       number = {6},
          eid = {063519},
        pages = {063519},
          doi = {10.1103/PhysRevD.73.063519},
archivePrefix = {arXiv},
       eprint = {astro-ph/0509418},
 primaryClass = {astro-ph},
       adsurl = {https://ui.adsabs.harvard.edu/abs/2006PhRvD..73f3519C}
}

@article{Springel:2005,
       author = {{Springel}, Volker},
        title = "{The cosmological simulation code GADGET-2}",
      journal = {MNRAS},
         year = 2005,
        month = dec,
       volume = {364},
       number = {4},
        pages = {1105-1134},
          doi = {10.1111/j.1365-2966.2005.09655.x},
archivePrefix = {arXiv},
       eprint = {astro-ph/0505010},
 primaryClass = {astro-ph},
       adsurl = {https://ui.adsabs.harvard.edu/abs/2005MNRAS.364.1105S}
}

@article{Linder:2005,
       author = {{Linder}, Eric V.},
        title = "{Cosmic growth history and expansion history}",
      journal = {Phys. Rev. D},
         year = 2005,
        month = aug,
       volume = {72},
       number = {4},
          eid = {043529},
        pages = {043529},
          doi = {10.1103/PhysRevD.72.043529},
archivePrefix = {arXiv},
       eprint = {astro-ph/0507263},
 primaryClass = {astro-ph},
       adsurl = {https://ui.adsabs.harvard.edu/abs/2005PhRvD..72d3529L}
}

@article{Hahn:2005,
       author = {{Hahn}, T.},
        title = "{CUBA{\textemdash}a library for multidimensional numerical integration}",
      journal = {Computer Physics Communications},
         year = 2005,
        month = jun,
       volume = {168},
       number = {2},
        pages = {78-95},
          doi = {10.1016/j.cpc.2005.01.010},
archivePrefix = {arXiv},
       eprint = {hep-ph/0404043},
 primaryClass = {hep-ph},
       adsurl = {https://ui.adsabs.harvard.edu/abs/2005CoPhC.168...78H}
}

@article{Gorski:2005,
       author = {{G{\'o}rski}, K.~M. and {Hivon}, E. and {Banday}, A.~J. and {Wandelt}, B.~D. and {Hansen}, F.~K. and {Reinecke}, M. and {Bartelmann}, M.},
        title = "{HEALPix: A Framework for High-Resolution Discretization and Fast Analysis of Data Distributed on the Sphere}",
      journal = {ApJ},
         year = 2005,
        month = apr,
       volume = {622},
       number = {2},
        pages = {759-771},
          doi = {10.1086/427976},
archivePrefix = {arXiv},
       eprint = {astro-ph/0409513},
 primaryClass = {astro-ph},
       adsurl = {https://ui.adsabs.harvard.edu/abs/2005ApJ...622..759G}
}

@article{Szapudi:2004,
       author = {{Szapudi}, Istv{\'a}n},
        title = "{Three-Point Statistics from a New Perspective}",
      journal = {ApJL},
         year = 2004,
        month = apr,
       volume = {605},
       number = {2},
        pages = {L89-L92},
          doi = {10.1086/420894},
archivePrefix = {arXiv},
       eprint = {astro-ph/0404476},
 primaryClass = {astro-ph},
       adsurl = {https://ui.adsabs.harvard.edu/abs/2004ApJ...605L..89S}
}

@article{Bernardeau:2002,
       author = {{Bernardeau}, F. and {Colombi}, S. and {Gazta{\~n}aga}, E. and {Scoccimarro}, R.},
        title = "{Large-scale structure of the Universe and cosmological perturbation theory}",
      journal = {Phys. Rep.},
         year = 2002,
        month = sep,
       volume = {367},
       number = {1-3},
        pages = {1-248},
          doi = {10.1016/S0370-1573(02)00135-7},
archivePrefix = {arXiv},
       eprint = {astro-ph/0112551},
 primaryClass = {astro-ph},
       adsurl = {https://ui.adsabs.harvard.edu/abs/2002PhR...367....1B}
}

@article{Hamilton:2000,
       author = {{Hamilton}, A.~J.~S.},
        title = "{Uncorrelated modes of the non-linear power spectrum}",
      journal = {MNRAS},
         year = 2000,
        month = feb,
       volume = {312},
       number = {2},
        pages = {257-284},
          doi = {10.1046/j.1365-8711.2000.03071.x},
archivePrefix = {arXiv},
       eprint = {astro-ph/9905191},
 primaryClass = {astro-ph},
       adsurl = {https://ui.adsabs.harvard.edu/abs/2000MNRAS.312..257H}
}

@article{Scoccimarro:1999,
       author = {{Scoccimarro}, Rom{\'a}n and {Couchman}, H.~M.~P. and {Frieman}, Joshua A.},
        title = "{The Bispectrum as a Signature of Gravitational Instability in Redshift Space}",
      journal = {ApJ},
         year = 1999,
        month = jun,
       volume = {517},
       number = {2},
        pages = {531-540},
          doi = {10.1086/307220},
archivePrefix = {arXiv},
       eprint = {astro-ph/9808305},
 primaryClass = {astro-ph},
       adsurl = {https://ui.adsabs.harvard.edu/abs/1999ApJ...517..531S}
}

@article{Szapudi:1998,
       author = {{Szapudi}, Istv{\'a}n and {Szalay}, Alexander S.},
        title = "{A New Class of Estimators for the N-Point Correlations}",
      journal = {ApJL},
         year = 1998,
        month = feb,
       volume = {494},
       number = {1},
        pages = {L41-L44},
          doi = {10.1086/311146},
       adsurl = {https://ui.adsabs.harvard.edu/abs/1998ApJ...494L..41S}
}

@INPROCEEDINGS{Hamilton:1998,
       author = {{Hamilton}, A.~J.~S.},
        title = "{Linear Redshift Distortions: a Review}",
    booktitle = {The Evolving Universe},
         year = 1998,
       editor = {{Hamilton}, Donald},
       series = {Astrophysics and Space Science Library},
       volume = {231},
        month = jan,
        pages = {185},
          doi = {10.1007/978-94-011-4960-0_17},
archivePrefix = {arXiv},
       eprint = {astro-ph/9708102},
 primaryClass = {astro-ph},
       adsurl = {https://ui.adsabs.harvard.edu/abs/1998ASSL..231..185H}
}

@book{Varshalovich:1988,
       author = {{Varshalovich}, D.~A. and {Moskalev}, A.~N. and {Khersonskii}, V.~K.},
        title = "{Quantum Theory of Angular Momentum}",
         year = 1988,
        publisher="{World Scientific}",
          doi = {10.1142/0270},
       adsurl = {https://ui.adsabs.harvard.edu/abs/1988qtam.book.....V}
}

@article{Kaiser:1987,
       author = {{Kaiser}, Nick},
        title = "{Clustering in real space and in redshift space}",
      journal = {MNRAS},
         year = 1987,
        month = jul,
       volume = {227},
        pages = {1-21},
          doi = {10.1093/mnras/227.1.1},
       adsurl = {https://ui.adsabs.harvard.edu/abs/1987MNRAS.227....1K}
}

@article{Linde:1982,
       author = {{Linde}, A.~D.},
        title = "{A new inflationary universe scenario: A possible solution of the horizon, flatness, homogeneity, isotropy and primordial monopole problems}",
      journal = {Physics Letters B},
         year = 1982,
        month = feb,
       volume = {108},
       number = {6},
        pages = {389-393},
          doi = {10.1016/0370-2693(82)91219-9},
       adsurl = {https://ui.adsabs.harvard.edu/abs/1982PhLB..108..389L}
}

@article{Sato:1981,
       author = {{Sato}, K.},
        title = "{First-order phase transition of a vacuum and the expansion of the Universe}",
      journal = {MNRAS},
         year = 1981,
        month = may,
       volume = {195},
        pages = {467-479},
          doi = {10.1093/mnras/195.3.467},
       adsurl = {https://ui.adsabs.harvard.edu/abs/1981MNRAS.195..467S}
}

@article{Guth:1980,
       author = {{Guth}, Alan H.},
        title = "{Inflationary universe: A possible solution to the horizon and flatness problems}",
      journal = {Phys. Rev. D},
         year = 1981,
        month = jan,
       volume = {23},
       number = {2},
        pages = {347-356},
          doi = {10.1103/PhysRevD.23.347},
       adsurl = {https://ui.adsabs.harvard.edu/abs/1981PhRvD..23..347G}
}

@article{Starobinsky:1980,
       author = {{Starobinsky}, A.~A.},
        title = "{A new type of isotropic cosmological models without singularity}",
      journal = {Physics Letters B},
         year = 1980,
        month = mar,
       volume = {91},
       number = {1},
        pages = {99-102},
          doi = {10.1016/0370-2693(80)90670-X},
       adsurl = {https://ui.adsabs.harvard.edu/abs/1980PhLB...91...99S}
}

@article{Alcock:1979,
       author = {{Alcock}, C. and {Paczynski}, B.},
        title = "{An evolution free test for non-zero cosmological constant}",
      journal = {Nature},
         year = 1979,
        month = oct,
       volume = {281},
        pages = {358},
          doi = {10.1038/281358a0},
       adsurl = {https://ui.adsabs.harvard.edu/abs/1979Natur.281..358A}
}

@article{Jackson:1972,
       author = {{Jackson}, J.~C.},
        title = "{A critique of Rees's theory of primordial gravitational radiation}",
      journal = {MNRAS},
         year = 1972,
        month = jan,
       volume = {156},
        pages = {1P},
          doi = {10.1093/mnras/156.1.1P},
archivePrefix = {arXiv},
       eprint = {0810.3908},
 primaryClass = {astro-ph},
       adsurl = {https://ui.adsabs.harvard.edu/abs/1972MNRAS.156P...1J}
}

@article{Moresco:2021,
       author = {{Moresco}, Michele and {Veropalumbo}, Alfonso and {Marulli}, Federico and {Moscardini}, Lauro and {Cimatti}, Andrea},
        title = "{C$^{3}$: Cluster Clustering Cosmology. II. First Detection of the Baryon Acoustic Oscillations Peak in the Three-point Correlation Function of Galaxy Clusters}",
      journal = {ApJ},
         year = 2021,
        month = oct,
       volume = {919},
       number = {2},
          eid = {144},
        pages = {144},
          doi = {10.3847/1538-4357/ac10c9},
archivePrefix = {arXiv},
       eprint = {2011.04665},
 primaryClass = {astro-ph.CO},
       adsurl = {https://ui.adsabs.harvard.edu/abs/2021ApJ...919..144M}
}

@article{Kuruvilla:2020,
       author = {{Kuruvilla}, Joseph and {Porciani}, Cristiano},
        title = "{The n-point streaming model: how velocities shape correlation functions in redshift space}",
      journal = {JCAP},
         year = 2020,
        month = jul,
       volume = {2020},
       number = {7},
          eid = {043},
        pages = {043},
          doi = {10.1088/1475-7516/2020/07/043},
archivePrefix = {arXiv},
       eprint = {2005.05331},
 primaryClass = {astro-ph.CO},
       adsurl = {https://ui.adsabs.harvard.edu/abs/2020JCAP...07..043K}
}

@ARTICLE{Moresco:2014,
       author = {{Moresco}, Michele and {Marulli}, Federico and {Baldi}, Marco and {Moscardini}, Lauro and {Cimatti}, Andrea},
        title = "{Disentangling interacting dark energy cosmologies with the three-point correlation function}",
      journal = {MNRAS},
         year = 2014,
        month = oct,
       volume = {443},
       number = {4},
        pages = {2874-2886},
          doi = {10.1093/mnras/stu1359},
archivePrefix = {arXiv},
       eprint = {1312.4530},
 primaryClass = {astro-ph.CO},
       adsurl = {https://ui.adsabs.harvard.edu/abs/2014MNRAS.443.2874M}
}

@article{Gaztanaga:2005,
       author = {{Gazta{\~n}aga}, E. and {Norberg}, P. and {Baugh}, C.~M. and {Croton}, D.~J.},
        title = "{Statistical analysis of galaxy surveys - II. The three-point galaxy correlation function measured from the 2dFGRS}",
      journal = {MNRAS},
         year = 2005,
        month = dec,
       volume = {364},
       number = {2},
        pages = {620-634},
          doi = {10.1111/j.1365-2966.2005.09583.x},
archivePrefix = {arXiv},
       eprint = {astro-ph/0506249},
 primaryClass = {astro-ph},
       adsurl = {https://ui.adsabs.harvard.edu/abs/2005MNRAS.364..620G}
}

@article{Pan:2005,
       author = {{Pan}, Jun and {Szapudi}, Istv{\'a}n},
        title = "{The monopole moment of the three-point correlation function of the two-degree Field Galaxy Redshift Survey}",
      journal = {MNRAS},
         year = 2005,
        month = oct,
       volume = {362},
       number = {4},
        pages = {1363-1370},
          doi = {10.1111/j.1365-2966.2005.09407.x},
archivePrefix = {arXiv},
       eprint = {astro-ph/0505422},
 primaryClass = {astro-ph},
       adsurl = {https://ui.adsabs.harvard.edu/abs/2005MNRAS.362.1363P}
}

@article{McBride:2011,
       author = {{McBride}, Cameron K. and {Connolly}, Andrew J. and {Gardner}, Jeffrey P. and {Scranton}, Ryan and {Newman}, Jeffrey A. and {Scoccimarro}, Rom{\'a}n and {Zehavi}, Idit and {Schneider}, Donald P.},
        title = "{Three-point Correlation Functions of SDSS Galaxies: Luminosity and Color Dependence in Redshift and Projected Space}",
      journal = {ApJ},
         year = 2011,
        month = jan,
       volume = {726},
       number = {1},
          eid = {13},
        pages = {13},
          doi = {10.1088/0004-637X/726/1/13},
archivePrefix = {arXiv},
       eprint = {1007.2414},
 primaryClass = {astro-ph.CO},
       adsurl = {https://ui.adsabs.harvard.edu/abs/2011ApJ...726...13M}
}

@ARTICLE{Moresco:2017,
       author = {{Moresco}, M. and {Marulli}, F. and {Moscardini}, L. and {Branchini}, E. and {Cappi}, A. and {Davidzon}, I. and {Granett}, B.~R. and {de la Torre}, S. and {Guzzo}, L. and {Abbas}, U. and {Adami}, C. and {Arnouts}, S. and {Bel}, J. and {Bolzonella}, M. and {Bottini}, D. and {Carbone}, C. and {Coupon}, J. and {Cucciati}, O. and {De Lucia}, G. and {Franzetti}, P. and {Fritz}, A. and {Fumana}, M. and {Garilli}, B. and {Ilbert}, O. and {Iovino}, A. and {Krywult}, J. and {Le Brun}, V. and {Le F{\`e}vre}, O. and {Ma{\l}ek}, K. and {McCracken}, H.~J. and {Polletta}, M. and {Pollo}, A. and {Scodeggio}, M. and {Tasca}, L.~A.~M. and {Tojeiro}, R. and {Vergani}, D. and {Zanichelli}, A.},
        title = "{The VIMOS Public Extragalactic Redshift Survey (VIPERS) . Exploring the dependence of the three-point correlation function on stellar mass and luminosity at $0.5 <z < 1.1$}",
      journal = {A\& A},
         year = 2017,
        month = aug,
       volume = {604},
          eid = {A133},
        pages = {A133},
          doi = {10.1051/0004-6361/201628589},
archivePrefix = {arXiv},
       eprint = {1603.08924},
 primaryClass = {astro-ph.CO},
       adsurl = {https://ui.adsabs.harvard.edu/abs/2017A&A...604A.133M}
}

@article{Padmanabhan:2008,
       author = {{Padmanabhan}, Nikhil and {White}, Martin},
        title = "{Constraining anisotropic baryon oscillations}",
      journal = {Phys. Rev. D},
         year = 2008,
        month = jun,
       volume = {77},
       number = {12},
          eid = {123540},
        pages = {123540},
          doi = {10.1103/PhysRevD.77.123540},
archivePrefix = {arXiv},
       eprint = {0804.0799},
 primaryClass = {astro-ph},
       adsurl = {https://ui.adsabs.harvard.edu/abs/2008PhRvD..77l3540P}
}

@book{Anderson:2003,
  title={An introduction to multivariate statistical analysis},
  author={Anderson, Theodore Wilbur},
  year={2003},
  publisher={Wiley New York}
}

@article{Peacock:2001,
       author = {{Peacock}, John A. and {Cole}, Shaun and {Norberg}, Peder and {Baugh}, Carlton M. and {Bland-Hawthorn}, Joss and {Bridges}, Terry and {Cannon}, Russell D. and {Colless}, Matthew and {Collins}, Chris and {Couch}, Warrick and {Dalton}, Gavin and {Deeley}, Kathryn and {De Propris}, Roberto and {Driver}, Simon P. and {Efstathiou}, George and {Ellis}, Richard S. and {Frenk}, Carlos S. and {Glazebrook}, Karl and {Jackson}, Carole and {Lahav}, Ofer and {Lewis}, Ian and {Lumsden}, Stuart and {Maddox}, Steve and {Percival}, Will J. and {Peterson}, Bruce A. and {Price}, Ian and {Sutherland}, Will and {Taylor}, Keith},
        title = "{A measurement of the cosmological mass density from clustering in the 2dF Galaxy Redshift Survey}",
      journal = {Nature},
         year = 2001,
        month = mar,
       volume = {410},
       number = {6825},
        pages = {169-173},
          doi = {10.1038/35065528},
archivePrefix = {arXiv},
       eprint = {astro-ph/0103143},
 primaryClass = {astro-ph},
       adsurl = {https://ui.adsabs.harvard.edu/abs/2001Natur.410..169P}
}

@article{Blot:2019,
       author = {{Blot}, Linda and {Crocce}, Martin and {Sefusatti}, Emiliano and {Lippich}, Martha and {S{\'a}nchez}, Ariel G. and {Colavincenzo}, Manuel and {Monaco}, Pierluigi and {Alvarez}, Marcelo A. and {Agrawal}, Aniket and {Avila}, Santiago and {Balaguera-Antol{\'\i}nez}, Andr{\'e}s and {Bond}, Richard and {Codis}, Sandrine and {Dalla Vecchia}, Claudio and {Dorta}, Antonio and {Fosalba}, Pablo and {Izard}, Albert and {Kitaura}, Francisco-Shu and {Pellejero-Ibanez}, Marcos and {Stein}, George and {Vakili}, Mohammadjavad and {Yepes}, Gustavo},
        title = "{Comparing approximate methods for mock catalogues and covariance matrices II: power spectrum multipoles}",
      journal = {MNRAS},
         year = 2019,
        month = may,
       volume = {485},
       number = {2},
        pages = {2806-2824},
          doi = {10.1093/mnras/stz507},
archivePrefix = {arXiv},
       eprint = {1806.09497},
 primaryClass = {astro-ph.CO},
       adsurl = {https://ui.adsabs.harvard.edu/abs/2019MNRAS.485.2806B}
}

@article{Colavincenzo:2019,
       author = {{Colavincenzo}, Manuel and {Sefusatti}, Emiliano and {Monaco}, Pierluigi and {Blot}, Linda and {Crocce}, Martin and {Lippich}, Martha and {S{\'a}nchez}, Ariel G. and {Alvarez}, Marcelo A. and {Agrawal}, Aniket and {Avila}, Santiago and {Balaguera-Antol{\'\i}nez}, Andr{\'e}s and {Bond}, Richard and {Codis}, Sandrine and {Dalla Vecchia}, Claudio and {Dorta}, Antonio and {Fosalba}, Pablo and {Izard}, Albert and {Kitaura}, Francisco-Shu and {Pellejero-Ibanez}, Marcos and {Stein}, George and {Vakili}, Mohammadjavad and {Yepes}, Gustavo},
        title = "{Comparing approximate methods for mock catalogues and covariance matrices - III: bispectrum}",
      journal = {MNRAS},
         year = 2019,
        month = feb,
       volume = {482},
       number = {4},
        pages = {4883-4905},
          doi = {10.1093/mnras/sty2964},
archivePrefix = {arXiv},
       eprint = {1806.09499},
 primaryClass = {astro-ph.CO},
       adsurl = {https://ui.adsabs.harvard.edu/abs/2019MNRAS.482.4883C}
}

@article{Buchner:2019,
       author = {{Buchner}, Johannes},
        title = "{Collaborative Nested Sampling: Big Data versus Complex Physical Models}",
      journal = {PASP},
         year = 2019,
        month = oct,
       volume = {131},
       number = {1004},
        pages = {108005},
          doi = {10.1088/1538-3873/aae7fc},
archivePrefix = {arXiv},
       eprint = {1707.04476},
 primaryClass = {stat.CO},
       adsurl = {https://ui.adsabs.harvard.edu/abs/2019PASP..131j8005B}
}

@article{Buchner:2014,
       author = {{Buchner}, J. and {Georgakakis}, A. and {Nandra}, K. and {Hsu}, L. and {Rangel}, C. and {Brightman}, M. and {Merloni}, A. and {Salvato}, M. and {Donley}, J. and {Kocevski}, D.},
        title = "{X-ray spectral modelling of the AGN obscuring region in the CDFS: Bayesian model selection and catalogue}",
      journal = {AAP},
         year = 2014,
        month = apr,
       volume = {564},
          eid = {A125},
        pages = {A125},
          doi = {10.1051/0004-6361/201322971},
archivePrefix = {arXiv},
       eprint = {1402.0004},
 primaryClass = {astro-ph.HE},
       adsurl = {https://ui.adsabs.harvard.edu/abs/2014A&A...564A.125B}
}

@article{Buchner:2021,
       author = {{Buchner}, Johannes},
        title = "{UltraNest - a robust, general purpose Bayesian inference engine}",
      journal = {The Journal of Open Source Software},
         year = 2021,
        month = apr,
       volume = {6},
       number = {60},
          eid = {3001},
        pages = {3001},
          doi = {10.21105/joss.03001},
archivePrefix = {arXiv},
       eprint = {2101.09604},
 primaryClass = {stat.CO},
       adsurl = {https://ui.adsabs.harvard.edu/abs/2021JOSS....6.3001B}
}

@article{Peebles:1974,
       author = {{Peebles}, P.~J.~E. and {Hauser}, M.~G.},
        title = "{Statistical Analysis of Catalogs of Extragalactic Objects. III. The Shane-Wirtanen and Zwicky Catalogs}",
      journal = {ApJs},
         year = 1974,
        month = nov,
       volume = {28},
        pages = {19},
          doi = {10.1086/190308},
       adsurl = {https://ui.adsabs.harvard.edu/abs/1974ApJS...28...19P}
}

@article{Gualdi:2020,
       author = {{Gualdi}, D. and {Verde}, L.},
        title = "{Galaxy redshift-space bispectrum: the importance of being anisotropic}",
      journal = {JCAP},
         year = 2020,
        month = jun,
       volume = {2020},
       number = {6},
          eid = {041},
        pages = {041},
          doi = {10.1088/1475-7516/2020/06/041},
archivePrefix = {arXiv},
       eprint = {2003.12075},
 primaryClass = {astro-ph.CO},
       adsurl = {https://ui.adsabs.harvard.edu/abs/2020JCAP...06..041G}
}

@article{Colless:2001,
       author = {{Colless}, Matthew and {Dalton}, Gavin and {Maddox}, Steve and {Sutherland}, Will and {Norberg}, Peder and {Cole}, Shaun and {Bland-Hawthorn}, Joss and {Bridges}, Terry and {Cannon}, Russell and {Collins}, Chris and {Couch}, Warrick and {Cross}, Nicholas and {Deeley}, Kathryn and {De Propris}, Roberto and {Driver}, Simon P. and {Efstathiou}, George and {Ellis}, Richard S. and {Frenk}, Carlos S. and {Glazebrook}, Karl and {Jackson}, Carole and {Lahav}, Ofer and {Lewis}, Ian and {Lumsden}, Stuart and {Madgwick}, Darren and {Peacock}, John A. and {Peterson}, Bruce A. and {Price}, Ian and {Seaborne}, Mark and {Taylor}, Keith},
        title = "{The 2dF Galaxy Redshift Survey: spectra and redshifts}",
      journal = {MNRAS},
         year = 2001,
        month = dec,
       volume = {328},
       number = {4},
        pages = {1039-1063},
          doi = {10.1046/j.1365-8711.2001.04902.x},
archivePrefix = {arXiv},
       eprint = {astro-ph/0106498},
 primaryClass = {astro-ph},
       adsurl = {https://ui.adsabs.harvard.edu/abs/2001MNRAS.328.1039C}
}

@article{Garilli:2008,
       author = {{Garilli}, B. and {Le F{\`e}vre}, O. and {Guzzo}, L. and {Maccagni}, D. and {Le Brun}, V. and {de la Torre}, S. and {Meneux}, B. and {Tresse}, L. and {Franzetti}, P. and {Zamorani}, G. and {Zanichelli}, A. and {Gregorini}, L. and {Vergani}, D. and {Bottini}, D. and {Scaramella}, R. and {Scodeggio}, M. and {Vettolani}, G. and {Adami}, C. and {Arnouts}, S. and {Bardelli}, S. and {Bolzonella}, M. and {Cappi}, A. and {Charlot}, S. and {Ciliegi}, P. and {Contini}, T. and {Foucaud}, S. and {Gavignaud}, I. and {Ilbert}, O. and {Iovino}, A. and {Lamareille}, F. and {McCracken}, H.~J. and {Marano}, B. and {Marinoni}, C. and {Mazure}, A. and {Merighi}, R. and {Paltani}, S. and {Pell{\`o}}, R. and {Pollo}, A. and {Pozzetti}, L. and {Radovich}, M. and {Zucca}, E. and {Blaizot}, J. and {Bongiorno}, A. and {Cucciati}, O. and {Mellier}, Y. and {Moreau}, C. and {Paioro}, L.},
        title = "{The Vimos VLT deep survey. Global properties of 20,000 galaxies in the I$_{AB}$ < 22.5 WIDE survey}",
      journal = {A\&A},
         year = 2008,
        month = aug,
       volume = {486},
       number = {3},
        pages = {683-695},
          doi = {10.1051/0004-6361:20078878},
archivePrefix = {arXiv},
       eprint = {0804.4568},
 primaryClass = {astro-ph},
       adsurl = {https://ui.adsabs.harvard.edu/abs/2008A&A...486..683G}
}

\end{document}